\documentclass[aps,rmp,reprint,amsmath,amssymb,longbibliography
]{revtex4-1}

\usepackage{bm,graphicx,wasysym}

\usepackage{color}

\usepackage[%
    colorlinks=true,
    linkcolor=blue,
    citecolor=blue,
    urlcolor=blue
	]{hyperref}

\allowdisplaybreaks[3]

\newcommand{\ud} {\mathrm{d}}

\newcommand{\dd} {\delta}
\newcommand{\nd} {\nabla}
\newcommand{\pd} {\partial}
\newcommand{\bfx} {\mathbf{x}}

\newcommand{\bfp} {\mathbf{p}}

\newcommand{\mc} {\mathcal}

\newcommand{\mn} {{\mu\nu}}

\newcommand{\ab} {{\alpha\beta}}
\newcommand{\ai}{{\alpha}}

\newcommand{\ri}{{\rho}}

\newcommand{\ep}{{\epsilon}}

\renewcommand{\L}{\mathcal{L}}

\newcommand{\stu}{St\"uckelberg }

\renewcommand{\d}{\mathrm{d}}
\renewcommand{\mn}{_{\mu \nu}}

\newcommand{\lp}{\left(}
\newcommand{\rp}{\right)}

\newcommand{\mpl}{M_\text{Pl}}
\newcommand{\p}{\partial}
\newcommand{\ie}{\emph{i.e.}}
\newcommand{\eg}{\emph{e.g.}}

\newcommand{\Rv}{r_\text{V}}
\newcommand{\Rs}{r_\text{S}}
\newcommand{\lambdag}{{\lambdabar_g}}
\newcommand{\unit}[1]{\text{~#1}}
\newcommand{\unitn}[1]{\text{#1}}
\newcommand{\eqn}{Eq.~}

\renewcommand{\vec}[1]{\boldsymbol{#1}}
\newcommand{\headrule}{}
\newcommand{\midrule}{\hline}
\newcommand{\btrule}{\hline}
\newcommand{\linerule}{}
\newcommand{\fig}{Fig.~}
\newcommand{\Sec}{Section~}
\newcommand{\ttbound}[4]{$\mathbf{#1}$ & $\mathbf{#2}$ & #3 & #4}
\newcommand{\ntbound}[4]{$#1$ & $#2$ & #3 & #4}
\newcommand{\ptbound}[4]{$\mathit{#1}$ & $\mathit{#2}$ & #3 & #4}
\newcommand{\bound}[2]{m_g<#1\unit{eV},\quad \lambdag>#2\unit{km}}
\newcommand{\pbound}[2]{m_g<#1\unit{eV},\quad \lambdag>#2\unit{km}\quad({projected})}

\begin{document}

\hfill{Preprint for {\it Reviews of Modern Physics}}

\title{Graviton Mass Bounds}
\author{Claudia de Rham}
\email{Claudia.deRham@case.edu}
\author{J.~Tate Deskins}
\email{Jennings.Deskins@case.edu}
\author{Andrew J.~Tolley}
\email{Andrew.J.Tolley@case.edu}
\author{Shuang-Yong Zhou}
\email{Shuangyong.Zhou@case.edu}

\affiliation{CERCA, Department of Physics, Case Western Reserve University, 10900 Euclid Ave, Cleveland, OH 44106, USA}
\date{\today}

\begin{abstract}
	Recently, aLIGO has announced the first direct detections of gravitational waves, a direct manifestation of the propagating degrees of freedom of gravity. The detected signals GW150914 and GW151226 have been used to examine the basic properties of these gravitational degrees of freedom, particularly setting an upper bound on their mass. It is timely to review what the mass of these gravitational degrees of freedom means from the theoretical point of view, particularly taking into account the recent developments in constructing consistent massive gravity theories. Apart from the GW150914 mass bound, a few other observational bounds have been established from the effects of the Yukawa potential, modified dispersion relation and fifth force that are all induced when the fundamental gravitational degrees of freedom are massive. We review these different mass bounds and examine how they stand in the wake of recent theoretical developments and how they compare to the bound from GW150914.
\end{abstract}

\maketitle

\tableofcontents

\section{Introduction and Summary}

Whether the propagating degrees of freedom for gravity have mass is a fundamental question which has profound consequences for many areas of physics. As we enter the age of gravitational wave observatories, this question has become even more pertinent. Not only can gravitational waves impose a direct bound on the mass, but the mass may also be linked with the existence of new gravitational wave polarizations.

A mass for the graviton\footnote{Technically speaking gravitons are quantum particles and aLIGO has shown evidence only of classical coherent propagating fields. However, our terminology `graviton' reflects the generally accepted point of view implied by low energy quantum effective field theory that associated with every propagating field is a quantum particle. It is with this mindset that we will utilize the term graviton throughout, taking as given that the majority of stated constraints are really on the mass of the classical propagating modes.} may arise from either a pole (hard mass) or a resonance (soft mass). It is generally believed that the masslessness of the graviton is guaranteed by diffeomorphism invariance as in General Relativity. However, as pointed out by Schwinger \cite{PhysRev.125.397}, gauge invariance does not always imply masslessness. Quantum effects from other fields may give rise to a graviton mass without breaking diffeomorphism invariance, a mechanism which has been realized on spacetimes with a negative cosmological constant \cite{Porrati:2001gx}. In extra dimensional models in which the effective volume of the extra dimensions is infinite, the four dimensional massless graviton may no longer be normalizable, leading necessarily to an effectively massive theory. In these models a resonance graviton may arise as a metastable state localized on a brane; the most well known example is the Dvali--Gabadadze--Porrati (DGP) model \cite{Dvali:2000rv,Dvali:2000hr}. In such models the effectively massive graviton arises without explicit breaking of diffeomorphism symmetry just as in the Schwinger mechanism.

Alternatively, one may imagine that a mass arises through a gravitational analogue of the Higgs mechanism. To date, no such explicit gravitational Higgs mechanism in a Lorentz invariant theory is known\footnote{Several papers claiming to have a Higgs mechanism actually describe a \stu\ mechanism, \ie~only the effective theory around the spontaneously broken state. A model which can achieve both an unbroken and a broken vacuum has yet to be described.}. However, it is well known how to give the graviton a mass and how to encode the additional degrees of freedom through a \stu\ formalism which would be the low energy effective theory of any possible gravitational Higgs mechanism. These \stu\ or Goldstone mode low energy effective theories for both Lorentz invariant and Lorentz violating massive gravities are now well known (see \cite{deRham:2014zqa} for a recent review). Many phenomenological implications may be inferred even in the absence of a known Higgs mechanism or UV completion.

With the recent direct detections of gravitational waves GW150914 \cite{Abbott:2016blz} and GW151226 \cite{Abbott:2016nmj}, the question of the massiveness of the graviton has become even more interesting. The analysis of the phasing of the GW150914 waveform by aLIGO has constrained the graviton mass to $m_g<1.2\times10^{-22}\unit{eV}$ \cite{TheLIGOScientific:2016src}. As we will discuss in more detail later, this is not the strongest bound on the graviton mass, but it is certainly a very solid bound for several reasons. Firstly, it is largely independent of the details of the underlying massive gravity model, mainly relying on the dispersion relation being of the standard relativistic form for a massive particle. Furthermore, it is significantly different from the previous bounds on the graviton mass in the sense that it directly measures the propagating degrees of freedom of the helicity--2 modes, while the previous bounds measure the auxiliary effects due to the existence of the the helicity--2 modes, which are inevitably more model dependent. The addition of the analysis of the GW151226 waveform does not significantly improve this bound \cite{TheLIGOScientific:2016pea}. (See \cite{Yunes:2016jcc} for various theoretical physics Implications of GW150914 as well as GW151226.) However, it is projected that incoming gravitational wave experiments such as eLISA can significantly improve the graviton mass bound along the same line of attack.

In this paper, we discuss what a graviton mass means in the framework of the latest theoretical developments, and review how the bound from GW150914 fits with the bound from other observations and experiments.

In comparing different bounds on the mass of the graviton, it is important to understand both what is the environment and what is physically probed. We usually define the mass by the dispersion relation for fluctuations around Minkowski spacetime; we shall refer to this as the bare mass. However, the actual mass may depend on the environment. For instance, a cosmological background or the background of a heavy object such as a black hole can dress the mass of the graviton making the effective mass smaller or larger than the bare mass. In addition, the graviton mass could be explicitly dependent upon extra fields which give rise to additional temporal or spatial variations (\eg~\cite{D'Amico:2011jj,Huang:2012pe}). Thus while some tests may not appear as constraining when stated as a numerical bound, they can still provide a new window on the effective graviton mass in a specific environment or epoch of the Universe.
For instance we anticipate the effective mass in the early Universe to be much larger than the bare mass and so a naively weaker bound in that regime could ultimately be the stronger constraint.

There are also distinctions between whether the bound is being placed on a Lorentz invariant or Lorentz violating massive gravity model, especially since this may affect the existence of the  helicity--0 mode.
In Lorentz breaking theories the helicity--0 mode may be absent and most of the bounds due to fifth forces, or at least those arising from the helicity--0 mode, can be evaded. In Lorentz invariant theories the helicity--0 mode is necessarily present and its interactions lead to a Vainshtein mechanism that screens, or suppresses the effect, of that mode in most astrophysical systems \cite{Vainshtein:1972sx,Deffayet:2001uk,Babichev:2013usa}.
In some extensions and generalized theories of massive gravity, the interactions of the helicity--0 mode may be suppressed via other means which also weakens the fifth force bounds.

We do not explicitly discuss models of bi-gravity which introduce an additional massless graviton explicitly \cite{Hassan:2011zd}. These models transition between massive theories in which the massive graviton dominates the interaction between matter, and theories in which the massless graviton dominates. When the massless graviton does dominate, these theories can be thought of as General Relativity coupled to an exotic form of spin--2 matter.  Our interest is in the case where the principal carrier of the gravitational force is massive. In bi--gravity models this corresponds to a region where the effective Planck mass of the massless graviton, $M_f$, is significantly greater than the true Planck mass, $\mpl$. A genuine bi--gravity regime will be one where $M_f \sim \mpl$ so that both the massless and massive gravitons contribute comparably to the gravitational force. In this case bounds on the mass of the massive graviton are more difficult to disentangle and deserve a separate discussion. Similar arguments hold for multi--gravity models \cite{Hinterbichler:2012cn}. It may be easily shown that in the limit where the additional Planck masses $M_f^I \gg \mpl$, multi--gravity models reduce to massive gravity.
Extra dimensional models with heavy massive Kaluza--Klein graviton modes will not be discussed since such modes will not be the principal contributors to the gravitational force at large distances and such massive spin-2 states need to be very massive to avoid existing particle physics constraints.

Although in a given model some of the strongest constraints on the graviton mass come from a given theory's implications for cosmology and in particular large scale structure and late time evolution, the majority of these constraints are highly model dependent. For instance, ghost--free massive gravity\footnote{Ghost--free massive gravity is sometimes referred to as `dRGT' massive gravity in the literature and we keep that terminology for consistency with the rest of the literature.} \cite{deRham:2010kj} and the DGP model have very similar phenomenological behavior at solar system and astrophysical scales, but have fundamentally different cosmological behavior. For this reason we concentrate mainly on the more universal mass bounds which are largely common to all such models. Some previous work has been done to categorize different bounds on the mass of the graviton in \cite{Yagi:2009zz,Yagi:2016jml,Will:2014kxa,Goldhaber:2008xy,Agashe:2014kda}.

This paper is organized as follows: In \Sec\ref{sec:ReviewMG}, we review the current theoretical understanding of the mass of the graviton in a general context and in various specific massive gravity models. We highlight the van Dam--Veltman--Zakharov (vDVZ) discontinuity and its resolution via the Vainshtein mechanism in many models of nonlinear massive gravity. We then discuss in \Sec\ref{sec:models of MG} the different theories of massive gravity that have been introduced in the literature and the relation between those.
One aspect of massive gravity is that the gravitational potential typically has a Yukawa type fall--off at the graviton Compton wavelength. The bounds due to this phenomenon are reviewed in \Sec\ref{sec:YP}. The bounds due to the modified dispersion relation in the presence of a nonzero mass are then reviewed in \Sec\ref{sec:MDR}. The bound from GW150914 belongs to this category. Finally, the bounds due to tests of the fifth force are  reviewed in \Sec\ref{sec:FF}. These three types of bounds are quite clean in that they use the ``bare minimum'' information required for a consistent massive gravity theory and thus can mostly be treated as ``model-independent'' bounds on the graviton mass. We then conclude in \Sec\ref{sec:clu}.

See Table \ref{tab:massbounds} for a summary of the current and projected competitive bounds on the graviton mass, the details of which will be filled in gradually in the following sections. 

Throughout this review, we use the $(-+++)$ signature, define $\eta\mn$ to be the flat Minkowski metric, and work in units where the reduced Planck constant and the speed of light are set to  $c=\hbar=1$. In natural units, we have
\begin{equation}
	1\unit{eV}\sim  \frac{1}{2 \times 10^{-10}\unit{km}}  \, .
\end{equation}
We will also use the reduced Compton wavelength
\begin{equation}
	\lambdag=\frac{\hbar}{m_gc}\,,
\end{equation}
which we often simply call the Compton wavelength. We define the (reduced) Planck mass $\mpl=1/\sqrt{8\pi G}$, where $G$ is Newton's gravitational constant. The helicity--0 mode of massive gravity or the Galileon scalar is denoted as $\pi$.

\begin{table*}[ht]
	\caption{Current and projected bounds on the graviton mass and its reduced Compton wavelength from three classes of massive graviton effects, the details of which will be explained accordingly in the following sections. These three classes, specifically the first two, are largely independent of the assumed massive gravity models, making them in some sense more robust or specific than typical bounds or constraints obtained from the cosmological considerations. Cosmological bounds or constraints on the graviton mass (except for the projected, clean bound from the CMB B--modes utilizing the modified graviton dispersion relation, see below) are not listed in this table or reviewed in this paper. The masses reported are upper bounds and the reduced Compton wavelengths lower bounds. Bold entries are the most model independent and rigorous, normal type face are for those measured with current data and italic for projected measurements.}
	\begin{ruledtabular}
		\begin{tabular}{cccp{.7\textwidth}}
			\multicolumn{4}{c}{\textbf{Yukawa}}\\ \headrule
				$m_g\,(\unitn{eV})$ & $\lambdag\,(\unitn{km})$ & Eq. &\\ \midrule
					\ttbound{7.2 \times 10^{-23}}{2.8\times 10^{12}}{\eqref{boundmars}}{A 2$\sigma$ bound from the precession of Mercury \cite{Talmadge:1988qz,Will:1997bb}.}\\ \linerule
					\ntbound{6\times10^{-32}}{3\times 10^{21}}{\eqref{eqn:ykw_weak}}{A 1$\sigma$ bound from weak lensing of a cluster at $z=1.2$ \cite{Choudhury:2002pu}. Sensitive to the dark matter distribution and cosmological model.}\\ \linerule
					\ntbound{10^{-29}}{10^{19}}{\eqref{eqn:ykw_virial}}{From observations of gravitationally bound clusters of $0.5\unit{Mpc}$ \cite{Goldhaber:1974wg,Hare:1973px}. Sensitive to the dark matter distribution.}\\ \linerule
			\btrule
			\multicolumn{4}{c}{\textbf{Dispersion Relation}}\\ \headrule
				$m_g\,(\unitn{eV})$ & $\lambdag\,(\unitn{km})$ & Eq. &\\ \midrule
					\ttbound{1.2\times 10^{-22}}{1.7\times 10^{12}}{\eqref{aLIGObound}}{A 90\% confidence bound two 30 $M_\odot$ bh-bh merger (GW150916) \cite{Will:1997bb,TheLIGOScientific:2016src}.}\\ \linerule
					\ntbound{7.6 \times 10^{-20}}{2.6\times 10^9}{\eqref{eqn:mdr_pulsar}}{From pulsar timing of PSR B1913+16 and PSR B1534+12 \cite{Finn:2001qi}}.\\ \linerule
					\ptbound{10^{-30}}{10^{20}}{\eqref{eqn:mdr_cmb}}{Observations of power in B--mode polarization in CMB at low $\ell$ \cite{Gumrukcuoglu:2012wt,Dubovsky:2009xk,Raveri:2014eea}.}\\ \linerule
					\ptbound{10^{-26}}{10^{16}}{\eqref{eLISAbound}}{A $10^4$ to $10^7$ $M_\odot$ merger by eLISA type experiment \cite{Will:1997bb}}. \\ \linerule
					\ptbound{10^{-24}}{10^{14}}{\eqref{eqn:eLISA_IBWD}}{A dual messenger observation of IBWD by eLISA type experiment \cite{Larson:1999kg,Cutler:2002ef,Cooray:2003cv}.}\\ \linerule
					\ptbound{10^{-23}}{10^{13}}{\eqref{eqn:mdr_pta}}{Pulsar timing array of 100ns accuracy with 10 year observation \cite{Lee:2010cg}.}\\
					\ptbound{10^{-20}}{10^{10}}{\eqref{eqn:mdr_SNE}}{Dual messenger observation of SNe gamma ray burst and gravitational waves \cite{Nishizawa:2014zna}.}\\ \linerule
			\btrule
			\multicolumn{4}{c}{\textbf{Fifth Force}}\\ \headrule
				$m_g\,(\unitn{eV})$ & $\lambdag\,(\unitn{km})$ & Eq. &\\ \midrule
					\ntbound{10^{-32}}{10^{22}}{\eqref{eqn:llrDGP}}{From earth-moon precession for cubic Galileon theories \cite{Dvali:2002vf}.}\\ \linerule
					\ntbound{10^{-32}}{10^{22}}{\eqref{eqn:5dDGP_precess}}{From precession in full 5D DGP in the Solar System \cite{Gruzinov:2001hp,Lue:2002sw}.}\\ \linerule
					\ntbound{10^{-30}}{10^{20}}{\eqref{eqn:4gal_moon}}{From earth-moon precession for quartic Galileon theories (dRGT-like) \cite{deRham:2014zqa}.}\\ \linerule
					\ntbound{10^{-27}}{10^{17}}{\eqref{eqn:cuGal_pulsar}}{From PSR B1913+16 pulsar in cubic Galileon theories (DGP) \cite{deRham:2012fw}.}\\ \linerule
					\ptbound{10^{-33}}{10^{23}}{\eqref{eqn:drgt_weak}}{A prospective 4$\sigma$ bound from weak lensing on next-gen surveys \cite{Wyman:2011mp,Park:2014aga}. Sensitive to alternative DM halo profiles.}\\ \linerule
					\ptbound{10^{-32}}{10^{22}}{\eqref{eqn:ff_structure}}{Observations of altered structure formation from fifth force \cite{Khoury:2009tk,Wyman:2011mp,Zu:2013joa,Park:2014aga}. Sensitive to the particular theory of massive gravity.}\\
		\end{tabular}
	\end{ruledtabular}
	\label{tab:massbounds}
\end{table*}

\section{Massive Graviton}
\label{sec:ReviewMG}

We start by clarifying what is usually meant by the mass of the graviton and briefly review the generic physics behind models where the graviton has a mass in a largely model independent fashion. See \cite{deRham:2014zqa,Hinterbichler:2011tt} for a recent review and more detailed discussions on theoretical aspects of massive gravity. We will discuss three generic implications of the graviton being massive: the implications for the finite range of gravity, the dispersion relation, and the existence of the fifth force. The graviton mass has other implications (for instance on the evolution of the Universe, formation of structure, etc.) but as mentioned previously we will focus on these three effects as they are relatively model independent (especially for the dispersion relation).

\subsection{Degrees of Freedom}
\label{sec:dofs}	

Particles can be classified by the irreducible representations of the Wigner's little group of the spacetime symmetry group \cite{Weinberg:1995mt}. General Relativity is Lorentz invariant and described by a massless spin--2 particle around Minkowski space with two helicity--2 degrees of freedom (or polarizations, or modes). The structure of theories of massive gravity depends significantly on whether or not the theories are required to be Lorentz invariant\footnote{Translation invariance is usually implicitly assumed, but not always.}. Resonances, which are often linked to large extra dimensions, have a continuous spectrum of degrees of freedom.

\subsubsection{Poincar\'e Invariant}

\begin{figure}[ht]
	\centering
	\includegraphics[width=\columnwidth]{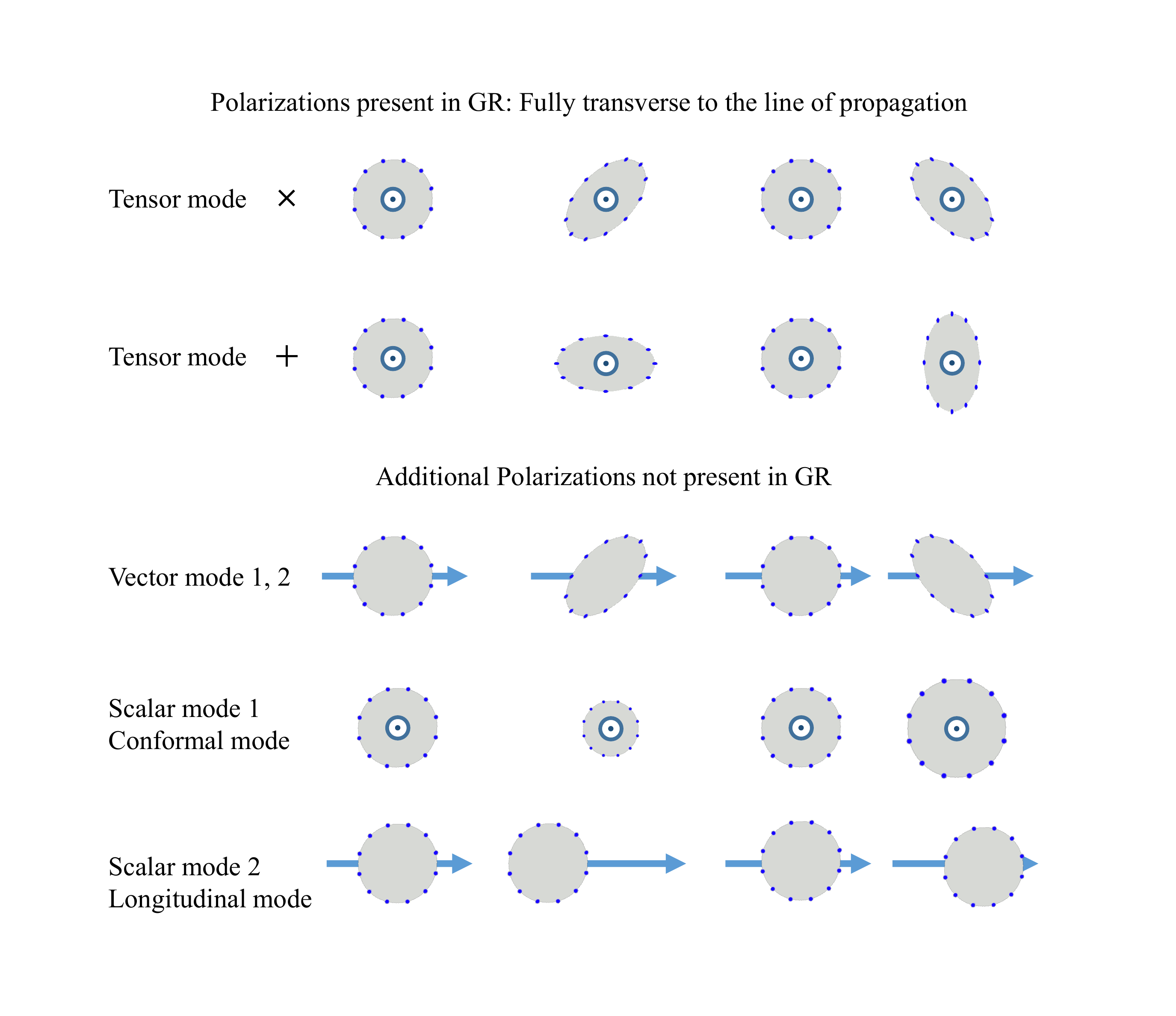}
	\caption{The six modes for a massive spin--2 field. The two tensor modes and the scalar conformal mode are propagating out of the page; the two vector modes and the longitudinal scalar mode are propagating to the right (taken from \cite{deRham:2014zqa}).}
	\label{fig:gmodes}	
\end{figure}

Assuming Lorentz invariance, or more precisely full Poincar\'e invariance, a massive graviton furnishes the spin--2 representation of SU(2), the little group of the Poincar\'e group, which has five degrees of freedom (two helicity--2, two helicity--1 and one helicity--0; see \Sec\ref{stufields}). This is three more than its massless counterpart. On other hand gravitational waves could have up to six polarizations (see Fig.~\ref{fig:gmodes}). In General Relativity only the two tensor modes, which are the polarizations strictly transverse to the line of propagation of the gravitational waves are allowed.

The simplest class of Poincar\'e invariant massive gravity models is to modify General Relativity by adding a graviton potential to the action. This potential consists of terms involving the metric and a reference Minkowski metric but without derivatives. A reference metric is necessary to construct a graviton potential for a local nonlinear massive gravity theory, and to preserve full Poincar\'e invariance the unique choice is the Minkowski metric.

We only consider this class of massive gravity models because modifying the kinetic structure by adding derivative terms will introduce ghost instabilities \cite{deRham:2013tfa,deRham:2015rxa,deRham:2015cha,Matas:2015qxa}.
A ghost is a field with negative kinetic energy. The existence of such a mode would make the vacuum extremely unstable, as the vacuum would then be able to decay into normal particles with positive energy and ghost particles with negative energy. In reality, modified kinetic terms may be added provided that the mass of the ghost is at or above the cutoff of the low energy effective theory, but this necessary implies the contributions of such terms will be suppressed, \ie~they should be treated as perturbative corrections.

In this class of massive gravity, gravitational waves could in principle carry all six polarizations depicted in Fig.~\ref{fig:gmodes}. However, the longitudinal scalar mode is always associated with a ghost instability known as the Boulware--Deser (BD) ghost \cite{Boulware:1973my} and therefore for massive gravity to make sense, there should be an additional constraint that prevents the propagation of one of the scalar polarizations. For instance at the linear level, the ghost can only be eliminated by the the unique Fierz--Pauli potential \cite{Fierz:1939ix}. Nonlinearly, there is a unique two--parameter family of nonlinear graviton potential, called ghost--free massive gravity or the de Rham--Gabadadze--Tolley (dRGT) model, \cite{deRham:2010ik,deRham:2010kj} which generalizes the linear Fierz--Pauli potential and entirely eliminates the BD ghost \cite{deRham:2010ik,deRham:2010kj,Hassan:2011hr,Hassan:2011ea}.  So in the dRGT model, there are five degrees of freedom, matching the number of degrees of freedom for a massive spin--2 particle\footnote{The degree of freedom and constraint counting is most clear in the Hamiltonian formulation. See \cite{deRham:2014zqa} for careful counting of them in massive gravity models.}.

Note that typically in Lorentz invariant massive gravity not all of the five degrees of freedom couple to the matter with the same strength. If they did, then fifth force tests (such as solar system tests or pulsar timing observations) would already rule out the model. For example, the existence of a scalar mode implies that binary systems can emit a monopole radiation. However since the scalar couples to matter much more weakly than the tensor modes, the power emitted in the monopole is very weak. Thus if there is a monopole signal accompanying the gravitational wave emission, it is typically expected to be much weaker (see the discussion on the Vainshtein mechanism in \Sec\ref{sec:nonvain} and on the fifth force constraints in \Sec\ref{sec:FF}).

\subsubsection{Lorentz Violating}

For Lorentz violating theories, the spacetime symmetry is usually assumed to be the Galilean Group, although more generally we may imagine any subgroup of the Poincar\'e or Galilean groups. The Galilean group has SU(2) as the little group for massive particles, and one can still define spins for massive particles. In this case one can construct many possible ghost--free Lagrangians \cite{Rubakov:2004eb,deRham:2014zqa} (see also \cite{Comelli:2010bj,Dubovsky:2004sg,Comelli:2014xga,Comelli:2015ksa,Lin:2013aha,DeFelice:2015hla}). A Lorentz violating massive graviton may carry between two and five degrees of freedom (potentially even six). Lorentz invariance violations have been tightly constrained by various experiments, particularly in the matter sector, and recently in the framework of the ``Standard Model Extension'' \cite{Kostelecky:2008ts,Bluhm:2005uj}, which parametrizes all possible Lorentz violations from the point of view of effective field theory. More recently, the GW150914 detection has also been used to impose a direct constraint on the Standard Model Extension and  the pure gravitational sector \cite{Kostelecky:2016kfm,Yunes:2016jcc}.

Nevertheless, assuming that Lorentz violations are sufficiently small to avoid constraints from the matter sector, these massive gravity models may still be viable. A common feature about Lorentz invariant and Lorentz violating theories is that both will necessarily include the two helicity--2 modes (\ie~transverse traceless) that will reproduce the graviton of General Relativity in the appropriate limits. Thus all such theories have a set of universal features determined by the helicity--2 modes alone, which can be used to establish some universal bounds on the graviton mass.

\subsection{Massless Graviton Propagator}

The helicity--2 degrees of freedom in General Relativity are sourced by the transverse and traceless projection of the spatial part of the stress--energy tensor $T_{ij}^\text{TT}$. Let us review how to understand this in a manner which will be useful for subsequent generalizations.
	
At an elementary level, a gravitational force is a force between two stress--energy tensors. In the weak field limit this is determined by the single graviton exchange amplitude between two sources, $T_1^{\mu \nu}(x)$ and $T_2^{\mu \nu}(y)$, which takes the form
\begin{equation}
	\label{GRamp}
	\mc{A} \sim  \frac{1}{2 \mpl^2} \int \d^4 x \int \d^4 y \, T_1^{\mu \nu}(x) G_{\mu \nu \alpha \beta}(x,y) T_2^{\alpha \beta}(y) \, ,
\end{equation}
where $G_{\mu \nu \alpha \beta}(x,y) $ is the graviton propagator
\begin{equation}
G_{\mu \nu \alpha \beta}(x,y) =  i \left\langle \hat T \left[h_{\mu\nu}(x) h_{\alpha \beta}(y)\right]\right \rangle,
\end{equation}
where $\hat T$ is the time ordering operator. At tree level in General Relativity,  we have
\begin{equation}
	 \label{GRprop}
	 G_{\mu\nu\alpha \beta}(x,y) =   \frac{f_{\mu\nu\alpha \beta}}{-\Box - i \epsilon } \delta^4(x-y)\,,
\end{equation}	
or more schematically
\begin{equation}
G_{\mu\nu\alpha \beta} =   \frac{f_{\mu\nu\alpha \beta}}{-\Box - i \epsilon }\,,
\end{equation}	
where $\Box=\eta^{\mu\nu}\p_\mu \p_\nu$ is the standard d'Alembertian in flat space. The polarization structure in Lorentz gauge is given by
\begin{align}
	\label{polarizationMassless}
	f_{\mu\nu\alpha \beta}&=  \tilde \eta_{\mu(\alpha}\tilde \eta_{|\nu| \beta)} - \frac 12 \tilde \eta_{\mu \nu} \tilde \eta_{\alpha \beta} , \\
	\text{with }& \tilde \eta\mn=\eta\mn -\frac{1}{\Box}\p_\mu \p_\nu.
\end{align}
where the symmetrization of indices is defined with the factorial in the front. This polarization structure of the propagator ensures that only the transverse traceless degrees of freedom are propagating. In particular, we note that only the transverse traceless part of the stress--energy tensor contributes to the imaginary part of the exchange amplitude
\begin{align}
	\text{Im}[ \mc{A} ]& \sim  \frac{\pi}{2 \mpl^2} \int\! \d^4 x \, T_1^{\mu \nu}(x) f_{\mu\nu\alpha \beta} \delta(\Box) T_2^{\alpha \beta}(x) \\
	&\sim   \frac{\pi}{2 \mpl^2} \int\! \d^4 x \, {T_1^\text{TT}}^{\mu \nu}(x) f_{\mu\nu\alpha \beta} \delta(\Box) {T_2^\text{TT}}^{\alpha \beta}(x) \, .
\end{align}
Explicitly demonstrating this requires writing this expression in momentum space and using the on-shell condition $\Box=-k^2=0$. Then, by the optical theorem only the transverse traceless or helicity--2 degrees of freedom are propagating particles.
				
The single graviton exchange amplitude is gauge invariant and uniquely determines the gravitational force in the weak field region and, as such, contains information on both the static part of the force and the radiative part through the pole structure of the propagator. For example, for two point masses $T^{\mu\nu}_1(x)=-M_1\dd(\bfx-\bfx_1)\dd^\mu_0 \dd^\nu_0$ and $T^{\mu\nu}_2(x)=-M_2\dd(\bfx-\bfx_2)\dd^\mu_0 \dd^\nu_0$, the static force between them is given by the Newtonian force
 \begin{equation}
  F_{12} \sim \frac{1}{T}\frac{\ud }{\ud r} \text{Re}[\mc{A}] \sim  \frac{M_1 M_2}{\mpl^2 r^2},~~\text{with}~ r=|\bfx_1-\bfx_2| .
\end{equation}
where $T$ is the total time integrated over. On the other hand, for the massive propagators considered later,  $T^{-1}\ud\text{Re}[\mc{A}] /\ud r$ is of the Yukawa form $\sim e^{-m_gr}/r^2$.

\subsection{Massive Graviton Propagators}

In a massive theory, the graviton propagator $ G_{\mu\nu\alpha \beta}$ pole structure is modified. Either it gains a pole at some finite mass $\Box=m^2$ or in the case of a resonance graviton it gains a branch cut and a pole on the second Riemann sheet. The emergence of the branch cut is clear in the spectral representation formula since a branch cut may be viewed as a continuum of poles. The pole lies on the second Riemann sheet since its energy should have a negative imaginary part $E_R-i \Gamma/2$ and is hence in the lower right quadrant of the complex $E$ plane. When constructing $s = - E^2$ the lower half of the complex $E$ plane becomes part of the second Riemann sheet. This modification of the pole structure is universal to both Lorentz invariant and Lorentz violating massive theories. Secondly, the polarization structure may be modified in a manner which allows for additional polarizations to be propagating. The precise details of this part can be model dependent.

\subsubsection{Hard Mass Graviton}
\label{sec:hmg}
	
At tree level, in a theory of a hard mass graviton which preserves rotational invariance, time translation invariance and time reverse invariance, the general structure of the propagator is
 \begin{equation}
 \label{eq:propmass}
 	  G^{(m)}_{\mu\nu\alpha \beta} = \sum_I \frac{  {f_I}^{(m)}_{\mu\nu\alpha \beta}}{\partial_t^2 -F_I[-\nabla^2]  +m_g^2- i \epsilon },
 \end{equation}
where the sum is performed over the different polarizations $I$ and ${f_I}^{(m)}_{\mu\nu\alpha \beta}$ may not necessarily be Lorentz invariant (\ie~not constructed solely out of $\eta_{\mu\nu}$ and $\partial_{\mu}$). The function $F_I$ accounts for the modified dispersion relation for propagating polarization $I$: $E^2-F_I[\vec p^2]=m_g^2$.  The different functions $F_I[-\nd^2]$ account for the fact that the different polarizations can have distinct dispersion relations.
It is usually assumed that at low energies $F_I[\vec p^2]$ has an analytic expansion $(c^I_s)^2 \bfp^2 + \bfp^4/\Lambda_I^2 + \ldots$. Many graviton mass bounds arising from dispersion relations implicitly assume $c_s^I=1$. \\

If time or space translation invariance is broken, for example in an FLRW Universe, then the spectral representation of the propagator is not so clean as it is necessary to solve the mode equations on the appropriate background. Even the definition of mass is ambiguous on a background which breaks translation invariance. The exception is the case of maximally symmetric spacetimes such as (anti) de Sitter, where there is an accepted definition of the mass for a spin 2 field based on the representation theory for the (anti) de Sitter group. However, these concerns are somewhat mute when putting bounds on the masses which are much larger than the associated curvature scale, since in this case spacetime is locally Minkowski, and the full propagator will for distances much less than the curvature scale approximate the above form. In other words, space or time dependence of the background, \ie~the breaking of space-time translation invariance, is only a concern for masses comparable to or smaller than the spacetime curvature scale at the time the bound is being placed.

\subsubsection{Resonance Graviton}	
	
At tree level in a theory of a resonance graviton which may be Lorentz violating, but preserves rotational, time translation and time reversal invariance, the general structure may be a superposition of the hard mass form
\begin{equation}
	G^{(m)}_{\mu\nu\alpha \beta} =  \sum_I \int_0^{\infty}   \frac{{f_I}^{(m)}_{\mu\nu\alpha \beta}(\mu) \rho_I(\mu)  \d \mu}{\partial_t^2 -F_I[-\nabla^2]  +\mu^2- i \epsilon }\,,
\end{equation}
where $\mu$ is the spectral mass, $ {f_I}^{(m)}_{\mu\nu\alpha \beta}(\mu)$ may be dependent on the spectral mass, $F_I[-\nabla^2]$ is a potentially Lorentz violating dispersion relation and $\rho_I(\mu)>0$ are the positive semi-definite spectral densities. This kind of propagator may arise in the 4D effective theory of higher dimensional models such as the DGP model \cite{Dvali:2000rv,Dvali:2000hr}. The different functions $F_I[-\nabla^2]$ account for the fact that the different polarizations can have distinct dispersion relations and spectral densities $\rho_I(\mu)$. The propagator reduces to that of a hard mass spin--2 field for $\rho(\mu) = \delta(\mu - m_g)$.

For a graviton resonance that is centered at a finite value and is not too wide, \ie~the width is much smaller than the mass bound itself, the bounds for the hard graviton mass may still apply. For a broad resonance, such as in the DGP model, the bounds are qualitatively similar, but quantitatively modified. Most noticeably, for a broad resonance, the large distance fall--off of the propagator may be much weaker than the exponential form, \eg~power law fall--off.

\subsubsection{Lorentz Invariant Polarization Structure}

At the linear level, the action for a single Lorentz invariant massive spin--2 field $h\mn$ on Minkowski was derived by Fierz and Pauli in 1939 \cite{Fierz:1939ix}
\begin{align}
	\label{Eq:FierzPauli}
	\L_\text{FP} =& -\frac {\mpl^2}4 h^{\mu\nu}\hat{\mathcal{E}}^{\alpha \beta}_{\mu\nu} h_\ab -\frac 1 8 m_g^2\mpl^2 \lp h\mn^2-h^2\rp
	 \nonumber \\
	& + \frac12 h_{\mu\nu} T^{\mu\nu} \,,
\end{align}
where $\hat{\mathcal{E}}$ represents the Lichnerowicz operator ($\hat{\mathcal{E}}^{\alpha \beta}_{\mu\nu} h_\ab$ being the linearized Einstein tensor) and $T^{\mu\nu}$ is again the matter stress-energy tensor. The structure of the Fierz--Pauli mass term  $\lp h\mn^2-h^2\rp $ is essential in avoiding a ghost instability to that order in perturbation theory, and this combination is unique assuming Lorentz invariance\footnote{The cosmological constant term $\sqrt{-g}\Lambda$ when expanded around $\eta_{\mu\nu}$ gives rise to a quadratic term $h\mn^2- h^2/2$, but these should not be confused with a mass term. Indeed, the expansion of $\sqrt{-g}\Lambda$ also gives rise to a tadpole term $h$, which indicates Minkowski space is not really a valid background for the theory, and therefore the quadratic term in $h$ cannot be thought of as a mass term in this case.}. In four spacetime dimensions, a Lorentz--invariant massive spin--2 field carries 5 polarizations. The propagator for a Lorentz invariant hard massive spin--2 field is
\begin{equation}
	\label{Eq:massive propagator}
	G^{(m)}_{\mu\nu\alpha \beta} = \frac{f^\text{(FP)}_{\mu\nu\alpha \beta}(m_g)}{-\Box + m_g^2-i\ep},
\end{equation}
where $f^\text{(FP)}_{\mu\nu\alpha \beta}(m_g)$ is now
\begin{equation}
	\label{Eq:tensorStructure}
	f^\text{(FP)}_{\mu\nu\alpha \beta}(m_g)= \tilde \eta_{\mu(\alpha}\tilde \eta_{|\nu| \beta)} - \frac13 \tilde \eta_{\mu \nu} \tilde \eta_{\alpha \beta} ,
\end{equation}
with
\begin{equation}
	\label{Eq:Effeta}
	\tilde \eta\mn=\eta\mn -\frac{1}{m_g^2}\p_\mu \p_\nu.
\end{equation}

Similarly the propagator for a Lorentz invariant resonance graviton is
\begin{equation}
	\label{Eq:softmassive propagator}
	G^{(m)}_{\mu\nu\alpha \beta} = \int_0^{\infty} \d \mu \, \rho(\mu) \frac{f^\text{(FP)}_{\mu\nu\alpha \beta}(\mu)}{-\Box + \mu^2-i\ep},
\end{equation}
where $\rho(\mu)$ is the positive semi--definite spectral density. Lorentz invariance does two things: one is it fixes the form of the dispersion relation, and hence the $1/({-\Box + \mu^2-i\ep})$ structure, and secondly it fixes the form of the polarization tensor $f^\text{(FP)}_{\mu\nu\alpha \beta}$.

\subsubsection{Linear vDVZ Discontinuity}
\label{lnvDVZ}

Comparing the massless and Lorentz invariant massive graviton propagators, we see that the single graviton exchange amplitude calculated with the $m_g\to 0$ limit of \eqn\eqref{Eq:massive propagator} does not reduce to the General Relativity case \eqn\eqref{GRamp}; there is a finite difference $\eta_{\mu \nu} \eta_{\alpha \beta}/6$ between $f^\text{(FP)}_{\mu\nu\alpha \beta}(m_g)$ and $f_{\mu\nu\alpha \beta}$ even when $m_g\to 0$. Note that in this comparison, since the stress--energy tensor is conserved $\pd_\mu T^{\mu\nu}=0$, one may replace $\tilde \eta\mn$ with $\eta\mn$, so the divergence of \eqn\eqref{Eq:Effeta} in the limit $m_g\to 0$ does not have physical significance. However, for a source with a traceless stress--energy tensor, which is the case for photons, the finite difference in the exchange amplitudes vanish. Thus this finite difference can not be compensated by redefining the Planck mass. This is for historical reasons dubbed the van Dam--Veltman--Zakharov (vDVZ) discontinuity \cite{Iwasaki:1971uz,vanDam:1970vg,Zakharov:1970cc}\footnote{Note that \cite{Iwasaki:1971uz} is no later than \cite{vanDam:1970vg,Zakharov:1970cc}.}. If the gravitational phenomena in the weak field limit, as in the solar system, were described by the propagator \eqn\eqref{Eq:massive propagator}, Lorentz invariant massive gravity with an infinitesimal graviton mass would have been ruled out observationally by this discontinuity. Indeed, the vDVZ discontinuity is sometimes wrongfully used to argue that the graviton mass is mathematically zero \cite{Agashe:2014kda}.

The resolution behind the previous apparent discontinuity lies in the Vainshtein mechanism which is related to whether or not the previous linear approximation is valid.
First it is worth emphasizing that already within the solar system, while the weak field approximation is a good one for General Relativity, we are already able to observe the non--linear effects of General Relativity, and therefore focusing solely on the previous linear approximation of either the massive or massless theory would lead to wrong predictions.

The real distinction between General Relativity and massive gravity is that the linear weak field approximation breaks down even earlier for massive gravity. If we take the example of the solar system, although the weak--field approximation is a good one for the helicity--2 mode, the helicity--1 and helicity--0 modes of the massive graviton are in the strong field regime. Therefore, it is not sufficient to use the linear approximation in these environments and the vDVZ discontinuity is an artifact of using that approximation beyond its regime of validity.

When breaking Lorentz invariance, it is possible to maintain a greater regime of validity for the linear approximation and the linear vDVZ discontinuity may even be avoided in some cases \cite{Rubakov:2008nh,DeFelice:2015hla}.

\subsubsection{\stu Fields and vDVZ Discontinuity}
\label{stufields}

The origin of the vDVZ discontinuity lies in the fact that the helicity--0 mode of the massive graviton couples to the matter source, specifically the trace of the stress energy, with a gravitational strength even in the limit of $m_g\to 0$. It is simple to see this in the \stu formulation of the linear Fierz--Pauli theory, by introducing the fields $A_\mu$ and $\pi$ with the replacement
\begin{equation}
	\label{hmnpi}
	h_{\mu\nu} \to h_{\mu\nu} +\pd_{(\mu} A_{\nu)} + \pd_{\mu}\pd_{\nu} \pi.
\end{equation}
The idea of this \stu formulation is to restore the same gauge invariance as the massless theory. Indeed, in this formulation, the Fierz--Pauli theory is invariant under the following gauge transformations: $\dd h_{\mu\nu} = \pd_{(\mu}\xi_{\nu)}$, provided the \stu fields are transformed appropriately, $\dd A_\mu = -\xi_\mu$. In addition the theory is also invariant under the following $U(1)$ gauge transformation of the \stu fields : $\dd A_\mu =\pd_\mu \Lambda$, with $\dd \pi = -\Lambda$.  By appropriate choices of gauge, $h_{\mu\nu}$, $A_\mu$ and $\pi$ become the helicity--2, helicity--1 and helicity--0 degrees of freedom in the high energy or massless limit\footnote{By a standard, but somewhat abuse of, terminology, $h_{\mu\nu}$, $A_\mu$ and $\pi$ are often called the helicity--2, helicity--1 and helicity--0 modes respectively even in the massive case, or after the kinetic diagonalization, or without gauge fixing.}. An important feature is that $\pi$ obtains its kinetic term by mixing with $h_{\mu\nu}$. After diagonalization $h_{\mu\nu}= \tilde h_{\mu\nu}+m_g^2\pi \eta_{\mu\nu}$ and then canonical normalization of the kinetic terms $\hat{h}_{\mu\nu}=\mpl \tilde h_{\mu\nu},~\hat{A}_\mu\sim m_g\mpl A_\mu,~\hat{\pi}\sim m_g^2\mpl\pi$, it turns out that, in the massless limit, the helicity--1 modes decouple from the matter source $T^{\mu\nu}$, while the helicity--2 and helicity--0 modes couple to $T^{\mu\nu}$ at the gravitational strength
\begin{align}
\label{eq:LinearFP}
	\mc{L}^{m_g\to0}_\text{FP} =& -\frac14 \hat{h}^{\mu\nu}\hat{\mathcal{E}}^{\alpha \beta}_{\mu\nu} \hat{h}_\ab - \pd_{[\mu}\hat{A}_{\nu]}\pd^{[\mu}\hat{A}^{\nu]} - \frac12 \pd_\mu \hat{\pi} \pd^\mu \hat{\pi} \nonumber\\
	& +\frac1{2\mpl} \hat{h}_{\mu\nu}T^{\mu\nu} + \frac{1}{2\sqrt{6}\mpl}\hat{\pi} T^\mu{}_\mu.
\end{align}
So compared to General Relativity, the single graviton exchange amplitude in Fierz-Pauli theory in the small $m_g$ limit has an extra contribution from the coupling $\hat{\pi} T^\mu{}_\mu/(4\sqrt{6}\mpl)$. This is the origin of the vDVZ discontinuity.
\subsection{Nonlinearities and Vainshtein Screening}
\label{sec:nonvain}

The propagators and the vDVZ discontinuity we have discussed so far, are linear properties of a massive gravity theory about the vacuum, relying only on the free (linearized) action. Just as Newtonian gravity is a good approximation for General Relativity in the weak field regime, the linear theory is only a good approximation for massive gravity  when the weak field approximation is justified (for instance about a single massive object, the weak field approximation is justified at sufficiently large distances from the object). In the case of Lorentz--invariant massive gravity such as the DGP and dRGT models, whose full nonlinear actions will be displayed in \Sec\ref{sec:models of MG}, the non--linearities of the theory are important much before what would be the case in General Relativity.

\subsubsection{Nonlinear Resolution of vDVZ Discontinuity}
\label{sec:vbinb}

The Vainshtein screening operates in very generic situations, but it is instructive to state the mechanism in a simple example. If we consider an isolated static mass like the Sun, in General Relativity, the non--linearities are important within the Schwarzschild radius which is of the order of $r_{S, \odot} \sim 3\unit{km}$. This means that beyond that distance, corrections to the linear weak field approximation are small (but can still be observable). In the DGP model and dRGT model, the non--linearities become important already at much larger distances of the order of the Vainshtein radius $r_{V}$, and for the Sun this is of astronomical orders,
\begin{align}
	r_{V, \odot} &= \left(\frac{M_\odot}{\mpl^2 m_g^2}\right)^{\frac13} \sim \left( r_{S,\odot} \lambdabar_g^{2} \right)^{\frac13}\\
	& \sim 10^7 \unit{km} \left(\frac{10^{-20}\unit{eV}}{m_g}\right)^{\frac23}\,,
\end{align}
where we have taken $m_g\sim 10^{-20}\unit{eV}$ as an arbitrary reference. As $m_g\to 0$, $r_{V, \odot}$ goes to $\infty$, which means that the non--linear corrections beyond the Fierz--Pauli action, \eqn\eqref{Eq:FierzPauli}, are relevant almost everywhere and the linear Fierz--Pauli approximation is never valid. The non--linear corrections beyond the Fierz--Pauli action are precisely what restore the smooth limit towards General Relativity in the massless limit. It has indeed been shown in some specific situations how one recovers observations which are very close to General Relativity once we take the nonlinear terms of massive gravity into account, \cite{Deffayet:2001uk,Babichev:2009jt}. Therefore, the vDVZ discontinuity is not physically present and is an artifact of using the linear Fierz--Pauli action beyond its regime of validity without accounting for the corrections that enter from the gravitational theory. The general mechanism by which General Relativity is recovered from nonlinearities is what we refer to as the {\it Vainshtein mechanism} \cite{Vainshtein:1972sx}.

\subsubsection{Vainshtein Redressing}
\label{sec:VSredress}
To get a better insight on how the Vainshtein mechanism works, it is useful to start with the linearized Lagrangian \eqref{eq:LinearFP}
and focus solely on the helicity--0 mode $\hat{\pi}$. Since $\hat \pi$ originates from the \stu replacement \eqref{hmnpi}, non--linearly, it will carry some derivative interactions at a scale $\Lambda \ll \mpl$ which is a geometrical mean between the Planck scale and the graviton mass. Then the non--linear Lagrangian for $\hat \pi$ will take the form (omitting dimensionless factors of order 1),
\begin{eqnarray}
\mc{L}_{\hat \pi} = - \frac12 \left(\p\hat{\pi} \right)^2 + \Lambda^4 G\left(\frac{\p \hat \pi}{\Lambda^2}, \frac{\p^2 \hat \pi}{\Lambda^3} \right) + \frac{1}{\mpl}\hat{\pi} T \,,
\end{eqnarray}
where $G$ captures the derivative self--interactions.

We may now consider the situation where the source $T$ can be decomposed into $T=T_0+\delta T$, where the scales involved in  $T_0$ are much larger than that involved in $\delta T$. This may for instance occur if we are interested in the gravitational force between two light objects encoded in $\delta T$ that are located in the vicinity of a large source (like the Sun or the Earth), encoded in $T_0$. Alternatively, this decomposition is useful for binary systems where the two masses can be split into a total mass at the center of mass $T_0$ and deviations from it encoded in $\delta T$.

Within that setup,  the field $\hat \pi$ acquires a non--trivial classical profile $\pi_0$ determined by $T_0$, with $\p \pi_0\gg \Lambda^2$ and $\p^2 \pi_0 \gg \Lambda^3$ in the vicinity of the large source, while $\delta T$ leads to small fluctuations $\delta \pi$ on top of that background.  Expanding the field about this profile,
$\hat \pi =\pi_0 + \delta \pi$ and considering the linearized theory to second order in the perturbed field $\delta \pi$ we have
\begin{eqnarray}
\mc{L}_{\delta \pi} = -\frac 12 Z^{\mu \nu } \p_\mu \delta \pi \p_\nu \delta \pi +\frac{1}{\mpl} \delta \pi \delta T\,,
\end{eqnarray}
where the effective metric $Z^{\mu \nu}$ depends on the background profile, $Z^{\mu \nu }=Z^{\mu \nu }\left(\frac{\p \hat \pi_0}{\Lambda^2}, \frac{\p^2 \hat \pi_0}{\Lambda^3}\right)$. Far away from the source $T_0$, we are in the weak field regime and the standard kinetic term for $\pi$ dominates and $Z^{\mu\nu}\sim \eta^{\mu\nu}$. In that regime, the force mediated by $\delta \pi$ is comparable to the gravitational Newtonian force.  However closer to the source $T_0$ (\ie\ within its Vainshtein radius), the interactions dominate $\p \pi_0\gg \Lambda^2$ and $\p^2 \pi_0 \gg \Lambda^3$, leading to $Z\gg 1$. To understand the effects it is easier to canonically normalize the field $\delta \pi$, symbolically, $\chi \sim \sqrt{Z}\delta \pi$ leading to
\begin{eqnarray}
\mc{L}_{\chi} = -\frac 12 \left(\p \chi \right)^2 +\frac{1}{\mpl \sqrt{Z}}\, \chi \delta\,  T\,.
\end{eqnarray}
For $Z\gg 1$, the coupling of the helicity--0 mode to the matter source $\delta T$ is strongly suppressed compared to the standard $\mpl$ gravitational coupling. It follows that within the Vainshtein region the helicity--0 mode mediates a weak force and effectively decouples. This the essence of the Vainshtein mechanism.

It is worth noting that in practice, the Vainshtein mechanism  does not necessarily need to involve a particular large source $T_0$. For instance just the effects from the vector modes may be sufficient to activate the Vainshtein mechanism and to decouple both the helicity--0 and --1 modes, \cite{deRham:2016plk,deRham:2015ijs}.

\subsubsection{Galileon Field Theory}
\label{sec:galileont}

In the previous discussion we considered the Vainshtein mechanism to be generated by arbitrary derivative interactions. In practice in any ghost--free theory of massive gravity or a resonance, (DGP or dRGT), those interactions are intimately intertwined with that of the Galileon \cite{Nicolis:2008in}, which is a scalar field invariant under the following nonlinearly realized shift symmetry
\begin{equation}
	\label{galsym}
	\pi \to {\pi} + c + b_\mu x^\mu\,,
\end{equation}
where $c$ and $b_\mu$ are constant and $x^\mu$ is the spacetime coordinate\footnote{The name stems from the resemblance of this field symmetry to the Galilean coordinate transformation.}. For example, the helicity--0 mode of the DGP and the dRGT model is a Galileon (see \Sec\ref{sec:models of MG}). The reason for this may be understood from the \stu  formulation of massive gravity, \eqn\eqref{hmnpi}, where the Goldstone scalar $\pi$ for a spin--2 particle always has two derivatives acting on it. For Lorentz violating massive gravity models that require a Vainshtein mechanism to recover General Relativity, a Goldstone scalar is also expected.

Imposing the Galileon symmetry \eqref{galsym} and the requirement of no high order derivatives in the equation of motion (to avoid Ostrogradsky ghosts), all possible Galileon interactions can be written as
\begin{equation}
	\label{galinteraction}
	\mc{L}^\text{Gal}_{I}(\pi)=\pd_{\mu_1}  {\pi} \pd^{[\mu_1}  {\pi} \pd_{\mu_2} \pd^{\mu_2}  {\pi} \cdots \pd_{\mu_{I-1}} \pd^{\mu_{I-1}]}  {\pi}\,.
\end{equation}
Note that for $I=2$ we recover the standard kinetic term  $(\pd {\pi})^2$ which naturally satisfies the Galileon symmetry (up to integrations by parts, that is at the level of the action but not the Lagrangian). In $n$ dimensional spacetime, there are only $n-1$ Galileon interactions (with $I>2$), as there are only $n$ spacetime indices. Even though $\mc{L}^\text{Gal}_{I}(\pi)$ seems to contain higher order derivatives, this is deceiving; indeed, the equations of motion are manifestly second order.

It is easy to show that the Vainshtein screening is a generic feature of Galileon field theory \cite{Nicolis:2008in}. Another interesting field theoretical property of all these Galileon interaction terms is that their coupling constants are not renormalized under loop corrections \cite{Luty:2003vm,Nicolis:2004qq,deRham:2012ew}. See \cite{Trodden:2011xh} for a review of Galileon field theory.
	
\subsection{Implications of a Massive Graviton}	

If the graviton has a mass, there will be many physical implications, some more model dependent than others. Broadly speaking they can be split into three categories: (1) those associated with the weakening of the force at large distance due to a Yukawa--like potential. (2)  Those associated with a strengthening of the gravitational force at intermediate scales due to the additional scalar mode $\hat \pi$ leading to a fifth force. (3) Finally there are those effects that probe the modified dispersion relation, \ie~the fact that gravitational influence no longer travels at the speed of light.  This does not exhaust possible physical effects, but the majority of observational constraints are associated with these three. Less obvious effects are that in certain theories, the massive gravitons could condense to form an effective negative pressure stress energy, potentially giving rise to the late time accelerated cosmic expansion. This type of self-accelerating mechanism, as was originally realized in the context of the DGP model \cite{Deffayet:2001pu}, is very model dependent.  It is still too early to claim that the observed late time cosmic acceleration has put a lower bound on the graviton mass, as dark energy can also be explained with models other than massive gravity. There is an on--going effort in cosmology to constrain and differentiate massive gravity \cite{deRham:2014zqa} and other dark energy models. In this paper we will review the mass bounds that are largely model independent which typically lie in one of the three stated categories.

\subsubsection{Yukawa Potential}
\label{sec:YPintro}

Forces from massless gauge bosons have the $1/r^2$ fall--off, but for a massive boson the force typically acquires an exponential Yukawa suppression. In the case of hard mass graviton theories, the  propagator \eqref{eq:propmass} obtained from the linear theory leads to a Yukawa type of potential. This can be seen by considering a static source $M$ localized at $\bfx=0$ with effective stress--energy tensor $T_1^{\mu\nu}=-M \delta^{3}(\bfx)\delta^\mu_0 \delta^\nu_0$, leading to a finite range Yukawa potential
\begin{equation}
    \label{Yukpot}
  	\Phi \sim \frac{M}{\mpl^2 r}e^{- m_g r}\,,
\end{equation}
simply because $G_s\sim e^{- m_g r}/ r$ is the static Green's function solution of $[-\nabla^2 +m_g^2] G_s = \delta^{3}(\bfx)$.

This is a generic feature of massive gravity theories, independent of the nonlinear interactions a theory may have, but does assume that the linear (weak field) approximation is applicable. In a Lorentz violating massive gravity where there is no vDVZ discontinuity, this is typically the case and in such theories gravity is weaker than General Relativity for the most part. Even in the cosmological context, gravitational fluctuations about the cosmological background are typically weak and the Yukawa suppression will be realized, albeit with a possible cosmological dressed mass.
In Lorentz invariant massive gravity the situation is more subtle, at intermediate scales, which may extend out as far as the Hubble horizon, the helicity--0 and helicity--1 modes cannot be treated linearly, and so the Yukawa form cannot be trusted. However, as we have discussed above, the nonlinearities of the Vainshtein mechanism screen the undesirable large effects from the helicity--0 and helicity--1 modes, while the linear approximation of the helicity--2 modes typically remains valid in conventional weak gravity regimes of General Relativity.

Therefore, whether Lorentz invariant or violating, the helicity--2 modes can typically be treated linearly in environments such as the solar system, and the Yukawa potential is expected to be applicable there.  Based on this linear approximation, if the graviton had a hard mass, the force of gravity would have a finite range of the order of the graviton's (reduced) Compton wavelength $\lambdag$. There may be a fifth force coming from the non--helicity--2 modes, but the fifth force will be largely screened by the Vainshtein mechanism so we expect the Yukawa fall--off to at least be significant around the graviton Compton wavelength.

Note, however, that in massive gravity models, there can be branches of nonlinear solutions which are asymptotically flat but will not exhibit this Yukawa suppression at large distances \cite{Comelli:2010bj,Nieuwenhuizen:2011sq,Berezhiani:2011mt,Gruzinov:2011mm}. Some of these solutions have problems such as horizon singularities and instabilities \cite{Berezhiani:2011mt,Berezhiani:2013dw,Berezhiani:2013dca}. Non--Yukawa fall--off is a known feature in massive gravity models augmented with additional degrees of freedom \cite{Tolley:2015ywa,Brito:2013xaa,Volkov:2014ooa,Wu:2016jfw}. Those solutions have a more promising fate in terms of theoretical consistency, due to the extra degrees of freedom. All of these features are a consequence of the massive gravitons condensing and acting as an effective stress energy which sources a slower fall--off.

More generally, even when working in the standard branch, in resonance graviton theories, the exponential fall--off of the Yukawa potential may be softened to a power law, albeit one with a stronger fall--off than $1/r^2$. The transition to this stronger power law, occurs around a scale determined by the effective mass of the resonance which may similarly be translated into a (reduced) Compton wavelength $\lambdag$. Thus while resonance graviton theories may lead to a weaker than Yukawa suppression, there will nevertheless be a suppression in the force from the helicity-2 modes at the scale $\lambdag$.

Some of the tightest current graviton mass bounds come from considering the Yukawa modification of the gravitational force in the solar system or larger structures; see \Sec\ref{sec:YP}.

\subsubsection{Modified Dispersion Relation}

Another feature universal to all the forms of the massive graviton propagator is the effect of the mass on the dispersion relation. This makes the speed of the gravitational waves depend on the wave frequency. As discussed in \Sec\ref{sec:hmg}, the minimal case $F_I[\vec p^2]=\vec p^2$ is usually assumed at least for the helicity--2 modes, which is the case for many massive gravity models. The modified dispersion relation for the helicity--2 modes is then given by
\begin{equation}
	\label{dpsrel}
	E^2-\bfp^2=m_g^2 .
\end{equation}
Equivalently, this can be formulated as the graviton traveling sub-luminally
\begin{equation}
	\label{disrel}
	v_g^2(E)=1-\frac{m_g^2}{E^2}\,,
\end{equation}
where $E$ is the graviton energy and $v_g$ is the group velocity. Non--helicity--2 modes may not exist in the case of some Lorentz violating massive gravity models. In the case of Lorentz invariant massive gravity they are necessarily present, and their dispersion relation is necessarily of the form of \eqn\eqref{dpsrel} perturbatively around the Minkowski vacuum. However, in a matter environment, the dispersion relation for the helicity-1 and helicity-0 modes may be significantly modified. Furthermore, in regions where the dispersion relation is modified, the nonlinear Vainshtein screening of the non--helicity--2 modes is expected so that their effects will be minimal. For example, the production of gravitational waves of helicity-1 and helicity-0 type is expected to be suppressed from dense sources. The end result is that one may only consider the modification for the helicity--2 modes but neglect the non--helicity--2 modes, which is the assumption for the mass bounds discussed in \Sec\ref{sec:MDR}. In this sense, the bounds from modified dispersion relation can be regarded as model independent.

The mass bound from the recent detection of GW150914 by aLIGO is of this kind. This relies on the fact that the gravitational wave frequency increases in the duration of GW150914. Since the velocity of a gravitational wave depends on its frequency or energy according to \eqn\eqref{disrel}, the tail of the signal will travel faster than the front if the graviton is massive, which makes the whole signal more ``squeezed'' than in General Relativity. This phasing difference leads to the bound quoted by \cite{TheLIGOScientific:2016src}. With the coming of gravitational wave astronomy, it is expected the mass bound from this simple consideration will improve in the future. Not only does the modified dispersion relation shape the phasing of directly detected gravitational waves, but it may also affect the production and evolution of primordial gravitational waves, among a few other effects.
The GW151226 event, from a lower mass merger, has lower GW frequencies and a lower signal--to--noise ratio, and thus it can only put a weaker bound on graviton mass. See \Sec\ref{sec:MDR} for more details.

\subsubsection{Fifth Force}
	
A large class of tests of General Relativity determine whether there exists a ``fifth force'' \cite{Will:2014kxa}. Indeed, many massive gravity models give rise to a fifth force of some sort. Unlike the effects from the Yukawa fall--off and the modified dispersion relation, which are basically based on information on the linear theory, one usually needs to consider all the nonlinear interactions to establish massive graviton bounds from the fifth force tests, particularly for massive gravity models that exhibit a Vainshtein mechanism.

The reason for this, as discussed in \Sec\ref{lnvDVZ}, is the existence of the vDVZ discontinuity. For example, for the tensor structure, \eqn\eqref{Eq:tensorStructure}, of the Lorentz invariant massive graviton propagator \eqref{Eq:massive propagator}, a factor of $-1/3$ enters the last combination instead of the $-1/2$ for a massless spin--2 field \eqref{polarizationMassless}. Within the linear theory, this corresponds to an order one correction compared to General Relativity, which does not disappear in the massless limit. However, the vDVZ discontinuity is not a physical discontinuity but a pure artifact of using the linear theory beyond its regime of validity, simply signaling the existence of additional polarizations. As discussed in \Sec\ref{sec:nonvain}, General Relativity is recovered to a good approximation in most conventional astronomical situations, thanks to the Vainshtein screening mechanism. However, the additional polarizations, particularly the helicity--0 mode,  still mediate a very small force, which can be constrained through fifth force experiments. Indeed, fifth forces can give rise to some of the tightest bounds on the graviton mass.

The nonlinear Vainshtein mechanism may differ in detail in specific models, but a common feature is that there will be a Galileon--like scalar that plays a major role (see \Sec\ref{sec:galileont}). In \Sec\ref{sec:FF}, we will restrict to this class of bounds for the DGP and dRGT models and the decoupling limit approximation of these models (see \Sec\ref{sec:models of MG}). Due to the common Galileon--like symmetries, it is expected that the mass bounds derived should be roughly applicable for all models where the helicity--0 mode does not decouple.

\section{Theories of Massive Gravity}
\label{sec:models of MG}

The form of the non--linearities and interactions in massive theories of gravity is significantly constrained by the necessity of preserving a ghost--free structure (at least up to a given energy scale). In higher dimensional soft--mass models such as the DGP model, this is achieved automatically by requiring the theory to be invariant under higher dimensional diffeomorphisms. In the case of Lorentz violating massive gravity, one has a little more flexibility to engineer a ghost--free structure \cite{Comelli:2014xga}. In the case of Lorentz--invariant hard mass gravity, this was successfully achieved for the first time in \cite{deRham:2010ik,deRham:2010kj}. The structure of this model is unique up to a couple of free parameters. In all of the Lorentz invariant models, and in certain Lorentz violating models, the non--linearities implement a Vainshtein mechanism \cite{Vainshtein:1972sx}, which is responsible for decoupling the additional polarizations of the graviton and hence strongly suppressing deviations from General Relativity. This was shown precisely in the context of soft massive gravity in \cite{Deffayet:2001uk}, and the implementation for a hard mass graviton \cite{Vainshtein:1972sx,Babichev:2010jd} is very similar  (see \cite{Babichev:2013usa} for a review). Also, as discussed in \Sec\ref{sec:vbinb}, nonlinear massive gravity allows for nonlinear backgrounds around which the vDVZ discontinuity is absent \cite{deRham:2016plk}.

Below we revisit some theories of massive gravity that have been proposed in the literature. Many of these theories have specific tests which may constrain their particular parameters in special ways. In this paper, we only review the generic constraints on the graviton mass which are applicable to most of these models, with a few specific exceptions. In particular, the mass bounds from the fifth force tests are for the DGP model and Lorentz invariant ghost--free massive gravity (namely, the dRGT model), or similar models whose implementation of the Vainshtein mechanism is approximated by the Galileon scalar \cite{Nicolis:2008in}. We refer to \cite{deRham:2014zqa} for a more complete review of these different models of massive gravity.

\subsection{Soft Massive Gravity}
\label{sec:DGPmodel}

Soft massive gravity models can arise in braneworld models where the graviton is not technically massless but a massive ``resonance'', \ie~a complex pole in the propagator. The DGP model \cite{Dvali:2000rv,Dvali:2000hr} is a simple example of such models where a brane is embedded in an infinitely large bulk and the bulk and the brane are both endowed with the corresponding Einstein-Hilbert term
\begin{align}
	S _\text{DGP} =& \frac{M_5^3}{2}\int \!\ud^5  x \sqrt{-g_5} \frac{R_5}{2} - M_5^3\int \!\ud^4  x \sqrt{-g} K \nonumber\\
	& + \frac{\mpl^2}{2} \int\! \ud^4 x  \sqrt{-g} \left( \frac{R}{2} +\mc{L}_m\right)  ,
\end{align}
where $K$ is the extrinsic curvature scalar of the brane, $\mpl$ and $M_5$ ($R$ and $R_5$) are the 4D and 5D Planck masses (Ricci scalars) respectively and $\mc{L}_m$ is the matter Lagrangian. (Note that some authors do not write the extrinsic curvature term explicitly.) At large distances the graviton behaves like a 5D massless spin--2 particle, while at short distances its behavior is like a 4D massless one, thus recovering General Relativity locally \cite{Deffayet:2001uk}. A simple dimensional analysis reveals that the cross--over scale is around $M_\text{cross} = M_5^3/\mpl^2$.

From the 4D point of view, the graviton of the DGP model looks like a resonance that is relatively broad, and its decay rate is of the order of the graviton mass itself \cite{Gregory:2000jc,Dvali:2000rv}
\begin{equation}
	m_g = M_\text{cross} = \frac{M_5^3}{\mpl^2}.
\end{equation}
Therefore, its dispersion relation can not be approximated by the standard massive particle's dispersion relation, and many of the mass bounds that rely on the dispersion relation $E^2-\bfp^2=m_g^2$ can not be directly applied to this model. Also, due to this broad width, it does not have the harsh cutoff behavior of the massive gravitational force at large distances. The associated potential is not of a Yukawa type but rather extrapolates between the standard 4D Newtonian potential $\Phi\sim r^{-1}$ at short distances and a 5D Newtonian potential $\Phi\sim r^{-2}$ at distances larger than the graviton Compton wavelength. The exact form of the DGP potential is \cite{Dvali:2000hr}
\begin{align}
\label{eq:DGPpot}
	\Phi(r)=& -\frac1{8\pi^2\mpl}\frac1r \biggl\{\sin\left(r m_g\right)\text{Ci}\left(r m_g\right)\nonumber\\
	&+\frac12\cos\left(r m_g\right)\left[ \pi-2\text{Si}\left(r m_g\right)\right]\biggr\},
\end{align}
where $\text{Si}(z) =\int_0^z\sin(t)/t\,dt$, $\text{Ci}(z)=\gamma+\ln(z)+\int_0^z(\cos(t)-1)/t\, dt$, and $\gamma\simeq 0.577$, the Euler Masceroni constant.

When viewed as a small correction in terms of $m_g r$, the leading order correction to the gravitational force enters at second order, the same as the hard mass case. Thus, the Yukawa type of mass bounds established in the perturbative regime can also be applied to the DGP model.

The DGP model contains two distinct branches. In FLRW, the `normal branch' and the opposite branch which is also the self--accelerating branch on FLRW. While the self--accelerating branch could in principle lead to a natural candidate for dark energy, it has been shown that it is not stable and either the helicity--0 or the helicity--1 mode is a ghost \cite{Koyama:2005tx}. The normal branch is stable but requires a more conventional source of dark energy. Thus while the normal branch of DGP is less interesting as an alternative explanation of late--time acceleration, it is interesting as a proof of principle of a model distinct from $\Lambda$CDM, which maintains many of the virtues of $\Lambda$CDM in addition to the graviton being effectively massive.

The DGP model has been generalized to higher dimensional constructions of soft massive gravity \cite{Gabadadze:2003ck} and Cascading Gravity \cite{deRham:2008zz,deRham:2007rw,deRham:2009wb}, where the potential may behave differently at large distances and even fall off faster than $r^{-2}$. These models also share many of the same features, including the presence of a graviton resonance. Some exact cosmological solutions have been found in  \cite{Niedermann:2014bqa,Eglseer:2015xla} which appear to either be unstable or phenomenologically uninteresting. However the analyses performed so far are not exhaustive and it is likely that solutions arbitrarily close to General Relativity should exist. 

Despite some fundamental differences, the DGP model is in many ways qualitatively similar to hard massive gravity, in particular at intermediate scales. It has in common the existence of additional polarizations, which give rise to a weak fifth force. This can be best seen in the decoupling limit we define below.

\subsubsection{DGP Decoupling Limit}
\label{sec:DGPDL}

The implementation of the nonlinear Vainshtein mechanism is very difficult in the full braneworld setup of DGP, but there exist a particular limit of the theory which to a great extent captures a lot of the important phenomenology. The existence of a new scale, the graviton mass, which is parametrically well below the Planck scales, implies a hierarchy of interaction scales which are distinct for the various helicity modes. It turns out that the Vainshtein mechanism is largely implemented in the helicity-0 mode sector whose interactions enter at an energy scale well below the Planck scale. The consequence of this is that the helicity-2 (and to a large extent the helicity-1) modes may be treated as linear in a regime where the Vainshtein mechanism is active. As discussed in \Sec\ref{sec:FF}, this may be well described by the decoupling limit which is the most practical tool for implementing fifth force tests of General Relativity.

This is realized as follows: for physics at distances much longer than the Planck length $\mpl^{-1}$ and much shorter than the cross--over length $m_g^{-1}$, which applies to most astronomical situations, the DGP model can be greatly simplified by defining the decoupling limit:
\begin{align}
	m_g\to 0,\quad\mpl \to &\infty,\quad T^{\mu\nu} \to \infty,\nonumber\\
	\label{decouplim}
	\Lambda_3 = (m_g^2\mpl)^{\frac13} \to \text{fixed}&,~~T^{\mu\nu}/\mpl  \to \text{fixed},
\end{align}
where $T^{\mu\nu}$ is the stress--energy tensor from $\L_m$ and $\Lambda_3$ is the strong coupling of the model, which is held fixed so as to capture the Vainshtein mechanism and accurately describe the physics around it.  In the decoupling limit approximation, omitting the contributions from the vector modes that decouple, the DGP model is given by the {\it local} 4D effective Lagrangian \cite{Luty:2003vm}
\begin{align}
	\label{dgpdlimit}
	\L^\text{dl}_\text{DGP} =& -\frac {1}4 \hat{h}^{\mu\nu}\hat{\mathcal{E}}^{\alpha \beta}_{\mu\nu} \hat{h}_\ab  + \frac1{2\mpl} \hat{h}_{\mu\nu} T^{\mu\nu}\nonumber\\
	& + \L^{\pi}_\text{DGP} + \frac{\hat{\pi}}{2\sqrt{6}\mpl} T^\mu{}_\mu \,,
\end{align}
with
\begin{equation}
	\L^{\pi}_\text{DGP} = -\frac12 (\pd\hat{\pi})^2 - \frac{1}{(\sqrt{6} \Lambda_3)^3} (\pd\hat{\pi})^2 \Box\hat{\pi},
\end{equation}
where $\hat{h}_{\mu\nu}$ and $\hat{\pi}$ are canonically normalized and $\hat{h}_{\mu\nu}$ is described by linearized General Relativity. From the braneworld point of view, $\hat{\pi}$ is roughly the brane bending mode as the extrinsic curvature goes like $K_{\mu\nu}\sim \pd_{\mu}\pd_{\nu}\hat\pi$. In this limit, all the nonlinearities are in the $\hat{\pi}$ sector $\L^{\pi}_\text{DGP}$, which satisfies the Galileon symmetry \eqref{galsym} and  serves as a good proxy for the full DGP model.

\subsection{Lorentz Invariant Hard Mass Gravity}

A non--linear Lorentz--invariant theory of massive gravity which eliminate the BD ghost to all orders and where the graviton has a finite hard mass was proposed in \cite{deRham:2010ik,deRham:2010kj}. This theory of ghost--free massive gravity (sometimes called dRGT) is the unique generalization of linear Fierz-Pauli theory. The action is
\begin{equation}
\label{dRGTdef}
	S_\text{dRGT} = \mpl^2 \int\!\ud^4 x \sqrt{-g} \left[ \frac{R}2  + m_g^2 \sum^4_{I=2} \ai_I \mc{U}_I(\mc{K}) \right],
\end{equation}
where $\ai_I$ are free parameters ($\ai_2=1$ can be chosen without loss of generality) and with
\begin{equation}
	\mc{U}_I(\mc{K}) = \mc{K}_{[\mu_1}^{\mu_1} \mc{K}_{\mu_2}^{\mu_2} \cdots \mc{K}_{\mu_I]}^{\mu_I},
\end{equation}
and
\begin{equation}
	\mc{K}_{\nu}^{\mu} = \dd^\mu_\nu - \mc{X}^\mu_\nu, \text{ and } \mc{X}^\mu_\nu= \left(\sqrt{g^{-1}\eta}\right)^\mu_{\, \nu},
\end{equation}
with the principal branch understood for the matrix square root and where $g^{-1}$ represents the inverse of the metric and $\eta$ the Minkowski metric $\eta\mn$.  Mathematically, there may be cases where the matrix square root is not well--defined in the real domain, but those correspond to unphysical solutions. Generally, there are several branches of solutions for the matrix square root, and as mentioned above the physical branch is the one where all the eigenvalues of the resulting matrix are non--negative.

The reference metric $\eta_{\ri\nu}$ explicitly breaks diffeomorphism invariance, but four nonlinear \stu fields $\phi^\ai$, which are four diffeomorphism scalars, can be introduced to restore diffeomorphism invariance with the replacement
\begin{equation}
	\mc{X}^\mu_\nu= \left(\sqrt{g^{-1}\eta}\right)^\mu_{\, \nu} \to \mc{X}^\mu_\nu= \left(\sqrt{g^{-1}\bar \eta}\right)^\mu_{\, \nu}\,,
\end{equation}
where we define the matrix $\bar \eta$ as $\bar \eta_{\mu\nu}=\p_\mu \phi^\alpha \p_\nu \phi^\beta \eta_{\alpha \beta}$
It is often useful to decompose $\phi^\ai$ as
\begin{equation}
	\phi^\ai = x^\ai + A^\ai + \pd^\ai\pi\,,
\end{equation}
when examining the effects from different helicities around the Minkowski vacuum, in which case the index $\ai$ here can be taken as the Lorentz index. Then the action is manifestly invariant under the Galileon symmetry for $\pi$, \eqn\eqref{galsym}.

As discussed in \Sec\ref{sec:dofs}, the BD ghost is eliminated because there is a primary second class constraint generated by the special graviton potential $\mc{U}_I(\mc{K})$, which in turn generates a secondary second class constraint. The generation of the primary constraint in the Lorentz invariant massive gravity is unique by construction.
This ultimately arises from the uniqueness of the Galileon interactions, \eqn\eqref{galinteraction}.

Although there is linear vDVZ discontinuity in this model, an active nonlinear Vainshtein mechanism is at work to screen the non--helicity--2 modes, recovering General Relativity in the $m_g \rightarrow 0$ limit in conventional situations.  All the three massive graviton features described in \Sec\ref{sec:ReviewMG}, namely, the Yukawa potential, modified dispersion relation and fifth force, can be used to put constraints on the graviton mass for this model.

\subsubsection{dRGT Decoupling Limit}
\label{sec:dRGTDL}

The dRGT model has many features in common with the soft--massive gravity models. Most notably, in a similarly defined decoupling limit, massive gravity theories reduce to a Galileon theory (plus a few extra interactions) just like the DGP model.
If we are interested in physics around the scale of $\Lambda_3$ we can define the decoupling limit of the dRGT model by taking the same limits we used for the DGP model in \eqn\eqref{decouplim}. The full decoupling limit is very complicated when the helicity--1 modes $A^\ai$ are included \cite{Ondo:2013wka}\cite{Gabadadze:2013ria}. But $A^\ai$ does not linearly couple to the matter source, and the vDVZ discontinuity arises because the helicity--0 mode linearly couples to the matter source. Therefore, a good understanding of the Vainshtein mechanism in the dRGT model can be obtained by neglecting the $A^\ai$ modes. The decoupling limit approximation of the dRGT model is (after omitting the vector modes),
\begin{align}
	\L^\text{dl}_\text{dRGT} =& -\frac {1}4 \hat{h}^{\mu\nu}\hat{\mathcal{E}}^{\alpha \beta}_{\mu\nu} \hat{h}_\ab  + \frac1{2\mpl} \hat{h}_{\mu\nu} T^{\mu\nu}  + \frac{a_1}{\mpl}\hat{\pi} T^\mu{}_\mu \nonumber \\
	&+ \frac{a_2}{\Lambda_3^3\mpl} \pd_\mu\hat{\pi}\pd_\nu\hat{\pi} T^{\mu\nu} + \frac{a_3}{\Lambda_3^6} \hat{h}_{\mu\nu} X_{(3)}^{\mu\nu} \nonumber \\
	& -\frac12 (\pd\hat{\pi})^2 + \sum_{I=3}^{5} \frac{b_I}{\Lambda_3^{3(I-2)}}  \mc{L}_I^\text{Gal}(\hat\pi),
\end{align}
where $\hat{h}_{\mu\nu}$ and $\hat{\pi}$ are canonically normalized, $a_I$ and $b_I$ are dimensionless constants depending only on the free parameters $\ai_I$ (see \cite{deRham:2014zqa} for detailed forms of $a_I$ and $b_I$), $\mc{L}_I^\text{Gal}(\hat\pi)$ are the Galileon terms defined in \eqn\eqref{galinteraction}, and  $X_{(3)}^{\mu}{}_{\nu}\equiv\dd^\mu_{[\nu}\pd_{\mu_1} \pd^{\mu_1} \hat{\pi}\pd_{\mu_2} \pd^{\mu_2} \hat{\pi}  \pd_{\mu_3]} \pd^{\mu_3} \hat{\pi}$,
which is nothing but the equation of motion term for $\mc{L}_4^\text{Gal}(\hat\pi)$. Thus, all the Galileon terms arise in the decoupling limit of the dRGT model\footnote{It seems that the Galileon symmetry is broken in $\L^\text{dl}_\text{dRGT}$ because of the couplings to matter. This is because we have redefined the helicity--2 modes, \ie~$\hat{h}_{\mu\nu}$ is not the original perturbative metric around Minkowski space. Rather it is a mixing between the original helicity--2 modes and $\hat{\pi}$.}. Two additional interaction terms arise beyond what is generally referred to as the Galileon: The $a_2$ term, which describes a non-minimal coupling to matter (sometimes referred to as a disformal coupling), and the $a_3$ term which cannot be diagonalized in a local way, but in this non-diagonal representation is a manifestly local, Galileon invariant interaction.
We will review the fifth force tests of the dRGT model based on this decoupling limit.

\subsubsection{Extended dRGT models}
\label{sec:EdRGT}

Once the cat has been let out the bag, and we allow for a single graviton to be massive, it is straightforward to extend this to theories of multiple gravitons which may in turn interact \cite{Hassan:2011zd,Hinterbichler:2012cn}. In these constructions, it is often considered that at least one graviton remains massless, which amounts to demanding that the action retains one overall copy of unbroken diffeomorphisms (without the introduction of \stu fields). However, this choice is not necessary since every massive graviton breaks one copy of diffeomorphism symmetry, and we may imagine a fully spontaneously broken state. What is forbidden, however, are multiple interacting massless gravitons, \ie,~more than one copy of unbroken diffeomorphisms per set of interacting fields. In theories in which a single massless graviton survives, that massless mode dominates the force at sufficiently large distances, in which case the theory effectively reduces to GR plus a cosmological constant. When looking at the physics within the Compton wavelength of the massive graviton, the contribution from the massless and massive modes may be comparable.  However, by making the associated Planck mass for the massless mode much larger than the physical Planck mass, it is possible to effectively decouple the massless mode so that the dominant contributor to the gravitational force are the massive gravitons. In this limit, although quantitatively different, all these models will have qualitative features in common with the theory of a single massive graviton, and so the associated observational constraints will apply, with some careful interpretation. Another line of extending the dRGT model is to introduce extra scalar degrees of freedom \cite{D'Amico:2012zv,Huang:2012pe,Huang:2013mha,DeFelice:2013dua,Andrews:2013ora,Mukohyama:2014rca,Huang:2015yga}, for which similar arguments apply when interpreting the mass bounds.

\subsection{Lorentz Violating Hard Mass Gravity}

Before the formulation of Lorentz invariant ghost--free massive gravity, it was believed that the only way to avoid the BD ghost issue in giving the graviton a hard mass (as opposed to a resonance) was to break Lorentz invariance \cite{Rubakov:2004eb} (see \cite{Rubakov:2008nh,deRham:2014zqa} for a review). Although Lorentz invariance is broken, rotational symmetry is usually kept intact, thus one may still classify particles into spins according to the induced representations of Galilean group. Breaking Lorentz invariance allows for considerably more freedom in the formulation of the theory, and also allows for models where not all five polarizations of the graviton are present; See \cite{Dubovsky:2004sg,Comelli:2014xga,Comelli:2015ksa,Lin:2013aha} and references therein.  Ref \cite{Dubovsky:2004ud} discussed the possibility of the Lorentz violating massive graviton as a cold dark matter candidate and the constraint on its mass in this scenario.

The simplest class of Lorentz violating massive gravity keeps the kinetic structure the same as that of General Relativity (Einstein--Hilbert term) and breaks Lorentz--invariance only through the graviton potential. Since the linear mass term does not have to be of the Lorentz invariant Fierz-Pauli form to avoid ghost instability, models with no vDVZ discontinuity can be constructed. For these models, the small mass limit can reduce to General Relativity, even without the need for the nonlinear Vainshtein mechanism. For this class of models, exclusively the helicity--0 mode or the helicity--1 modes may be absent. Recently, a new ``minimal theory of massive gravity'' has been proposed in \cite{DeFelice:2015hla} where the kinetic term also breaks Lorentz invariance and the theory only contains the two helicity--2 modes of the graviton, while the other helicity modes, and thus the fifth force, are completely absent.

In Lorentz violating massive gravity, the breaking of Lorentz invariance is expected to be only present in the gravitational sector. That breaking may propagate into the matter sector (the standard model of particle physics) where Lorentz invariance is extremely well constrained \cite{Kostelecky:2008ts,Bluhm:2005uj}, but those effects are expected to be suppressed by the graviton mass and the Planck scale. Lorentz violation in the gravitational sector has also been tightly constrained \cite{Will:2014kxa,Kostelecky:2016kfm,Yagi:2013qpa,Yagi:2013ava}.

For most massive gravity models in this class, the mass bounds from the Yukawa potential should be directly applicable. If a model's dispersion relation for the helicity--2 modes is the standard one, it is also expected that the mass bounds from \Sec\ref{sec:MDR} should also be directly applied. For fifth force tests on the other hand, the situation is less straightforward. In most models of massive gravity that break Lorentz invariance, the helicity--0 mode is either absent or does not directly couple to matter and fifth force types of experiments do not constrain those models. There can be however Lorentz--breaking models of massive gravity that involve a helicity--0 mode that couples to matter and these models are then also constrained by fifth--force experiments. In most models of Lorentz--invariant massive gravity (soft or hard) and their extensions, the helicity--0 mode is present and takes a Galileon--like form. For all these models, we hence expect the fifth force constrained to be approximately similar and applicable.

\subsection{Non--local Massive Gravity}

There has been recent interest in non--local theories of massive gravity that can be formulated without any reference metric but typically involve $\Box=g^{\mu\nu}\nd_\mu \nd_\nu$ in the denominator in the field equation which may arise when integrating out some degrees of freedom in a local theory \cite{Jaccard:2013gla,Modesto:2013jea,Cusin:2016nzi}.

Lorentz invariance is usually kept intact in these models. As formulated, these theories should not (and cannot) be treated as fundamental since the non--localities are indicative of additional degrees of freedom that are present already at low--energy. In other words, these theories cannot be directly quantized as low energy effective theories but may be considered as some emergent classical description of some deeper underlying theories. This implies that these theories may be used as an effective description for classical processes (like the production of gravitational waves by binary black hole mergers and their subsequent propagation) but cannot be used for the description of quantum processes like the production of primordial gravitational waves during inflation which is a genuine quantum mechanical process.

Putting these subtleties  aside, for those models where the helicity--2 pole structure is close to that of General Relativity it is expected that the mass bounds from the Yukawa potential and modified dispersion relation should be applicable and therefore the following discussions of \S~\ref{sec:YP} and \S~\ref{sec:MDR} are relevant to these models. The scalar mode on the other hand is expected to behave very differently than in standard massive gravity \cite{Jaccard:2013gla} and the bounds from \S~\ref{sec:FF} are not applicable, but an alternative analysis was provided in \cite{Kehagias:2014sda} and a series of other cosmological tests have been provided in \cite{Maggiore:2013mea,Foffa:2013vma,Dirian:2014ara}.

\subsection{Non--Fierz--Pauli Structure}
\label{sec:nonFP}

The vDVZ discontinuity for linearized massive gravity, observed in Fierz--Pauli theory, can be artificially removed by considering a slightly different mass term of the form $ h\mn^2- h^2/2$ \cite{Visser:1997hd,Finn:2001qi} on Minkowski. (Note that here the Minkowski space is assumed to be a valid background, as compared to the cosmological constant case discussed in footnote 6, where  the Minkowski space is not a valid background.) The different coefficient in front of the $h^2$ term as opposed to that in the Fierz--Pauli action \eqref{Eq:FierzPauli} directly implies that the theory suffers from a pathological ghost--like instability. Actually, the existence of this ghost implies a repulsive force which, at the linear level exactly cancels that of the standard helicity--0 mode and explains the absence of linear vDVZ discontinuity. Unfortunately, this cancellation will not carry through at higher order. Consequently not only does this theory exhibit a fifth force, it is also plagued by a fundamental instability.

Nonetheless when focused strictly on the helicity--2 modes the bounds put on the graviton mass in this theory are applicable to any other theory of massive gravity.

Interestingly, a non--Fierz Pauli extension of massive gravity was very recently considered in \cite{Konnig:2016idp} which does carry a ghost but at a sufficiently large scale for it not to be an issue within the region where the theory is explored. This leads to interesting phenomenology, particularly for cosmology. Similar extensions were also considered in the coupling with matter in \cite{deRham:2014naa,deRham:2014fha}. It is expected that all the three categories of tests and bounds provided below (Yukawa Potential, Modified Dispersion Relation and Fifth Forces) are applicable for these types of non--Fierz--Pauli extensions that carry a ghost a sufficiently large scale.

\section{Yukawa Potential}
\label{sec:YP}

As stated previously, when the graviton is given a mass, a generic feature is that the gravitational potential of a static point-like source $M$ changes from the standard Newtonian form $M/(\mpl^2 r)$ to the Yukawa potential
    \begin{equation}
    	\Phi \sim \frac{M}{\mpl^2 r}e^{- m_g r}\, .
    \end{equation}
In other words this changes the left hand side of the Poisson equation from $\nd^2 \Phi$ to  $(\nd^2 -m^2 )\Phi$. In this section, we will review the bounds put on the graviton mass based on this simple change.

Given a Yukawa potential with a small mass scale, we would naturally have to probe length scales associated with the Compton wavelength of the mass to see large deviations from a $1/r$ potential.   Alternatively, we could look for very small deviations to the gravitational force at distances less than the Compton wavelength. When treated perturbatively, the leading order correction, $\mc{O}(m_g r)^2$, is of the same magnitude for all known models of massive gravity. This is independent of whether the potential falls down exponentially at large distances or as power law as in soft--massive gravity (see \Sec\ref{sec:DGPmodel}).

As noted in \Sec\ref{sec:ReviewMG}, the Yukawa potential approximation is only valid in the linearized regime of massive gravity theories. Thus the most trusted bounds for the graviton mass from the Yukawa potential will come from systems firmly in this regime like the Solar System. Although gravity may be weaker for larger astronomical structures like galaxies, the components and dynamics of those structures are less understood.

\subsection{Bounds from the Solar System}

During the last century, General Relativity has been tested to an extremely high accuracy in the Solar System, especially under the Parametrized Post--Newtonian (PPN) formalism \cite{Will:2014kxa}. Particularly, the orbits of the inner planets agree with the predictions of General Relativity to an accuracy of $\mc{O}(10^{-8})$. The first order correction to the Newtonian form of the gravitational force from a massive graviton is of order $(m_gr)^2=(r/\lambdag)^2$. Thus, an order of magnitude estimate shows that $\lambdag$ has to be $10^4$ times bigger than the astronomical units, which gives a good estimate bound \cite{Will:1997bb},
\begin{equation}
	\label{PPNYB}
	\bound{10^{-23}}{10^{12}}
\end{equation}
which agrees with the more careful considerations below.

In the small $m_g$ limit, the effect of a hard mass graviton on the gravitational force  can be captured by simply modifying the standard gravitational parameter $\mu$ from the radius independent expression $GM_\odot=M_\odot/(8\pi\mpl^2)$ in General Relativity to having a radius dependence given by
\begin{equation}
	\label{mktl}
	\mu(r)=\frac{M_\odot}{8\pi\mpl^2}\left[1-\frac{(r m_g)^2}2+\mathcal O\left((rm_g)^3\right)\right],
\end{equation}
in massive gravity,
where $M_\odot$ is the mass of the Sun\footnote{For a resonance, the precise factor in front of the correction $(r m_g)^2$ may differ slightly but will still be negative and of order one, so the test proposed here is also perfectly applicable to resonances.}. This means that in massive gravity, Kepler's third law includes a radius dependence,
\begin{eqnarray}
a^3= \frac{T^2\mu(a)}{(2\pi)^2}\,,
\end{eqnarray}
where $a$ and $T$ are the semi--major axis and the period of the planet orbit respectively.

In General Relativity, (neglecting the effects from the other objects in the solar system, and without accounting for the relativistic corrections), the ratio  $a^3/T^2$ is the same for any planet of the solar system. In massive gravity on the other hand, that ratio differs for every planet since the effect of the mass on the orbits depends on the distance to the Sun. This dependence can be  parameterized by $\eta$ as follows:
\begin{eqnarray}
1+\eta \equiv \frac{a}{a_{\oplus}}\left(\frac{T_{\oplus}}{T(a)}\right)^{2/3}=\left( \frac{\mu(a)}{\mu_{\oplus}}\right)^{1/3}\,,
\end{eqnarray}
where the subscript $\oplus$ designates quantities associated with the Earth, so naturally, $a_{\oplus}=1\unit{AU}$, $T_{\oplus}=1\unit{year}$ and $\mu_{\oplus}$ is the parameter $\mu$ measured at the radius of the Earth. The other quantities in the previous expressions are computed for  a different planet in the solar system.
A given bound on $\eta$ then directly translates into a bound on the  mass $m_g$ as follows,
\begin{equation}
	\label{eqn:talmadge}
	m_g<\sqrt{\left|\frac{6 \eta}{(1\unit{AU})^2-a^2}\right|}+\mathcal O(\eta).
\end{equation}

In \cite{Talmadge:1988qz} a similar estimation was used when  considering a Kaluza--Klein contribution, \ie~an  additional  fifth force of the Yukawa type in addition to the standard Coulomb gravitational potential. The bounds on the mass of the Kaluza--Klein mode are similar to \eqref{eqn:talmadge}. Using the data from the inner planets, one can establish a two parameter bound on the strength and the range of this fifth force. We emphasize however that the situation considered in \cite{Talmadge:1988qz} is different than what we have in mind, \ie,~constraining a massive graviton where no massless mode is present.  The case of a single massive graviton (not an additional  Kaluza--Klein mode) was analyzed in  \cite{Will:1997bb}. Using the data given in \cite{Talmadge:1988qz}, it was  found that the best bound on the graviton mass comes from comparing the parameter $\mu$ from that of the planet Mars, yielding at the $2\sigma$ level
\begin{equation}
\label{boundmars}
	\bound{7.2 \times 10^{-23}}{2.8\times 10^{12}}\,.
\end{equation}
This is one of the best and most rigorous model independent bounds on the graviton mass. One could update this bound by using the ephemeris data to compute $\mu(a)$ as experienced by various planetary bodies.

In addition to the modification of Kepler's third law, the Yukawa potential also induces other effects, the dominant one of which is an additional precession of perihelion of the planet orbit on top of the precession predicted by General Relativity \cite{Talmadge:1988qz}. For a near circular orbit whose semi major axis $a$ is much less than the graviton Compton wavelength the precession to lowest order in $(a/\lambdag)$ goes as
\begin{equation}
\label{eqn:ykw_precess}
\dd \phi \simeq \pi (a/\lambdag)^2 .
\end{equation}
The graviton mass bounds from the anomalous precession are comparable, only a factor of two weaker than the bound from the modified Kepler's third law in \eqref{boundmars} \cite{Will:1997bb}.

Recently,  a new mass bound has been derived from comparing the observational data and the simulations of the S2 (or Source 2) star orbit at the Galactic Center \cite{Zakharov:2016lzv},
\begin{equation}
	\bound{2.9 \times 10^{-21}}{6.9\times 10^{10}}\,,
\end{equation}
assuming a Yukawa fall--off of the gravitational force. The S2 star is a bright star very close to a structure called Sagittarius A*, which is believed to be a supermassive black hole. This is of course not a bound from the Solar System, but the distance scale of the S2 orbit is of the same order as the Solar System.

\subsection{Bounds from Clusters}

As mentioned above, to get a tighter bound on the graviton mass, one may look for a longer ``ruler'' than the solar system. For this, one naturally looks for galaxies,  galaxy clusters, or even superclusters and filaments. These systems can allow us to probe more directly the gravitational force at distances comparable to the graviton's Compton wavelength. Indeed some of the tightest graviton mass bounds come from these structures. However, our understanding of the gravitational interactions for those systems is less robust than the solar system. Even at the level of galaxies, there is an on--going debate concerning the precise distribution of dark matter. Thus, these mass bounds should be taken \emph{cum grano salis}.

In \cite{Goldhaber:1974wg} it was noted that clusters, which typically have a size of 1-10\unit{Mpc}, are virialized and thus gravitationally bound. Thus they give a direct lower bound on the Compton wavelength of the graviton. Assuming that at least an inner region of a cluster of the size $\sim 0.5\unit{Mpc}$ is effectively gravitationally bound, this simple estimate yields the graviton mass bound
\begin{equation}
\label{eqn:ykw_virial}
	\bound{10^{-29}}{10^{19}}.
\end{equation}
An early similar consideration for galaxies gives a slightly weaker bound at $m_g<10^{-27}$eV \cite{Hare:1973px}. If the same argument can be applied to superclusters or filaments, which were not known in the 1970s, the mass bound may be raised by a few orders of magnitude.

\subsection{Weak Lensing}

Weak gravitational lensing  can be used to effectively probe the large scale structure of the Universe \cite{Bartelmann:1999yn}. Giving the graviton a mass changes the propagator from the Newtonian to the Yukawa which differ by a factor of $k^2/(k^2+m_g^2)$. Thus the power spectra of effective convergence (which arises from the two point function) gets an additional multiplicative factor of $k^2/(k^2+m_g^2)$ over the standard one.

Using data from a cluster of stars with average redshift $z=1.2$, \cite{VanWaerbeke:2001nh}, and assuming the standard $\Lambda$CDM cosmological model, one can fit against  the variance of the modified power spectrum  in massive gravity, as performed in \cite{Choudhury:2002pu}.
Requiring that the theoretical values for the model do not fall out of the $1\sigma$ error bars of the data leads to a bound on the graviton mass of
\begin{equation}
\label{eqn:ykw_weak}
	\bound{6\times10^{-32}}{3\times 10^{21}}\,.
\end{equation}
This is a remarkably strong bound, but one should bare in mind that it requires trusting the statistical methods used as well as the strong dependence on dark matter distributions inferred from a standard $\Lambda$CDM cosmological model hence assuming that the graviton mass has little effect on the cosmological evolution. We are hence approaching the type of cosmological tests that may be quite dependent on the specific model involved.

\subsection{Non--Yukawa Potentials}

As we discussed in \Sec\ref{sec:YPintro} not all massive gravity theories necessarily have a Yukawa potential form in the weak field limit. The departure from the Yukawa potential can come from several features.

First, the mass of the graviton is redressed by its environment, so when considering a massive gravity theory in a curved space time, the mass gets redressed by the background making it space-time dependent. As an example when considering (dRGT) massive gravity on FLRW, the graviton mass is redressed as \cite{Fasiello:2012rw}
\begin{align}
\label{eq:redressing}
	\tilde m_g^2(H) = m_g^2\frac{H}{H_0}\Bigg[c_0 +c_1\frac{H}{H_0}+ c_2\frac{H^2}{H_0^2}\Bigg],
\end{align}
where $\tilde m_g^2(H)$ is the redressed graviton mass, $H$ (resp. $H_0$) is the Hubble parameter of the FLRW background (resp. reference metric) and the coefficients $c_i$ are order 1 dimensionless coefficients which depend on the parameters $\alpha_3$ and $\alpha_4$ (see Eq.~(\ref{dRGTdef}) for the definition of $\alpha_3$ and $\alpha_4$). This is an exact example of how the graviton mass gets redressed by its environment. In the case of a localized star,  the star curves spacetime in a way which also redresses the graviton mass, and we therefore expect the effective graviton mass to depend on the mass of the star and on its distance to it. These redressing effects may be subtle and could in principle contribute to a departure from the Yukawa potential.

Second, in soft mass theories although the potential is weaker than $1/r$ it may not quite be Yukawa.  This we can see from the asymptotic behavior of the DGP potential, \eqn\eqref{eq:DGPpot} which is only $1/r^2$ \cite{Dvali:2000hr}.

Finally,  many black hole solutions in modified gravity have also non--Yukawa asymptotic form \cite{Tolley:2015ywa,Brito:2013xaa,Volkov:2014ooa,Wu:2016jfw,Berezhiani:2011mt,Berezhiani:2013dca}. This may arise either because such solutions rely on different branches which do not connect with the Minkowski vacuum solution, or because the linear weak--field approximation is never an appropriate description of such models and one never recovers a regime with a Yukawa--type of behavior. Interestingly, in many of the exact Black Hole  solutions found so far in massive gravity, the potential resembles more a Coulomb potential than a Yukawa potential for the helicity--2 mode and the departures  from General Relativity are suppressed in the region near the Black Holes. However much work still needs to be done to determine whether this is truly a generic feature of non--linear massive gravity.

\section{Modified Dispersion Relation}
\label{sec:MDR}

It is usually assumed that the helicity--2 modes of the massive graviton have a dispersion relation
\begin{equation}
\label{mfdssrl}
	E^2 = {\vec{k}}^2 + m_g^2  ~~~\text{or}~~~v_g^2(E)=1-\frac{m_g^2}{E^2}.
\end{equation}
To lowest order in ${\vec{k}}^2$, this is the only possible dispersion relation in a Lorentz invariant, hard mass massive gravity theory that is ghost--free, since there are no allowed modifications to the kinetic term beyond the Einstein--Hilbert one to that order in derivatives \cite{deRham:2013tfa}. For resonance models, a similar statement may be made at the level of the spectral representation. 
In Lorentz violating massive gravity, if the kinetic term remains in the Einstein--Hilbert form, the relation \eqn\eqref{mfdssrl} is also applicable. In general more complicated dispersion relations may be considered in Lorentz violating theories \cite{Kostelecky:2016kfm}. If the other helicity modes are present, they are usually Vainshtein screened, making it significantly more difficult to find observational effects of their dispersion relation.

In this section, we will review the graviton mass bounds based on the modified dispersion relation (see  \cite{Hare:1973px} for an early discussion of such a bound). By dispersion relation, we do not merely restrict to the numerical relation \eqn\eqref{mfdssrl}, rather it is meant for the broad context of modifying the propagators with a mass term, and considering dynamical processes and radiative effects.

\subsection{Bounds from Direct Gravitational Waves Detection}

\subsubsection{Gravitational Waves Alone: aLIGO, eLISA, etc.}

Gravitational waves generated by a strong gravity system can be observed directly by ground--based gravitational wave detectors such as aLIGO \cite{Harry:2010zz,Abbott:2016blz} or space--based eLISA \cite{AmaroSeoane:2012km}. For an inspiralling compact binary, such as GW150914 and GW151226, the frequency of gravitational waves emitted increases during the process. According to the dispersion relation \eqn\eqref{disrel}, the lower frequency gravitational waves, which are emitted earlier, travel slightly slower than the higher frequency gravitational waves. This difference in the propagating speed would distort the shape (or phasing) of the observed gravitational waveform, that is, the gravitational wave signal would be shorter or more squeezed than that in General Relativity \cite{Will:1997bb}.  In particular,  via the matched filtering technique, the signal duration when emitted $\Delta t_e$ can be determined very accurately \cite{Will:1997bb}, which can then be compared to the signal duration when observed $\Delta t_a$. Including the possible Hubble redshift effect, the ``time difference'' is $\Delta t = \Delta t_a - (1+z)\Delta t_e$, where $z$ is the redshift. For the observed gravitational waveform, this arrival difference contributes an extra gravitational wave phase term
\begin{equation}
	\Phi_\text{MG}(f) = -{D }/[4\pi{\lambdag^2(1+z)f}],
\end{equation}
where $f$ is the gravitational wave frequency and $D=(1+z)\int^{t_a}_{t_e} a(t)/a(t_a) \ud t$, where $a(t)$ is the scale factor at time $t$ and $t_e$ and $t_a$ are the emission and arrival time of the signal \cite{Will:1997bb}. Note that $D$ agrees with the standard luminosity distance $D_L$ when $z\ll 1$. Despite the apparent dependence on the distance of the gravitational wave source in this phase term, the mass bound obtained in this way only depends weakly on the distance (an $\mc{O}(1)$ factor). This is because while the distortion of the signal shape is enhanced with the distance the signal strength decreases with the distance \cite{Will:1997bb}.

 By using the dispersion relation alone, a rule--of--thumb estimate for the graviton Compton wavelength is given by
\begin{equation}
	\label{gcwlb}
	\lambdag > 5\times 10^{11}\unit{km} \left( \frac{D}{200\unit{Mpc}} \frac{100\unit{Hz}}{f} \frac{1}{f\Delta t} \right)^\frac12\,,
\end{equation}
where $D$ can be taken to be the usual luminosity distance. For experiments like aLIGO or eLISA, the phase distortion $f\Delta t$ can be measured up to $1/\rho$, where $\rho$ is the signal--to--noise ratio, so we may take $f\Delta t\sim 1/\rho$. For GW150914 with $D\sim 400 \unit{Mpc}$, $f\sim 100\unit{Hz}$ and $\rho\sim 23$, we have $\lambdag \gtrsim 10^{12}\unit{km}$, which is roughly the number quoted by aLIGO \cite{TheLIGOScientific:2016src}. A more careful calculation gives at 90\% confidence \cite{Will:1997bb,TheLIGOScientific:2016src} yields
\begin{equation}
	\label{aLIGObound}
	\bound{1.2\times 10^{-22}}{1.7\times 10^{12}}\,.
\end{equation}
For the event GW151226 the distance is roughly similar, $D\sim 400\unit{Mpc}$ but the Black Hole involved are lighter leading to a higher frequency, hence making the modifications to the dispersion relation in massive gravity less significant. In addition the signal--to--noise ratio is lower, $\rho\sim 13$  \cite{Abbott:2016nmj,TheLIGOScientific:2016pea}, so the bound on the graviton mass inferred by GW151226 are not as competitive as those inferred by GW150914 and do not significantly improve the bound \eqref{aLIGObound}. Eq.~(\ref{aLIGObound}) was also confirmed by Ref.~\cite{Yunes:2016jcc} which also further constrained the generic corrections on the graviton dispersion relation from the events GW150914 and GW151226.

For space--based laser--interferometer detectors such as eLISA, much higher signal--to--noise ratios can be achieved and much lower GW frequencies can be measured. From black hole binaries with masses of $10^4$ to $10^7M_\odot$, one would expect $D\sim 3\unit{Gpc}$, $f\sim 10^{-3}\unit{Hz}$, and $\rho\sim 10^3$, so that eLISA could typically constrain the graviton mass at
\begin{equation}
\label{eLISAbound}
	\pbound{10^{-26}}{10^{16}}.
\end{equation}
See Table I of \cite{Will:1997bb} for more details.

This predicted bound would be weakened by about an order of magnitude for massive black holes with aligned, non--precessing spins \cite{Scharre:2001hn,Berti:2004bd}; however, in a generic case once the effect of spin precession is incorporated, the bound can be restored \cite{Stavridis:2009mb}. For very massive black hole binaries, the bound can be raised by an order of magnitude when the gravitational waveform templates are improved by higher harmonics \cite{Arun:2009pq,Huwyler:2011iq}. For an ensemble of gravitational wave events, the combined mass bound can be improved by a factor greater than the square root of the number of events. For example, the bound can be raised by an order of magnitude by detecting about 50 events in a two year space--based mission \cite{Berti:2011jz}. More careful bounds are available by taking into account the specifics of Einstein Telescope, LISA, DECIGO and BBO\footnote{The specifics of eLISA, which has inherited and replaced LISA, may be different.} \cite{Will:2004xi,Berti:2004bd,Keppel:2010qu,Yagi:2013du} and by using eccentric binaries \cite{Jones:2004uy,Yagi:2009zm}. More sophisticated Bayesian statistics have been employed to model-select between massive gravity and General Relativity based on the gravitational waveform phasing difference and estimate the model parameters using the nested sampling (Monte Carlo) algorithm \cite{DelPozzo:2011pg}. This is claimed to slightly improve the mass bound compared to the Fisher information matrix approach used by \cite{Will:1997bb} and others.

Much insight can already been gained using a fully analytical waveform template based on the perturbative Post--Newtonian formalism that can model the inspiral stage of the coalescence process,  \cite{Will:1997bb}.
However recent progress in numerical and analytical relativity \cite{Keppel:2010qu} have led to a full improved inspiral--merger--ringdown waveform template. This allowed \cite{Keppel:2010qu} to fully take into account the whole waveform signal, which is expected to improve the mass bound by an order of magnitude. The only massive graviton effect considered in \cite{Keppel:2010qu} is the modified dispersion relation after the gravitational waveform is generated, \ie~when the gravitational waves are propagating. This approach has been justified by \cite{Yunes:2016jcc}, which estimates that the propagation effects are about 18 orders of magnitude stronger than
generation effects assuming a non--Fierz--Pauli structure for the graviton potential. In massive gravity theories, the black hole solutions and their horizons could be modified to become hairy and the merger (nonlinear strong gravity) and ringdown (quasi--normal modes) periods are particularly sensitive to the near horizon geometry. For a complete treatment, the hairy structure of the black hole should be included, which would then be model--dependent.

\subsubsection{Multi--Messenger Detection}

Another way to constrain the graviton mass is by detecting a \emph{multi--messenger} event, meaning that photons or neutrinos are detected alongside the gravitational waves. Then, the arrival phase or time difference between the two kinds of signals can be used to estimate the mass bound using relation \eqn\eqref{gcwlb}.

Interacting binary white dwarfs (IBWDs) are one of these systems where multi--messenger detection is feasible \cite{Larson:1999kg,Cutler:2002ef}. In a stellar binary system which emits both gravitational and electromagnetic waves like IBWDs, the energy of the emitted gravitons is much smaller than the energy of optical photons. Thus according to the dispersion relation \eqref{disrel} the photon velocity is much closer to $c=1$ than the graviton velocity even if the photon mass was much larger than the graviton mass.  This means that this method does not need to presume the photon is exactly massless or even have a better bound than the graviton; indeed, the current bounds for the photon mass are much weaker than those of the graviton mass (see the ``$\gamma$ mass'' section of \cite{Agashe:2014kda}).

A particularly interesting class of IBWDs are helium cataclysmic variables (HeCVs). HeCVs are binary systems where a white dwarf accretes hydrogen rich matter from its companion, creating unstable accretion disks. These disks create outbursts when part of them fall into the white dwarf. During these outbursts the stored potential energy is released through both gravitational and electromagnetic waves. One of the best known example of such systems is the AM Canum Venaticorum variable binary in the constellation of Canes Venatici.
A few HeCVs have been detected and have been studied optically. Many more are predicted to exist in the Galaxy and detectable by future space--based gravitational wave detectors such as eLISA \cite{Larson:1999kg}. To estimate the arrival phase difference, one needs to know the initial phase difference between the gravitational and electromagnetic wave signal. This can be achieved by making two observations of the source half a year apart, and can be more accurately determined by optically imaging those binaries' detailed structure \cite{Larson:1999kg}. The former method has the disadvantage of exchanging the galactic distance for the much smaller 2AU distance, which degrades the mass bound. The latter method yields a projected typical mass bound of
\begin{align}
\label{eqn:eLISA_IBWD}
	\pbound{10^{-24}}{10^{14}}.
\end{align}
For close orbiting white dwarf binaries, a mass bound of the same order of magnitude is also expected \cite{Cooray:2003cv}.

The method of bounding the graviton mass from short gamma--ray bursts (SGRBs) and supernovae (SNe) is very similar to the one described previously for IBWDs, \ie~by comparing the propagation time difference between multiple messengers. In the cases of SGRBs and SNe, the second messenger can be the photon, neutrino, or both. The seemingly non--trivial factor to determine is again the messengers' initial emission time difference. The model building and the numerical simulations of SGRBs and SNe have reached such a level of maturity that the initial time delays can be accurately determined, giving rise to some competitive graviton mass bounds. In the event where the detection of a SGRB or SN was accompanied with a direct gravitational wave detection, the resulting bound on the graviton mass could reach \cite{Nishizawa:2014zna}
\begin{equation}
\label{eqn:mdr_SNE}
	\pbound{10^{-20}}{10^{10}}.
\end{equation}

The possibility of real--time multi--messenger detections by a satellite mission such as LISA from supermassive black hole (SMBH) binary coalescences in the presence of gas has been considered in \cite{Kocsis:2007yu}. The accompanying electromagnetic signal may significantly help LISA/eLISA locate the gravitational wave source. A graviton mass bound can again be established by comparing the arrival times of the two messengers. However, for a SMBH, the timing uncertainty is limited by the dynamical scale of the SMBH merger which is one final gravitational wave cycle. This is sizable in constraining the graviton mass. For a SMBH with $10^6M_\odot$ at a cosmic distance, this translates to a detection limit of $m_g\sim 10^{-25}$eV. However, a better understanding of the dynamical process of the SMBH coalescence with gas may lower the detection limit by making use of the relation between the phasing of the two messengers \cite{Kocsis:2007yu}. 
It is also worth emphasizing that the previous results assumes that the gravitational and electromagnetic waves are emitted at the same time, an assumption which may lead to some systematic errors. 
A better understanding of the initial phasing difference between the gravitational and electromagnetic wave signals would therefore be useful.

\subsection{Bounds from Primordial Gravitational Waves}

It is naturally expected that physics at length scales comparable to the graviton Compton wavelength should be most sensitive to the scale of the graviton mass. For a mass of the cosmological scale, one would expect that cosmology gives the best bounds. The cosmic microwave background (CMB) probes the physics at the recombination epoch and has been precisely measured since the 1990s, ushering in the so-called era of ``precision cosmology''. Indeed, it has been shown that if the B--mode polarizations of the CMB were observed\footnote{Such observations are the targets of many experiments such as BICEP, Spider, POLARBEAR, CMBPol, etc. \cite{Ade:2014gua,Crill:2008rd,Ade:2014afa,Baumann:2008aq}.}, a massive graviton could give rise to a characteristic B--mode power spectrum at low $\ell$ if its mass was around the Hubble scale at that time of recombination \cite{Dubovsky:2009xk}. This is just four orders of magnitude bigger than the current Hubble constant.

A massive graviton modifies the tensor mode equation of motion by adding a mass term in the tensor mode evolution equation of the form \cite{Dubovsky:2009xk}
\begin{equation}
\label{eq:modTensorProp}
	\mc{D}''_q(\tau) + 2\frac{a'}{a} \mc{D}'_q(\tau)  + (q^2 + m_g^2 a^2)   \mc{D}_q(\tau) = J_q(\tau)  ,
\end{equation}
where $\mc{D}_q(\tau)$ is the tensor mode amplitude of the comoving momentum mode $q$, $\tau$ is the conformal time, $a$ is the scale factor and $J_q(\tau)$ is the external source. We stress that in \eqref{eq:modTensorProp}, the mass scale $m_g$ should be thought of as the effective graviton mass which may be redressed similarly as in \eqref{eq:redressing} and hence be significantly different from the bare mass in the early  Universe.

\begin{figure}[hb]
	\centering
	\includegraphics[width=.9\linewidth]{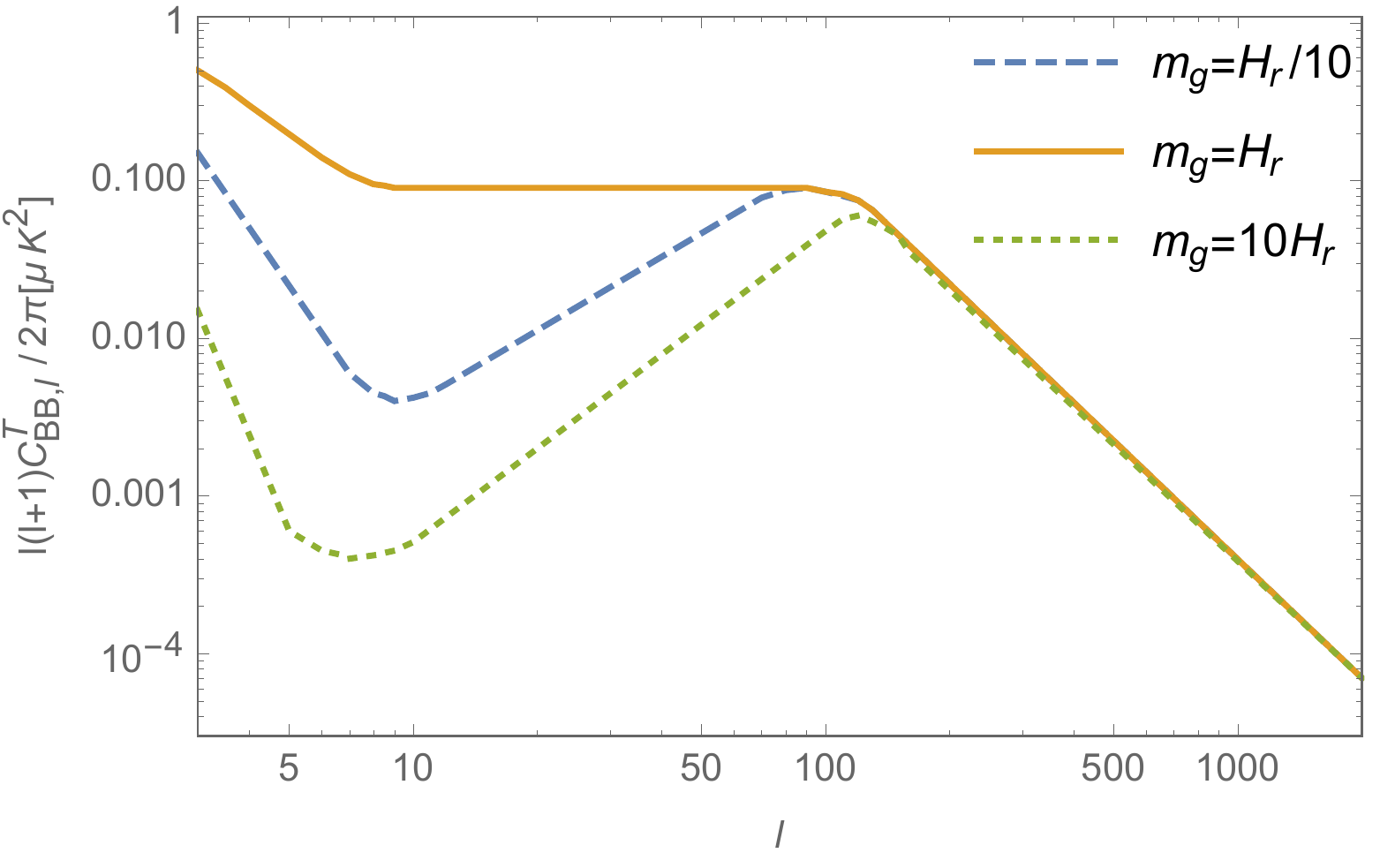}
	\caption{A schematic sketch of B--mode power spectra for a range of graviton masses (See \fig3 of \cite{Dubovsky:2009xk} for more details).  We use $H_r$ as the Hubble parameter at recombination. A characteristic plateau appears at $\ell<100$ when the graviton mass is of the order of the Hubble parameter at recombination. When $m_g$ is much bigger or smaller than $H_r$, the low $\ell$ power spectrum is, on the other hand, suppressed. }
	\label{fig:cmbgm}
\end{figure}

At high momenta $q\gg m_g a$, the additional mass $m_g$ is negligible, and the modes behave as in General Relativity.  In General Relativity, the tensor modes are frozen if their wavelengths are bigger than the Hubble horizon at the time, which does not contribute to the CMB temperature anisotropies. In a massive theory, on the other hand, the superhorizon wavelength modes will start to oscillate when the Hubble parameter drops below the graviton mass, which generates temperature anisotropy quadrupoles that can be converted to B--mode polarizations during the recombination and reionization epochs\footnote{This effect will also modify the gravitational wave power spectrum itself by generating a peak around the graviton mass \cite{Gumrukcuoglu:2012wt}. If observed by future low frequency detectors such as eLISA, DECIGO or pulsar timing arrays it can also be used to bound the mass of the graviton.} \cite{Dubovsky:2009xk}.  The characteristic signature of this massive graviton effect is that there will be a raised plateau for the $\ell<100$ modes of the B--mode spectrum if the mass is around the Hubble scale of the recombination or a couple of orders of magnitude above it \cite{Dubovsky:2009xk}. The amplitude of the raised plateau actually oscillates when $m_g$ varies, which can be linked to the local isotropy of the early Universe and is similar to the well--known acoustic peaks of the scalar modes \cite{Dubovsky:2009xk}. For an effective mass that is much bigger than the recombination Hubble scale, the relevant modes oscillate rapidly and the end result averages out, so the low $\ell$ B--modes get strongly suppressed. See \fig\ref{fig:cmbgm} for a schematic sketch. Observation of such a raised plateau at low $\ell$ in the ongoing and further CMB polarization experiments would strongly support that the graviton is massive, and a null observation can place a bound on the massive graviton:
\begin{equation}
\label{eqn:mdr_cmb}
	\pbound{10^{-30}}{10^{20}}.
\end{equation}
Note that this B--mode signature is very clean as the primordial gravitational waves decouple from the environment after their production. Furthermore other secondary B--mode sources are negligible at $\ell<100$. Also, it seems to be difficult to come up with inflation models or other mechanisms to mock up such a signature \cite{Dubovsky:2009xk}. For example, even a significant modification of the gravitational wave sound speed from one would mostly just shift the B--mode spectrum slightly \cite{Raveri:2014eea} (see also \cite{Amendola:2014wma}). Detection of primordial polarization would hence provide a clear bound on the effective graviton mass at the time of recombination. Effects coming from a squeezed vacuum state on B mode polarization in Lorentz--breaking massive gravity were also recently considered in \cite{Malsawmtluangi:2016agy} and could potentially provide additional bounds for these theories. Recently, the improvement of the graviton mass bound from proposed future CMB experiments has been studied, and it is projected, assuming a tensor-to-scalar ratio of $r=0.01$, that the bound can be pushed down to $m_g\sim 10^{-32}{\rm eV}$ (which is comparable to the graviton mass that may produce the late time cosmic acceleration) for the experiments such as COrE (Cosmic Origins Explorer), CMB Stage-IV and PIXIE (Primordial Inflation Explorer) \cite{Lin:2016gve}.

%%%%%%%%%%%%%%%%%%%%%%%%%%%%%%%%

%%%%%%%%%%%%%%%%%%%%%%%%%%%%%%%%%

\subsection{Bounds from Indirect Gravitational Wave Detection}

\subsubsection{Pulsar Timing}

The timing of the periodic pulses of binary pulsars can be measured by radio telescopes with extraordinary precision. The gravitational field around a binary pulsar is relatively strong, so its gravitational wave emission can induce sizable changes for the period of the binary pulsar. The measured changes can be used to precisely constrain gravitational theories.

As mentioned in \Sec\ref{sec:ReviewMG}, gravitational waves typically have more polarizations in massive gravity as compared to General Relativity in which there are only two transverse tensor modes. The existence of these additional polarizations in massive gravity theories could lead to new channels of gravitational wave radiation which can be constrained by observations. Since this effect arises from the new polarizations, its precise effect is discussed in \Sec\ref{sec:BinaryPulsar Galileon} as part of the discussion on fifth force effects. In this section we focus solely on the fact that the two transverse tensor modes already present in General Relativity could have a mass. This leads to a different gravitational wave power emitted in binary pulsar systems and thus a bound on the graviton mass.

This is the rationale followed by \cite{Finn:2001qi}. As mentioned in \Sec\ref{sec:nonFP}, the precise model considered in \cite{Finn:2001qi} is not physically viable as it suffers from an incurable ghost--like instability. Nevertheless, the behavior of the tensor modes is the same as that in a healthy theory of massive gravity and the bounds provided by \cite{Finn:2001qi} are in principle applicable to binary pulsar systems. The calculation of \cite{Finn:2001qi} follows closely that of General Relativity, but the field equation is modified to
\begin{equation}
	(\Box - m_g^2) \lp h_{\mu\nu}-\frac12 h\eta_{\mu\nu} \rp  = - 16\pi G T_{\mu\nu}   ,
\end{equation}
where $T_{\mu\nu}$ is the matter energy momentum tensor. After computing the gravitational wave power emission and comparing that to the data from PSR B1913+16 and PSR B1534+12, the combined bound at $90\%$ confidence is
\begin{equation}
	\label{eqn:mdr_pulsar}
	\bound{7.6 \times 10^{-20}}{2.6\times 10^9}.
\end{equation}

There is a theoretical ambiguity in making only the tensor modes massive. In the massless case, the tensor mode $h^\text{TT}_{ij}$ is given by the transverse and traceless part of $h_{ij}$. When generalizing to have a mass term for the tensor mode, one may choose to keep the transverse and traceless projection and thus have a mass term proportional to $h^\text{TT}_{ij} h^\text{TT}{}^{ij}$. However this breaks Lorentz invariance. The approach taken by  \cite{Finn:2001qi} is to take the equations for linear General Relativity and simply add a mass term for the tensor mode by replacing $\Box \to \Box - m_g^2$. These two approaches ``minimally'' generalize the massless case in a different way. They both formerly reduce to General Relativity in the small $m_g$ limit, however the two approaches differ by an $\mc{O}(m_g^2)$ correction. The bounds from the binary pulsar systems as derived in  \cite{Finn:2001qi}  precisely relies on the $\mc{O}(m_g^2)$ effect, which is affected by this ambiguity. A fully consistent calculation in a consistent Lorentz invariant theory has not yet been done, however it is likely that the \cite{Finn:2001qi} gives the correct bound within a factor of order unity and thus despite its theoretical problems may still be taken seriously.

\subsubsection{Pulsar Timing Arrays}

A passing gravitational wave can shift the frequency of the very precise pulsar timing. An array of widely spaced millisecond pulsars can hence be used as a nano--Hertz gravitational wave detector by timing the pulsars together coordinately \cite{Hobbs:2009yy}. By analyzing the angular correlation between the timing residual of pulsar pairs, it is possible to detect a stochastic nano--Hertz gravitational wave background from massive black hole binaries \cite{Lee:2010cg,Lee:2014awa}. If the graviton is massive, the dispersion relation \eqref{gcwlb} indicates that a gravitational wave with frequency less than $m_g$ attenuates or does not propagate. This can significantly change the angular correlation of the timing residuals by flattening out the power spectrum for a graviton mass no bigger than the inverse of the observing time \cite{Lee:2010cg}. At least 60 pulsars are required to make such a measurement. For an array of 300 pulsars with 100\unit{ns} timing accuracy a 10 year observation would bound the graviton mass \cite{Lee:2010cg}
\begin{equation}
\label{eqn:mdr_pta}
	\pbound{3\times 10^{-23}}{7\times 10^{12}}.
\end{equation}

\subsection{Bounds from Graviton Decay}

\begin{figure}[h]
		\centering
		% \begin{fmffile}{gravdecay}
		% 	\fmfframe(10,10)(10,10){
		% 	\begin{fmfgraph*}(120,80)
		%        \fmfstraight
		% 		\fmfleft{i1} \fmfright{o1,o2}
		% 		\fmf{gluon,tension=2}{i1,v1}
		% 		\fmf{photon}{o1,v1,o2}
		% 		\fmflabel{graviton}{i1}
		% 		\fmflabel{photon}{o2}
		% 		\fmflabel{photon}{o1}
		% 	\end{fmfgraph*}
		% 	}
		% \end{fmffile}
		%\includegraphics[width=0.9\linewidth]{gravdecay.eps}
		\includegraphics[width=0.9\linewidth]{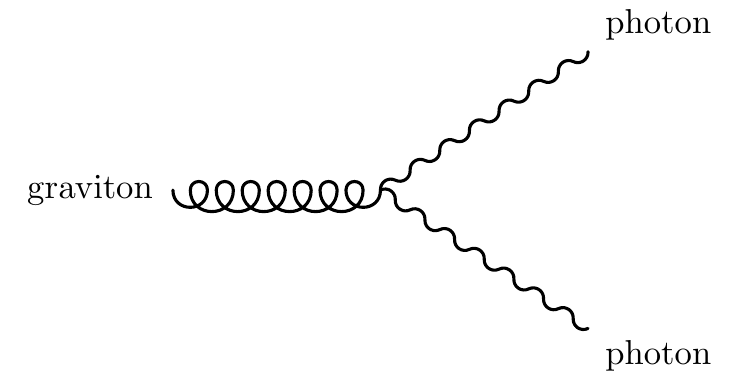}
		\caption{Allowed tree level graviton decay diagram if the graviton is massive (and one can go in its rest frame).}
		\label{fig:fenymdiag1}
	\end{figure}

If the graviton is not the lightest particle, then it will inevitably be unstable since it necessarily couples to the lightest particle. In particular, if in a given theory the photon remains massless, then the graviton will be unstable to decay into two photons, see Fig.~\ref{fig:fenymdiag1}. In addition, in theories where there exist multiple light scalar fields (\eg~axions, possibly in a hidden sector and thus not subject to Standard Model constraints), the graviton may decay into many different channels by an amount which grows with the number of lighter degrees of freedom. The observation of gravitational waves from aLIGO puts a very minimal constraint that the decay time for the gravitational waves is longer than the time taken for the waves to reach us from their binary black hole merger source. More generally, one may worry that if the decay time for the graviton were shorter than a Hubble time, a given massive gravity theory may run into conflict with current cosmological observations. A conservative bound on the graviton decay rate is then
\begin{equation}
	\Gamma \ll H_\text{today},
\end{equation}
that is
\begin{equation}
	\text{Im}[m^2_g] \ll H_\text{today} \sqrt{\text{Re}[m^2_g]}.
\end{equation}
In models where the graviton is a resonance, such as the DGP model, the graviton already has a finite lifetime even without taking into account its possible decay into photons. In particular, for the DGP model one may easily show that the decay rate is of the order of the graviton mass and so this conservative bound would imply $m_g \lesssim H_\text{today} $, which is indeed necessary for the DGP model to produce early time cosmology correctly. On the other hand, in a theory of hard mass gravity, at tree level in the graviton propagator $\text{Im}[m^2_g]=\Gamma=0$, and we must look at loop effects on the graviton self--energy, which by the optical theorem correspond to tree level decays of gravitons into lighter particles. Since any such decay is necessarily Planck suppressed, then on dimensional grounds alone we may estimate the magnitude of such a decay to be
\begin{equation}
	\Gamma \sim  N \frac{m_g^3}{\mpl^2} \, ,
\end{equation}
where $N$ is the number of lighter species. This conservative constraint then amounts to
\begin{equation}
	m_g \lesssim (H_\text{today} \mpl^2)^{1/3} N^{-1/3} \sim 10^7 \text{eV} \times N^{-1/3} \, .
\end{equation}
The number of lighter species must necessarily be exponentially large for this bound to become in any way competitive with any of the other bounds discussed. \\

%%%%%%%%%%%%%%%%%%%%%%%%%%%%%%%%%
	
%%%%%%%%%%%%%%%%%%%%%%%%%%%%%%%%%

A slight variant on this bound is given in  \cite{Hare:1973px}, where the author considers a gravitational wave being detected at a given frequency $\omega$. In the rest frame of the receiver, the bound from the decay is then weakened by the associated Lorentz boost factor $\gamma=\omega/m_g$, i.e.~in the rest frame of the receiver $\Gamma \sim N m_g^4/(\omega \mpl^2)$. If the distance travelled by light from its emitter is $\tau$, then the bound becomes
\begin{equation}
m_g \lesssim \gamma^{1/3}(\tau^{-1} \mpl^2)^{1/3} N^{-1/3}  \, .
\end{equation}
Since $\gamma \ge 1$ and $\tau \lesssim H_{\rm today}^{-1}$ this bound is always weaker than the previous one\footnote{Note that \cite{Hare:1973px} claims a stronger bound, but the associated Lorentz boost $\gamma$ for the masses considered would be less than unity and hence unachievable.}.
\\

We note that an interesting conundrum arises in the calculation of the decay rate in a hard mass gravity theory. The decay rate is determined by the imaginary part of the self--energy in the graviton propagator. In a hard mass theory, this imaginary part arises only at the loop level, \eg~by an intermediate photon loop. This is consistent with the optical theorem whereby the imaginary part is determined by the squared modulus of the tree level process for a graviton to decay into two photons. The conundrum is that if photons couple to gravity in a diffeomorphism invariant way, which is usually assumed in massive gravity theories, then their loop effects cannot break diffeomorphism invariance. Naively a diffeomorphism invariant operator is not expected to generate a mass term, and therefore is not expected to correct the self--energy, which would lead to a contradiction with the optical theorem. The resolution is that in a theory in which diffeomorphism invariance is already broken, so a non--zero mass $m_g$ is already present, the addition of a diffeomorphism invariant operator (\eg~$R \ln[-\Box] R$, of the type that arise from integrating out massless particles) will shift the pole in the propagator by an amount that tends to zero as $m_g \rightarrow 0$, \ie~$\Delta m_g \sim m_g^3/\mpl^2$. This is then consistent with the recovery of the symmetry as $m_g \rightarrow 0$. This argument is closely analogous to the technical naturalness argument used to imply that the graviton mass is not expected to receive large quantum corrections due to the reemergence of the symmetry as $m_g \rightarrow 0$ \cite{deRham:2012ew,deRham:2013qqa}. In other words the graviton decay width is guaranteed to be small by the same naturalness arguments, if it is already small at tree level. As we have seen the exception to this is if the number of particles contributing in the loop is itself exponentially large.

\section{Fifth Force}
\label{sec:FF}

In this section, we review the bounds on the graviton mass from the fifth force\footnote{By fifth force, we mean an interaction that is different from the four known interactions (the three Standard Model interactions plus the GR interaction). In this language, we will consider both static and radiative contributions to the force.} constraints on the additional modes that arise in massive gravity models. In doing so we have in mind principally Lorentz invariant models or Lorentz violating models that continue to have a linear vDVZ discontinuity. Lorentz violating models with no vDVZ discontinuity do not by construction exhibit significant fifth forces. Our central focus will be the helicity--0 mode. In principle, the helicity--1 modes can induce some ``fifth force'', but in the decoupling limit, the helicity--1 modes usually do not couple to matter directly, whereas the helicity--0 modes do. In a given nonlinear massive gravity model, a nonzero coupling will be turned on for the helicity--1 mode (and of course interactions are generated at loop level), however the helicity--1 interactions will be inevitably significantly suppressed relative to the helicity--0 interactions provided only that $m_g \ll \mpl$. Although at the linear level, the helicity--0 modes appear to couple to matter with equal strength as the helicity--2 modes, as mentioned in \Sec\ref{sec:nonvain}, this is not the case in the presence of matter thanks to the nonlinear Vainshtein screening mechanism. Nevertheless, they do continue to induce small fifth forces, which are tightly constrained in the solar system and beyond.

The mass bounds in this section are some of the best bounds on the graviton mass, reaching or approaching the order of the Hubble scale. These bounds may appear to be model dependent, as the discussion involves explicit models such as DGP \cite{Dvali:2000rv,Dvali:2000hr} and dRGT \cite{deRham:2010ik,deRham:2010kj}, and often makes explicit use of their nonlinear interactions in the decoupling limit, which are of the Galileon form. But we emphasize that these bounds are arguably less model dependent than they seem, because in models containing the helicity--0 mode, this mode is naturally Galileon--like, as in the \stu decomposition two indices of the metric become two partial derivatives which respect the Galileon symmetry (see \eqn\eqref{hmnpi} and \Sec\ref{sec:galileont}).
Indeed the one universal feature of all Lorentz invariant massive gravity models is the Galileon field.
This also occurs in other models of massive gravity like Cascading Gravity \cite{deRham:2008zz,deRham:2007rw}  as well as three--dimensional New Massive Gravity \cite{Bergshoeff:2009hq,deRham:2011ca}.

In the dRGT case we shall focus on those models for which $a_3=0$, since including $a_3$ typically leads to instabilities \cite{Berezhiani:2013dca}.

\subsection{Bounds from the Solar System}

\subsubsection{Lunar Laser Ranging Experiments}

The lunar laser ranging experiment \cite{Williams:2004qba} has become an important tool to constrain gravitational models. The measurements of the perihelion precession of the lunar orbit have reached a sensitivity of $10^{-11}$ \cite{Merkowitz:2010kka}, and strong bounds on the graviton mass can be derived from them. As mentioned above, the decoupling limits of the DGP and dRGT models (see \Sec\ref{sec:DGPDL} and \Sec\ref{sec:dRGTDL}) are used for the bounds reviewed here. The decoupling limit of these theories are good approximations of the full DGP and dRGT models in the context of the lunar laser ranging experiment, assuming that the Compton wavelength $\lambdag$ is much longer than the distance between the Earth and the Moon.

\paragraph{DGP}

In the decoupling limit, the DGP model reduces to a linearized helicity--2 sector plus a cubic Galileon field theory, and the graviton mass manifests in this limit in the strong coupling scale $\Lambda_3=(m_g^2\mpl)^{1/3}$. Matter couples to the following metric perturbation about flat spacetime
\begin{equation}
	h_{\mu\nu}=h_{\mu\nu}^\text{GR}+\pi\eta_{\mu\nu}
\end{equation}
where $\pi$ is the Galileon field, playing the role of the additional helicity--0 mode and giving rise to the fifth force. The change in the gravitational potential of the earth is given by $\delta\Phi=\Phi-\Phi^\text{GR}=-\pi$, where $\Phi^\text{GR}={r_{S,\oplus}  \mpl}/{r}$, with $r_{S,\oplus}$ the Schwarzschild radius of the earth, and $\pi$ is obtained by solving the static and spherically symmetric Galileon field equation derived from \eqn\eqref{dgpdlimit}. In solving the Galileon equation, it is important to recognize the nonlinear scale of the equation, that is, the Vainshtein radius of the earth, which comes in at $r_{V,\oplus}=\left({r_{S,\oplus}}/{m_g^2}\right)^{1/3}$ (\eg~\cite{Dvali:2002vf}). The Vainshtein radius is huge for the range of $m_g$ we are exploring here, and the moon is firmly within that radius. In this limit $r\ll r_{V,\oplus}$, the Galileon equation is dominated by the highest order term, and can be easily solved to give
\begin{equation}
	\partial_r\pi\sim\frac{r_{S,\oplus}  \mpl}{r_{V,\oplus}^{3/2}r^{1/2}}\,.
\end{equation}

The fractional shift in the gravitational potential induces an anomalous perihelion precession of the lunar orbit given by
\begin{align}
	\delta\phi&=\pi a\partial_r\left[r^2\partial_r\left(r^{-1}\frac{\delta\Phi}{\Phi^\text{GR}}\right)\right]_{r\to a}\,,
	\label{eq:Galpre}
\end{align}
where $a$ is the semi--major axis of the lunar orbit. Evaluating this derivative we find the precession in DGP to be
\begin{equation}
	\label{eq:ShiftCuGal}
	\delta\phi\sim\frac{3\pi}{2}\left(\frac a{r_{V,\oplus}}\right)^{3/2}\,.
\end{equation}
Solving this for a bound on the graviton mass yields
\begin{equation}
	\label{eq:mCuGal}
	m_g<\frac{4}{3\pi}\delta\phi\left(\frac{r_{S,\oplus} }{a^3}\right)^{1/2}.
\end{equation}
With the sensitivity of the lunar laser range experiments $\delta\phi\sim 10^{-11}$, the earth--moon system thus gives a mass bound \cite{Dvali:2002vf}\footnote{The original logic in this paper is inverse of establishing this bound. There, in the hight of dark energy frenzy, $m_g$ is assumed to be around the Hubble scale, and it was argued that the lunar laser range experiments were about to reach the sensitivity to detect a fifth force from the DGP model.}
\begin{equation}
	\label{eqn:llrDGP}
	\bound{10^{-32}}{10^{22}}
\end{equation}
which is interestingly around the current Hubble scale.

\paragraph{dRGT}

The full decoupling limit of the dRGT model also involves the helicity--1 modes \cite{Gabadadze:2013ria,Ondo:2013wka}, but the helicity--1 modes do not couple to the matter directly. If the helicity--1 modes are neglected along with higher order helicity--2 to matter couplings, it is just a linearized helicity--2 sector plus the full Galileon field theory, which may well capture the correct order of magnitude of the mass bound, and the graviton mass again appears in the strong coupling scale $\Lambda_3=(m_g^2\mpl)^{1/3}$. In this simplified theory, the only difference, compared to the DGP model, is that now one needs to solve a quartic static and spherically symmetric Galileon equation, which is equally easy to do in the limit $r\ll r_{V,\oplus}$,
\begin{equation}
	\partial_r\pi\sim\frac{r_{S,\oplus}\mpl}{r_{V,\oplus}^2}.
\end{equation}
Again, evaluating \eqn\eqref{eq:Galpre} we find the precession in this decoupling limit
\begin{equation}
	\label{eq:ShiftQuadGal}
	\delta\phi\sim\pi\left(\frac a{r_{V,\oplus}}\right)^{2} ,
\end{equation}
yielding the bound on the graviton mass
\begin{equation}
\label{eq:mQuadGal}
	m_g<\left(\frac{\delta\phi}{\pi}\right)^{3/4} \left(\frac{r_{S,\oplus} }{a^{3}}\right)^{1/2}.
\end{equation}
Note that, compared to the DGP case, the graviton mass is less sensitive to the perihelion in the dRGT model, as it comes in as $\delta\phi^{3/4}$. Using the sensitivity of the lunar laser range experiments the graviton mass bound from the perihelion precession of the lunar orbit is given by
\begin{equation}
\label{eqn:4gal_moon}
	\bound{10^{-30}}{10^{20}}\,,
\end{equation}
which is slightly weaker than the bound for the DGP model \eqref{eqn:llrDGP}.

\subsubsection{Planetary Orbits}

There have also been attempts to bound the graviton mass from the full theory of the DGP model using the planet orbits in the solar system.
In \cite{Gruzinov:2001hp} the corrections to the Schwarzschild solution in the full five--dimensional DGP theory were computed. In the small $m_g$ limit the corrections behave as
\begin{equation}
	\frac{\delta \Phi}{\Phi^\text{GR}}=-m_g\left(\frac{8r^3}{r_{S,\odot} }\right)^{\frac12}.
\end{equation}
This is also consistent with another bound also derived from a full five--dimensional picture for the DGP model by \cite{Lue:2002sw}, where the additional perihelion precession over General Relativity is calculated to be $\delta\phi=3\pi m_g\left({r^3}/\Rs\right)^{1/2}/2^{3/2}$. Both of these calculations lead to the graviton mass being proportional to
\begin{equation}
	\label{eqn:dgp_precess_mass}
	m_g<\delta\phi\left(\frac\Rs{r^3}\right)^{1/2}.
\end{equation}
In either calculation the fractional $\delta\phi$ deviation is then required to be less $10^{-8}$ at $r\sim 5\unit{AU}$ based on the precision of Jupiter's orbit from \cite{Talmadge:1988qz}. This bounds the graviton mass at the $2\sigma$ level to  \cite{Gruzinov:2001hp}
\begin{equation}
\label{eqn:5dDGP_precess}
	\bound{10^{-32}}{10^{22}}.
\end{equation}
Note that this bound is the same order as the bound from the decoupling limit of the DGP model using the lunar laser range experiments in \eqref{eqn:llrDGP}. Furthermore it has the same suppression in terms of $\delta\phi$, the Schwarzschild radius, and the semi--major axis as from the Lunar precession in \eqref{eq:mCuGal}. This suggests that for the dRGT theory we would expect, via a similar calculation, that the Solar System provides a bound that schematically goes as what we found in \eqn\eqref{eq:mQuadGal},
\begin{equation}
	m_g<\delta\phi^{3/4} \left(\frac{\Rs }{r^{3}}\right)^{1/2}.
\end{equation}

\subsection{Bounds from Binary Pulsars}
\label{sec:BinaryPulsar Galileon}

As discussed in \Sec\ref{sec:ReviewMG}, in massive gravity theories gravitational waves will have additional polarizations. In the DGP and dRGT model the extra polarization from the scalar Galileon mode would also radiate power from Binary Pulsar systems. Systems like the Hulse--Taylor Pulsar \cite{Hulse:1974eb,Weisberg:2004hi,Taylor:1989sw} can therefore give bounds for the graviton mass in Galileon theories based on the limits on extra power that can be emitted in these Galileon modes \cite{deRham:2012fw}. To date, the Hulse--Taylor system still remains the best pulsar system to bound the graviton mass. This is mainly because most of the other binary pulsar systems are more relativistic, leading to a more efficient Vainshtein mechanism which suppresses the power emitted into the helicity--0 mode.
Observations of the Hulse--Taylor system can be used to put a bound on the graviton mass. In particular, for any theory which is well  captured by a cubic Galileon in its decoupling limit one gets
\cite{deRham:2012fw}
\begin{equation}
\label{eqn:cuGal_pulsar}
	\bound{10^{-27}}{10^{17}}\,.
\end{equation}

When including the quartic Galileon terms that arise in some massive gravity theories, like in the dRGT model,  the perturbation theory about the static and spherically symmetric configurations used to derive this bound breaks down \cite{deRham:2012fg}. This breakdown should really be understood as a failure of the approximation used. In other words, perturbations are non--linear and thus should contribute to their own Vainshtein suppression \cite{deRham:2012fg}. Numerical work is currently underway  to solve this quartic Galileon system exactly \cite{NumericalPulsar}.

\subsection{Bounds from Structure and Lensing}

The existence of the helicity--0 mode of the graviton (the Galileon mode) can in principle strengthen lensing potentials. In the decoupling limit of the DGP model, the helicity--0 mode couples to the trace of the stress--energy tensor and therefore decouples from the photon in that limit. In the dRGT model, however, the theory also involves a coupling of the form $\partial_\mu \pi \partial_\nu \pi T^{\mu\nu}$ which implies that photon are sensitive to the helicity--0 mode.
 For dRGT massive gravity, the lensing potential $\Phi_L=(\Phi-\Psi)/2$ ($\Phi$ and $\Psi$ being the usual scalar metric perturbations in Newtonian gauge) would be modified by adding
\begin{equation}
	\Delta\Phi=\frac{1}{m_g^2\mpl}(\partial_r\pi)^2
\end{equation}
to $\Phi$ as worked out in \cite{Wyman:2011mp}. For a spherically symmetric body, this will yield a fractional change in the lensing potential of
\begin{equation}
	\frac{\Delta\Phi_L}{\Phi_L^\text{GR}}=\frac{r}{4\Rv}  \left(\frac{r}{\Rv}-\sqrt[3]{\left(\frac{r}{\Rv}\right)^3+1}\right)^2.	
\end{equation}
This peaks at 4\% at $r/\Rv\sim0.33$. As stated in \cite{Wyman:2011mp,Park:2014aga}, putting an observational constraint on this would best be done with galaxy--galaxy weak lensing because of the intermediate length scale at which this effect peaks. Current surveys like the Sloan Digital Sky Survey III \cite{Alam:2015mbd} do not have the fine mass binning to put constraints on the graviton mass but a survey with four mass bins within a range of $10^{13}M_\odot-10^{14}M_\odot$ could in principle put a $4\sigma$ constraint on the graviton mass of
\begin{equation}
\label{eqn:drgt_weak}
	\pbound{10^{-33}}{10^{23}}.
\end{equation}
Although this projected constraint is insensitive to the concentration of the NFW halo, it may be sensitive to different halo profiles that could arise in this modified gravity theory.

Future experiments may also be able to put a bound on the graviton mass from the change in structure formation from the fifth force. Such observations are predicted in \cite{Khoury:2009tk,Wyman:2011mp,Zu:2013joa,Park:2014aga,Barreira:2016ovx} to give bounds on the order of
\begin{equation}
\label{eqn:ff_structure}
	\pbound{10^{-32}}{10^{22}}\,.
\end{equation}

\section{Discussions}
\label{sec:clu}

We have presented a list of competitive bounds on the graviton mass which is summarized in Table~\ref{tab:massbounds}.   For bounds from the same method, we only listed the typical, representative mass bound. In addition to the constraints presented so far, there are a few additional effects which could provide model--dependent constraints. For instance the cosmological evolution of the Universe is expected to be very sensitive to the graviton mass.  If the graviton mass is of the current Hubble scale, then the late time cosmological observations should be good probes to establish a tighter bound. Indeed, there is a vast literature on these, most of which, however, are very much model dependent. For instance, the DGP model affects cosmology in a significantly different way as compared to the dRGT model or Lorentz--breaking massive gravity. The review of these bounds would warrant a separate review and we defer to the review \cite{deRham:2014zqa}.

Moreover, there could be additional constraints coming from the existence of different polarization of gravitational waves in Lorentz--invariant theories which we have not explicitly discussed here. In the case of the recent direct detections of Gravitational Waves GW150914 and GW151226, the data did not allow a precise comment on the potential existence of other polarizations. In the future, when Virgo is included or for multi--messenger detections with companion electromagnetic or neutrino signals, a much better handle on the other polarizations could be achieved since the direction of the gravitational wave signal may be inferred from the other messenger. The existence of additional polarizations would automatically imply new physics beyond General Relativity and could be a strong hint for massive gravity or other models of modified gravity. However the reverse situation where no additional polarizations are observed would likely put little to no bound on the graviton mass. This is because during black hole or neutron star mergers, the Vainshtein mechanism is expected to be highly active and therefore to significantly suppress the amount of radiation emitted in these additional modes (see for instance \cite{deRham:2012fw} for related discussions.)

Another aspect we have not considered is Cherenkov radiation.  For Lorentz violating theories of gravity (including Lorentz breaking theories of massive gravity) the equivalent of Cherenkov radiation is possible where particles that travel faster than the graviton may radiate gravitational waves \cite{Blas:2016qmn}. This process is kinetically forbidden in Lorentz invariant theories of gravity but can occur if Lorentz invariance is broken in the gravitational sector or in the coupling between gravity and matter. This process can be used to put a bound on the difference between the speed of light and gravity from high energy cosmic rays \cite{Moore:2001bv, Caves:1980jn} and constrain modifications of the dispersion relation for gravitational waves \cite{Kiyota:2015dla}. This bound on the difference in speeds translates into a very weak bound for the graviton mass even if Lorentz violation is purely coming from the graviton mass. A bound on the speed of propagation of gravitational waves was also derived in the context of Galileon scalar-tensor theories at the level of $10^{-2}$ via binary pulsar observations \cite{Jimenez:2015bwa}.

A Yukawa gravitational potential with a small graviton mass would also be able to explain the observed lack of power in the CMB low $\ell$. The relative weakening of the gravitational potential at large scales would modify structure growth at large angles and thus decrease the power at these same scales \cite{Kahniashvili:2014wua}.

\begin{acknowledgments}
	CdR is supported by a Department of Energy grant DE-SC0009946. AJT and SYZ are supported by Department of Energy Early Career Award DE-SC0010600. We would like to thank Clifford Will and Leo Stein for comments.
\end{acknowledgments}

\bibliography{refs}

%merlin.mbs apsrmp4-1.bst 2010-07-25 4.21a (PWD, AO, DPC) hacked
%Control: author (3) reversed first dotless
\begin{thebibliography}{165}%
\makeatletter
\providecommand \@ifxundefined [1]{%
 \@ifx{#1\undefined}
}%
\providecommand \@ifnum [1]{%
 \ifnum #1\expandafter \@firstoftwo
 \else \expandafter \@secondoftwo
 \fi
}%
\providecommand \@ifx [1]{%
 \ifx #1\expandafter \@firstoftwo
 \else \expandafter \@secondoftwo
 \fi
}%
\providecommand \natexlab [1]{#1}%
\providecommand \enquote  [1]{``#1''}%
\providecommand \bibnamefont  [1]{#1}%
\providecommand \bibfnamefont [1]{#1}%
\providecommand \citenamefont [1]{#1}%
\providecommand \href@noop [0]{\@secondoftwo}%
\providecommand \href [0]{\begingroup \@sanitize@url \@href}%
\providecommand \@href[1]{\@@startlink{#1}\@@href}%
\providecommand \@@href[1]{\endgroup#1\@@endlink}%
\providecommand \@sanitize@url [0]{\catcode `\\12\catcode `\$12\catcode
  `\&12\catcode `\#12\catcode `\^12\catcode `\_12\catcode `\%12\relax}%
\providecommand \@@startlink[1]{}%
\providecommand \@@endlink[0]{}%
\providecommand \url  [0]{\begingroup\@sanitize@url \@url }%
\providecommand \@url [1]{\endgroup\@href {#1}{\urlprefix }}%
\providecommand \urlprefix  [0]{URL }%
\providecommand \Eprint [0]{\href }%
\providecommand \doibase [0]{http://dx.doi.org/}%
\providecommand \selectlanguage [0]{\@gobble}%
\providecommand \bibinfo  [0]{\@secondoftwo}%
\providecommand \bibfield  [0]{\@secondoftwo}%
\providecommand \translation [1]{[#1]}%
\providecommand \BibitemOpen [0]{}%
\providecommand \bibitemStop [0]{}%
\providecommand \bibitemNoStop [0]{.\EOS\space}%
\providecommand \EOS [0]{\spacefactor3000\relax}%
\providecommand \BibitemShut  [1]{\csname bibitem#1\endcsname}%
\let\auto@bib@innerbib\@empty
%</preamble>
\bibitem [{\citenamefont {Abbott}\ \emph
  {et~al.}(2016{\natexlab{a}})\citenamefont {Abbott} \emph
  {et~al.}}]{TheLIGOScientific:2016pea}%
  \BibitemOpen
  \bibfield  {author} {\bibinfo {author} {\bibnamefont {Abbott}, \bibfnamefont
  {B~P}},  \emph {et~al.} (\bibinfo {collaboration} {Virgo, LIGO Scientific})}
  (\bibinfo {year} {2016}{\natexlab{a}}),\ \bibfield  {title} {\enquote
  {\bibinfo {title} {{Binary Black Hole Mergers in the first Advanced LIGO
  Observing Run}},}\ }\href {\doibase 10.1103/PhysRevX.6.041015} {\bibfield
  {journal} {\bibinfo  {journal} {Phys. Rev.}\ }\textbf {\bibinfo {volume}
  {X6}}~(\bibinfo {number} {4}),\ \bibinfo {pages} {041015}},\ \Eprint
  {http://arxiv.org/abs/1606.04856} {arXiv:1606.04856 [gr-qc]} \BibitemShut
  {NoStop}%
%%CITATION = ARXIV:1606.04856;%%
\bibitem [{\citenamefont {Abbott}\ \emph
  {et~al.}(2016{\natexlab{b}})\citenamefont {Abbott} \emph
  {et~al.}}]{Abbott:2016nmj}%
  \BibitemOpen
  \bibfield  {author} {\bibinfo {author} {\bibnamefont {Abbott}, \bibfnamefont
  {B~P}},  \emph {et~al.} (\bibinfo {collaboration} {Virgo, LIGO Scientific})}
  (\bibinfo {year} {2016}{\natexlab{b}}),\ \bibfield  {title} {\enquote
  {\bibinfo {title} {{GW151226: Observation of Gravitational Waves from a
  22-Solar-Mass Binary Black Hole Coalescence}},}\ }\href {\doibase
  10.1103/PhysRevLett.116.241103} {\bibfield  {journal} {\bibinfo  {journal}
  {Phys. Rev. Lett.}\ }\textbf {\bibinfo {volume} {116}}~(\bibinfo {number}
  {24}),\ \bibinfo {pages} {241103}},\ \Eprint
  {http://arxiv.org/abs/1606.04855} {arXiv:1606.04855 [gr-qc]} \BibitemShut
  {NoStop}%
%%CITATION = ARXIV:1606.04855;%%
\bibitem [{\citenamefont {Abbott}\ \emph
  {et~al.}(2016{\natexlab{c}})\citenamefont {Abbott} \emph
  {et~al.}}]{Abbott:2016blz}%
  \BibitemOpen
  \bibfield  {author} {\bibinfo {author} {\bibnamefont {Abbott}, \bibfnamefont
  {B~P}},  \emph {et~al.} (\bibinfo {collaboration} {Virgo, LIGO Scientific})}
  (\bibinfo {year} {2016}{\natexlab{c}}),\ \bibfield  {title} {\enquote
  {\bibinfo {title} {{Observation of Gravitational Waves from a Binary Black
  Hole Merger}},}\ }\href {\doibase 10.1103/PhysRevLett.116.061102} {\bibfield
  {journal} {\bibinfo  {journal} {Phys. Rev. Lett.}\ }\textbf {\bibinfo
  {volume} {116}}~(\bibinfo {number} {6}),\ \bibinfo {pages} {061102}},\
  \Eprint {http://arxiv.org/abs/1602.03837} {arXiv:1602.03837 [gr-qc]}
  \BibitemShut {NoStop}%
%%CITATION = ARXIV:1602.03837;%%
\bibitem [{\citenamefont {Abbott}\ \emph
  {et~al.}(2016{\natexlab{d}})\citenamefont {Abbott} \emph
  {et~al.}}]{TheLIGOScientific:2016src}%
  \BibitemOpen
  \bibfield  {author} {\bibinfo {author} {\bibnamefont {Abbott}, \bibfnamefont
  {B~P}},  \emph {et~al.} (\bibinfo {collaboration} {Virgo, LIGO Scientific})}
  (\bibinfo {year} {2016}{\natexlab{d}}),\ \bibfield  {title} {\enquote
  {\bibinfo {title} {{Tests of general relativity with GW150914}},}\ }\href
  {\doibase 10.1103/PhysRevLett.116.221101} {\bibfield  {journal} {\bibinfo
  {journal} {Phys. Rev. Lett.}\ }\textbf {\bibinfo {volume} {116}}~(\bibinfo
  {number} {22}),\ \bibinfo {pages} {221101}},\ \Eprint
  {http://arxiv.org/abs/1602.03841} {arXiv:1602.03841 [gr-qc]} \BibitemShut
  {NoStop}%
%%CITATION = ARXIV:1602.03841;%%
\bibitem [{\citenamefont {Ade}\ \emph {et~al.}(2014{\natexlab{a}})\citenamefont
  {Ade} \emph {et~al.}}]{Ade:2014afa}%
  \BibitemOpen
  \bibfield  {author} {\bibinfo {author} {\bibnamefont {Ade}, \bibfnamefont
  {P~A~R}},  \emph {et~al.} (\bibinfo {collaboration} {POLARBEAR})} (\bibinfo
  {year} {2014}{\natexlab{a}}),\ \bibfield  {title} {\enquote {\bibinfo {title}
  {{A Measurement of the Cosmic Microwave Background B-Mode Polarization Power
  Spectrum at Sub-Degree Scales with POLARBEAR}},}\ }\href {\doibase
  10.1088/0004-637X/794/2/171} {\bibfield  {journal} {\bibinfo  {journal}
  {Astrophys. J.}\ }\textbf {\bibinfo {volume} {794}}~(\bibinfo {number} {2}),\
  \bibinfo {pages} {171}},\ \Eprint {http://arxiv.org/abs/1403.2369}
  {arXiv:1403.2369 [astro-ph.CO]} \BibitemShut {NoStop}%
%%CITATION = ARXIV:1403.2369;%%
\bibitem [{\citenamefont {Ade}\ \emph {et~al.}(2014{\natexlab{b}})\citenamefont
  {Ade} \emph {et~al.}}]{Ade:2014gua}%
  \BibitemOpen
  \bibfield  {author} {\bibinfo {author} {\bibnamefont {Ade}, \bibfnamefont
  {P~A~R}},  \emph {et~al.} (\bibinfo {collaboration} {BICEP2})} (\bibinfo
  {year} {2014}{\natexlab{b}}),\ \bibfield  {title} {\enquote {\bibinfo {title}
  {{BICEP2 II: Experiment and Three-Year Data Set}},}\ }\href {\doibase
  10.1088/0004-637X/792/1/62} {\bibfield  {journal} {\bibinfo  {journal}
  {Astrophys. J.}\ }\textbf {\bibinfo {volume} {792}}~(\bibinfo {number} {1}),\
  \bibinfo {pages} {62}},\ \Eprint {http://arxiv.org/abs/1403.4302}
  {arXiv:1403.4302 [astro-ph.CO]} \BibitemShut {NoStop}%
%%CITATION = ARXIV:1403.4302;%%
\bibitem [{\citenamefont {Alam}\ \emph {et~al.}(2015)\citenamefont {Alam} \emph
  {et~al.}}]{Alam:2015mbd}%
  \BibitemOpen
  \bibfield  {author} {\bibinfo {author} {\bibnamefont {Alam}, \bibfnamefont
  {Shadab}},  \emph {et~al.} (\bibinfo {collaboration} {SDSS-III})} (\bibinfo
  {year} {2015}),\ \bibfield  {title} {\enquote {\bibinfo {title} {{The
  Eleventh and Twelfth Data Releases of the Sloan Digital Sky Survey: Final
  Data from SDSS-III}},}\ }\href {\doibase 10.1088/0067-0049/219/1/12}
  {\bibfield  {journal} {\bibinfo  {journal} {Astrophys. J. Suppl.}\ }\textbf
  {\bibinfo {volume} {219}}~(\bibinfo {number} {1}),\ \bibinfo {pages} {12}},\
  \Eprint {http://arxiv.org/abs/1501.00963} {arXiv:1501.00963 [astro-ph.IM]}
  \BibitemShut {NoStop}%
%%CITATION = ARXIV:1501.00963;%%
\bibitem [{\citenamefont {Amaro-Seoane}\ \emph {et~al.}(2013)\citenamefont
  {Amaro-Seoane} \emph {et~al.}}]{AmaroSeoane:2012km}%
  \BibitemOpen
  \bibfield  {author} {\bibinfo {author} {\bibnamefont {Amaro-Seoane},
  \bibfnamefont {Pau}},  \emph {et~al.}} (\bibinfo {year} {2013}),\ \bibfield
  {title} {\enquote {\bibinfo {title} {{eLISA/NGO: Astrophysics and cosmology
  in the gravitational-wave millihertz regime}},}\ }\href@noop {} {\bibfield
  {journal} {\bibinfo  {journal} {GW Notes}\ }\textbf {\bibinfo {volume} {6}},\
  \bibinfo {pages} {4--110}},\ \Eprint {http://arxiv.org/abs/1201.3621}
  {arXiv:1201.3621 [astro-ph.CO]} \BibitemShut {NoStop}%
%%CITATION = ARXIV:1201.3621;%%
\bibitem [{\citenamefont {Amendola}\ \emph {et~al.}(2014)\citenamefont
  {Amendola}, \citenamefont {Ballesteros},\ and\ \citenamefont
  {Pettorino}}]{Amendola:2014wma}%
  \BibitemOpen
  \bibfield  {author} {\bibinfo {author} {\bibnamefont {Amendola},
  \bibfnamefont {Luca}}, \bibinfo {author} {\bibfnamefont {Guillermo}\
  \bibnamefont {Ballesteros}}, \ and\ \bibinfo {author} {\bibfnamefont
  {Valeria}\ \bibnamefont {Pettorino}}} (\bibinfo {year} {2014}),\ \bibfield
  {title} {\enquote {\bibinfo {title} {{Effects of modified gravity on B-mode
  polarization}},}\ }\href {\doibase 10.1103/PhysRevD.90.043009} {\bibfield
  {journal} {\bibinfo  {journal} {Phys. Rev.}\ }\textbf {\bibinfo {volume}
  {D90}},\ \bibinfo {pages} {043009}},\ \Eprint
  {http://arxiv.org/abs/1405.7004} {arXiv:1405.7004 [astro-ph.CO]} \BibitemShut
  {NoStop}%
%%CITATION = ARXIV:1405.7004;%%
\bibitem [{\citenamefont {Andrews}\ \emph {et~al.}(2013)\citenamefont
  {Andrews}, \citenamefont {Goon}, \citenamefont {Hinterbichler}, \citenamefont
  {Stokes},\ and\ \citenamefont {Trodden}}]{Andrews:2013ora}%
  \BibitemOpen
  \bibfield  {author} {\bibinfo {author} {\bibnamefont {Andrews}, \bibfnamefont
  {Melinda}}, \bibinfo {author} {\bibfnamefont {Garrett}\ \bibnamefont {Goon}},
  \bibinfo {author} {\bibfnamefont {Kurt}\ \bibnamefont {Hinterbichler}},
  \bibinfo {author} {\bibfnamefont {James}\ \bibnamefont {Stokes}}, \ and\
  \bibinfo {author} {\bibfnamefont {Mark}\ \bibnamefont {Trodden}}} (\bibinfo
  {year} {2013}),\ \bibfield  {title} {\enquote {\bibinfo {title} {{Massive
  Gravity Coupled to Galileons is Ghost-Free}},}\ }\href {\doibase
  10.1103/PhysRevLett.111.061107} {\bibfield  {journal} {\bibinfo  {journal}
  {Phys. Rev. Lett.}\ }\textbf {\bibinfo {volume} {111}}~(\bibinfo {number}
  {6}),\ \bibinfo {pages} {061107}},\ \Eprint {http://arxiv.org/abs/1303.1177}
  {arXiv:1303.1177 [hep-th]} \BibitemShut {NoStop}%
%%CITATION = ARXIV:1303.1177;%%
\bibitem [{\citenamefont {Arun}\ and\ \citenamefont
  {Will}(2009)}]{Arun:2009pq}%
  \BibitemOpen
  \bibfield  {author} {\bibinfo {author} {\bibnamefont {Arun}, \bibfnamefont
  {K~G}}, \ and\ \bibinfo {author} {\bibfnamefont {Clifford~M.}\ \bibnamefont
  {Will}}} (\bibinfo {year} {2009}),\ \bibfield  {title} {\enquote {\bibinfo
  {title} {{Bounding the mass of the graviton with gravitational waves: Effect
  of higher harmonics in gravitational waveform templates}},}\ }\href {\doibase
  10.1088/0264-9381/26/15/155002} {\bibfield  {journal} {\bibinfo  {journal}
  {Class. Quant. Grav.}\ }\textbf {\bibinfo {volume} {26}},\ \bibinfo {pages}
  {155002}},\ \Eprint {http://arxiv.org/abs/0904.1190} {arXiv:0904.1190
  [gr-qc]} \BibitemShut {NoStop}%
%%CITATION = ARXIV:0904.1190;%%
\bibitem [{\citenamefont {Babichev}\ \emph {et~al.}(2009)\citenamefont
  {Babichev}, \citenamefont {Deffayet},\ and\ \citenamefont
  {Ziour}}]{Babichev:2009jt}%
  \BibitemOpen
  \bibfield  {author} {\bibinfo {author} {\bibnamefont {Babichev},
  \bibfnamefont {E}}, \bibinfo {author} {\bibfnamefont {C.}~\bibnamefont
  {Deffayet}}, \ and\ \bibinfo {author} {\bibfnamefont {R.}~\bibnamefont
  {Ziour}}} (\bibinfo {year} {2009}),\ \bibfield  {title} {\enquote {\bibinfo
  {title} {{Recovering General Relativity from massive gravity}},}\ }\href
  {\doibase 10.1103/PhysRevLett.103.201102} {\bibfield  {journal} {\bibinfo
  {journal} {Phys. Rev. Lett.}\ }\textbf {\bibinfo {volume} {103}},\ \bibinfo
  {pages} {201102}},\ \Eprint {http://arxiv.org/abs/0907.4103} {arXiv:0907.4103
  [gr-qc]} \BibitemShut {NoStop}%
%%CITATION = ARXIV:0907.4103;%%
\bibitem [{\citenamefont {Babichev}\ \emph {et~al.}(2010)\citenamefont
  {Babichev}, \citenamefont {Deffayet},\ and\ \citenamefont
  {Ziour}}]{Babichev:2010jd}%
  \BibitemOpen
  \bibfield  {author} {\bibinfo {author} {\bibnamefont {Babichev},
  \bibfnamefont {E}}, \bibinfo {author} {\bibfnamefont {C.}~\bibnamefont
  {Deffayet}}, \ and\ \bibinfo {author} {\bibfnamefont {R.}~\bibnamefont
  {Ziour}}} (\bibinfo {year} {2010}),\ \bibfield  {title} {\enquote {\bibinfo
  {title} {{The Recovery of General Relativity in massive gravity via the
  Vainshtein mechanism}},}\ }\href {\doibase 10.1103/PhysRevD.82.104008}
  {\bibfield  {journal} {\bibinfo  {journal} {Phys. Rev.}\ }\textbf {\bibinfo
  {volume} {D82}},\ \bibinfo {pages} {104008}},\ \Eprint
  {http://arxiv.org/abs/1007.4506} {arXiv:1007.4506 [gr-qc]} \BibitemShut
  {NoStop}%
%%CITATION = ARXIV:1007.4506;%%
\bibitem [{\citenamefont {Babichev}\ and\ \citenamefont
  {Deffayet}(2013)}]{Babichev:2013usa}%
  \BibitemOpen
  \bibfield  {author} {\bibinfo {author} {\bibnamefont {Babichev},
  \bibfnamefont {Eugeny}}, \ and\ \bibinfo {author} {\bibfnamefont {C\'edric}\
  \bibnamefont {Deffayet}}} (\bibinfo {year} {2013}),\ \bibfield  {title}
  {\enquote {\bibinfo {title} {{An introduction to the Vainshtein
  mechanism}},}\ }\href {\doibase 10.1088/0264-9381/30/18/184001} {\bibfield
  {journal} {\bibinfo  {journal} {Class. Quant. Grav.}\ }\textbf {\bibinfo
  {volume} {30}},\ \bibinfo {pages} {184001}},\ \Eprint
  {http://arxiv.org/abs/1304.7240} {arXiv:1304.7240 [gr-qc]} \BibitemShut
  {NoStop}%
%%CITATION = ARXIV:1304.7240;%%
\bibitem [{\citenamefont {Barreira}\ \emph {et~al.}(2016)\citenamefont
  {Barreira}, \citenamefont {Sánchez},\ and\ \citenamefont
  {Schmidt}}]{Barreira:2016ovx}%
  \BibitemOpen
  \bibfield  {author} {\bibinfo {author} {\bibnamefont {Barreira},
  \bibfnamefont {Alexandre}}, \bibinfo {author} {\bibfnamefont {Ariel~G.}\
  \bibnamefont {Sánchez}}, \ and\ \bibinfo {author} {\bibfnamefont {Fabian}\
  \bibnamefont {Schmidt}}} (\bibinfo {year} {2016}),\ \bibfield  {title}
  {\enquote {\bibinfo {title} {{Validating estimates of the growth rate of
  structure with modified gravity simulations}},}\ }\href {\doibase
  10.1103/PhysRevD.94.084022} {\bibfield  {journal} {\bibinfo  {journal} {Phys.
  Rev.}\ }\textbf {\bibinfo {volume} {D94}}~(\bibinfo {number} {8}),\ \bibinfo
  {pages} {084022}},\ \Eprint {http://arxiv.org/abs/1605.03965}
  {arXiv:1605.03965 [astro-ph.CO]} \BibitemShut {NoStop}%
%%CITATION = ARXIV:1605.03965;%%
\bibitem [{\citenamefont {Bartelmann}\ and\ \citenamefont
  {Schneider}(2001)}]{Bartelmann:1999yn}%
  \BibitemOpen
  \bibfield  {author} {\bibinfo {author} {\bibnamefont {Bartelmann},
  \bibfnamefont {Matthias}}, \ and\ \bibinfo {author} {\bibfnamefont {Peter}\
  \bibnamefont {Schneider}}} (\bibinfo {year} {2001}),\ \bibfield  {title}
  {\enquote {\bibinfo {title} {{Weak gravitational lensing}},}\ }\href
  {\doibase 10.1016/S0370-1573(00)00082-X} {\bibfield  {journal} {\bibinfo
  {journal} {Phys. Rept.}\ }\textbf {\bibinfo {volume} {340}},\ \bibinfo
  {pages} {291--472}},\ \Eprint {http://arxiv.org/abs/astro-ph/9912508}
  {arXiv:astro-ph/9912508 [astro-ph]} \BibitemShut {NoStop}%
%%CITATION = ASTRO-PH/9912508;%%
\bibitem [{\citenamefont {Baumann}\ \emph {et~al.}(2009)\citenamefont {Baumann}
  \emph {et~al.}}]{Baumann:2008aq}%
  \BibitemOpen
  \bibfield  {author} {\bibinfo {author} {\bibnamefont {Baumann}, \bibfnamefont
  {Daniel}},  \emph {et~al.} (\bibinfo {collaboration} {CMBPol Study Team})}
  (\bibinfo {year} {2009}),\ \bibfield  {title} {\enquote {\bibinfo {title}
  {{CMBPol Mission Concept Study: Probing Inflation with CMB Polarization}},}\
  }\bibfield  {booktitle} {\emph {\bibinfo {booktitle} {{Theory and
  foregrounds: CMBPol mission concept study. Proceedings, CMB Polarization
  Workshop, Batavia, USA, June 23-26, 2008}}},\ }\href {\doibase
  10.1063/1.3160885} {\bibfield  {journal} {\bibinfo  {journal} {AIP Conf.
  Proc.}\ }\textbf {\bibinfo {volume} {1141}},\ \bibinfo {pages} {10--120}},\
  \Eprint {http://arxiv.org/abs/0811.3919} {arXiv:0811.3919 [astro-ph]}
  \BibitemShut {NoStop}%
%%CITATION = ARXIV:0811.3919;%%
\bibitem [{\citenamefont {Beltran~Jimenez}\ \emph {et~al.}(2016)\citenamefont
  {Beltran~Jimenez}, \citenamefont {Piazza},\ and\ \citenamefont
  {Velten}}]{Jimenez:2015bwa}%
  \BibitemOpen
  \bibfield  {author} {\bibinfo {author} {\bibnamefont {Beltran~Jimenez},
  \bibfnamefont {Jose}}, \bibinfo {author} {\bibfnamefont {Federico}\
  \bibnamefont {Piazza}}, \ and\ \bibinfo {author} {\bibfnamefont {Hermano}\
  \bibnamefont {Velten}}} (\bibinfo {year} {2016}),\ \bibfield  {title}
  {\enquote {\bibinfo {title} {{Evading the Vainshtein Mechanism with Anomalous
  Gravitational Wave Speed: Constraints on Modified Gravity from Binary
  Pulsars}},}\ }\href {\doibase 10.1103/PhysRevLett.116.061101} {\bibfield
  {journal} {\bibinfo  {journal} {Phys. Rev. Lett.}\ }\textbf {\bibinfo
  {volume} {116}}~(\bibinfo {number} {6}),\ \bibinfo {pages} {061101}},\
  \Eprint {http://arxiv.org/abs/1507.05047} {arXiv:1507.05047 [gr-qc]}
  \BibitemShut {NoStop}%
%%CITATION = ARXIV:1507.05047;%%
\bibitem [{\citenamefont {Berezhiani}\ \emph {et~al.}(2012)\citenamefont
  {Berezhiani}, \citenamefont {Chkareuli}, \citenamefont {de~Rham},
  \citenamefont {Gabadadze},\ and\ \citenamefont {Tolley}}]{Berezhiani:2011mt}%
  \BibitemOpen
  \bibfield  {author} {\bibinfo {author} {\bibnamefont {Berezhiani},
  \bibfnamefont {L}}, \bibinfo {author} {\bibfnamefont {G.}~\bibnamefont
  {Chkareuli}}, \bibinfo {author} {\bibfnamefont {C.}~\bibnamefont {de~Rham}},
  \bibinfo {author} {\bibfnamefont {G.}~\bibnamefont {Gabadadze}}, \ and\
  \bibinfo {author} {\bibfnamefont {A.~J.}\ \bibnamefont {Tolley}}} (\bibinfo
  {year} {2012}),\ \bibfield  {title} {\enquote {\bibinfo {title} {{On Black
  Holes in Massive Gravity}},}\ }\href {\doibase 10.1103/PhysRevD.85.044024}
  {\bibfield  {journal} {\bibinfo  {journal} {Phys. Rev.}\ }\textbf {\bibinfo
  {volume} {D85}},\ \bibinfo {pages} {044024}},\ \Eprint
  {http://arxiv.org/abs/1111.3613} {arXiv:1111.3613 [hep-th]} \BibitemShut
  {NoStop}%
%%CITATION = ARXIV:1111.3613;%%
\bibitem [{\citenamefont {Berezhiani}\ \emph
  {et~al.}(2013{\natexlab{a}})\citenamefont {Berezhiani}, \citenamefont
  {Chkareuli}, \citenamefont {de~Rham}, \citenamefont {Gabadadze},\ and\
  \citenamefont {Tolley}}]{Berezhiani:2013dca}%
  \BibitemOpen
  \bibfield  {author} {\bibinfo {author} {\bibnamefont {Berezhiani},
  \bibfnamefont {L}}, \bibinfo {author} {\bibfnamefont {G.}~\bibnamefont
  {Chkareuli}}, \bibinfo {author} {\bibfnamefont {C.}~\bibnamefont {de~Rham}},
  \bibinfo {author} {\bibfnamefont {G.}~\bibnamefont {Gabadadze}}, \ and\
  \bibinfo {author} {\bibfnamefont {A.~J.}\ \bibnamefont {Tolley}}} (\bibinfo
  {year} {2013}{\natexlab{a}}),\ \bibfield  {title} {\enquote {\bibinfo {title}
  {{Mixed Galileons and Spherically Symmetric Solutions}},}\ }\href {\doibase
  10.1088/0264-9381/30/18/184003} {\bibfield  {journal} {\bibinfo  {journal}
  {Class. Quant. Grav.}\ }\textbf {\bibinfo {volume} {30}},\ \bibinfo {pages}
  {184003}},\ \Eprint {http://arxiv.org/abs/1305.0271} {arXiv:1305.0271
  [hep-th]} \BibitemShut {NoStop}%
%%CITATION = ARXIV:1305.0271;%%
\bibitem [{\citenamefont {Berezhiani}\ \emph
  {et~al.}(2013{\natexlab{b}})\citenamefont {Berezhiani}, \citenamefont
  {Chkareuli},\ and\ \citenamefont {Gabadadze}}]{Berezhiani:2013dw}%
  \BibitemOpen
  \bibfield  {author} {\bibinfo {author} {\bibnamefont {Berezhiani},
  \bibfnamefont {Lasha}}, \bibinfo {author} {\bibfnamefont {Giga}\ \bibnamefont
  {Chkareuli}}, \ and\ \bibinfo {author} {\bibfnamefont {Gregory}\ \bibnamefont
  {Gabadadze}}} (\bibinfo {year} {2013}{\natexlab{b}}),\ \bibfield  {title}
  {\enquote {\bibinfo {title} {{Restricted Galileons}},}\ }\href {\doibase
  10.1103/PhysRevD.88.124020} {\bibfield  {journal} {\bibinfo  {journal} {Phys.
  Rev.}\ }\textbf {\bibinfo {volume} {D88}},\ \bibinfo {pages} {124020}},\
  \Eprint {http://arxiv.org/abs/1302.0549} {arXiv:1302.0549 [hep-th]}
  \BibitemShut {NoStop}%
%%CITATION = ARXIV:1302.0549;%%
\bibitem [{\citenamefont {Bergshoeff}\ \emph {et~al.}(2009)\citenamefont
  {Bergshoeff}, \citenamefont {Hohm},\ and\ \citenamefont
  {Townsend}}]{Bergshoeff:2009hq}%
  \BibitemOpen
  \bibfield  {author} {\bibinfo {author} {\bibnamefont {Bergshoeff},
  \bibfnamefont {Eric~A}}, \bibinfo {author} {\bibfnamefont {Olaf}\
  \bibnamefont {Hohm}}, \ and\ \bibinfo {author} {\bibfnamefont {Paul~K.}\
  \bibnamefont {Townsend}}} (\bibinfo {year} {2009}),\ \bibfield  {title}
  {\enquote {\bibinfo {title} {{Massive Gravity in Three Dimensions}},}\ }\href
  {\doibase 10.1103/PhysRevLett.102.201301} {\bibfield  {journal} {\bibinfo
  {journal} {Phys. Rev. Lett.}\ }\textbf {\bibinfo {volume} {102}},\ \bibinfo
  {pages} {201301}},\ \Eprint {http://arxiv.org/abs/0901.1766} {arXiv:0901.1766
  [hep-th]} \BibitemShut {NoStop}%
%%CITATION = ARXIV:0901.1766;%%
\bibitem [{\citenamefont {Berti}\ \emph {et~al.}(2005)\citenamefont {Berti},
  \citenamefont {Buonanno},\ and\ \citenamefont {Will}}]{Berti:2004bd}%
  \BibitemOpen
  \bibfield  {author} {\bibinfo {author} {\bibnamefont {Berti}, \bibfnamefont
  {Emanuele}}, \bibinfo {author} {\bibfnamefont {Alessandra}\ \bibnamefont
  {Buonanno}}, \ and\ \bibinfo {author} {\bibfnamefont {Clifford~M.}\
  \bibnamefont {Will}}} (\bibinfo {year} {2005}),\ \bibfield  {title} {\enquote
  {\bibinfo {title} {{Estimating spinning binary parameters and testing
  alternative theories of gravity with LISA}},}\ }\href {\doibase
  10.1103/PhysRevD.71.084025} {\bibfield  {journal} {\bibinfo  {journal} {Phys.
  Rev.}\ }\textbf {\bibinfo {volume} {D71}},\ \bibinfo {pages} {084025}},\
  \Eprint {http://arxiv.org/abs/gr-qc/0411129} {arXiv:gr-qc/0411129 [gr-qc]}
  \BibitemShut {NoStop}%
%%CITATION = GR-QC/0411129;%%
\bibitem [{\citenamefont {Berti}\ \emph {et~al.}(2011)\citenamefont {Berti},
  \citenamefont {Gair},\ and\ \citenamefont {Sesana}}]{Berti:2011jz}%
  \BibitemOpen
  \bibfield  {author} {\bibinfo {author} {\bibnamefont {Berti}, \bibfnamefont
  {Emanuele}}, \bibinfo {author} {\bibfnamefont {Jonathan}\ \bibnamefont
  {Gair}}, \ and\ \bibinfo {author} {\bibfnamefont {Alberto}\ \bibnamefont
  {Sesana}}} (\bibinfo {year} {2011}),\ \bibfield  {title} {\enquote {\bibinfo
  {title} {{Graviton mass bounds from space-based gravitational-wave
  observations of massive black hole populations}},}\ }\href {\doibase
  10.1103/PhysRevD.84.101501} {\bibfield  {journal} {\bibinfo  {journal} {Phys.
  Rev.}\ }\textbf {\bibinfo {volume} {D84}},\ \bibinfo {pages} {101501}},\
  \Eprint {http://arxiv.org/abs/1107.3528} {arXiv:1107.3528 [gr-qc]}
  \BibitemShut {NoStop}%
%%CITATION = ARXIV:1107.3528;%%
\bibitem [{\citenamefont {Blas}\ \emph {et~al.}(2016)\citenamefont {Blas},
  \citenamefont {Ivanov}, \citenamefont {Sawicki},\ and\ \citenamefont
  {Sibiryakov}}]{Blas:2016qmn}%
  \BibitemOpen
  \bibfield  {author} {\bibinfo {author} {\bibnamefont {Blas}, \bibfnamefont
  {Diego}}, \bibinfo {author} {\bibfnamefont {Mikhail~M.}\ \bibnamefont
  {Ivanov}}, \bibinfo {author} {\bibfnamefont {Ignacy}\ \bibnamefont
  {Sawicki}}, \ and\ \bibinfo {author} {\bibfnamefont {Sergey}\ \bibnamefont
  {Sibiryakov}}} (\bibinfo {year} {2016}),\ \bibfield  {title} {\enquote
  {\bibinfo {title} {{On constraining the speed of gravitational waves
  following GW150914}},}\ }\href {\doibase 10.1134/S0021364016100040,
  10.7868/S0370274X16100039} {\bibfield  {journal} {\bibinfo  {journal} {Pisma
  Zh. Eksp. Teor. Fiz.}\ }\textbf {\bibinfo {volume} {103}}~(\bibinfo {number}
  {10}),\ \bibinfo {pages} {708--710}},\ \bibinfo {note} {[JETP
  Lett.103,no.10,624(2016)]},\ \Eprint {http://arxiv.org/abs/1602.04188}
  {arXiv:1602.04188 [gr-qc]} \BibitemShut {NoStop}%
%%CITATION = ARXIV:1602.04188;%%
\bibitem [{\citenamefont {Bluhm}(2006)}]{Bluhm:2005uj}%
  \BibitemOpen
  \bibfield  {author} {\bibinfo {author} {\bibnamefont {Bluhm}, \bibfnamefont
  {Robert}}} (\bibinfo {year} {2006}),\ \bibfield  {title} {\enquote {\bibinfo
  {title} {{Overview of the SME: Implications and phenomenology of Lorentz
  violation}},}\ }\bibfield  {booktitle} {\emph {\bibinfo {booktitle} {{339th
  WE Heraeus Seminar on Special Relativity: Will It Survive the Next 100 Years?
  Potsdam, Germany, February 13-18, 2005}}},\ }\href {\doibase
  10.1007/3-540-34523-X_8} {\bibfield  {journal} {\bibinfo  {journal} {Lect.
  Notes Phys.}\ }\textbf {\bibinfo {volume} {702}},\ \bibinfo {pages}
  {191--226}},\ \bibinfo {note} {[,191(2005)]},\ \Eprint
  {http://arxiv.org/abs/hep-ph/0506054} {arXiv:hep-ph/0506054 [hep-ph]}
  \BibitemShut {NoStop}%
%%CITATION = HEP-PH/0506054;%%
\bibitem [{\citenamefont {Boulware}\ and\ \citenamefont
  {Deser}(1972)}]{Boulware:1973my}%
  \BibitemOpen
  \bibfield  {author} {\bibinfo {author} {\bibnamefont {Boulware},
  \bibfnamefont {D~G}}, \ and\ \bibinfo {author} {\bibfnamefont {Stanley}\
  \bibnamefont {Deser}}} (\bibinfo {year} {1972}),\ \bibfield  {title}
  {\enquote {\bibinfo {title} {{Can gravitation have a finite range?}}}\ }\href
  {\doibase 10.1103/PhysRevD.6.3368} {\bibfield  {journal} {\bibinfo  {journal}
  {Phys. Rev.}\ }\textbf {\bibinfo {volume} {D6}},\ \bibinfo {pages}
  {3368--3382}}\BibitemShut {NoStop}%
%%CITATION = PHRVA,D6,3368;%%
\bibitem [{\citenamefont {Brito}\ \emph {et~al.}(2013)\citenamefont {Brito},
  \citenamefont {Cardoso},\ and\ \citenamefont {Pani}}]{Brito:2013xaa}%
  \BibitemOpen
  \bibfield  {author} {\bibinfo {author} {\bibnamefont {Brito}, \bibfnamefont
  {Richard}}, \bibinfo {author} {\bibfnamefont {Vitor}\ \bibnamefont
  {Cardoso}}, \ and\ \bibinfo {author} {\bibfnamefont {Paolo}\ \bibnamefont
  {Pani}}} (\bibinfo {year} {2013}),\ \bibfield  {title} {\enquote {\bibinfo
  {title} {{Black holes with massive graviton hair}},}\ }\href {\doibase
  10.1103/PhysRevD.88.064006} {\bibfield  {journal} {\bibinfo  {journal} {Phys.
  Rev.}\ }\textbf {\bibinfo {volume} {D88}},\ \bibinfo {pages} {064006}},\
  \Eprint {http://arxiv.org/abs/1309.0818} {arXiv:1309.0818 [gr-qc]}
  \BibitemShut {NoStop}%
%%CITATION = ARXIV:1309.0818;%%
\bibitem [{\citenamefont {Caves}(1980)}]{Caves:1980jn}%
  \BibitemOpen
  \bibfield  {author} {\bibinfo {author} {\bibnamefont {Caves}, \bibfnamefont
  {C~M}}} (\bibinfo {year} {1980}),\ \bibfield  {title} {\enquote {\bibinfo
  {title} {{GRAVITATIONAL RADIATION AND THE ULTIMATE SPEED IN ROSEN'S BIMETRIC
  THEORY OF GRAVITY}},}\ }\href {\doibase 10.1016/0003-4916(80)90117-7}
  {\bibfield  {journal} {\bibinfo  {journal} {Annals Phys.}\ }\textbf {\bibinfo
  {volume} {125}},\ \bibinfo {pages} {35--52}}\BibitemShut {NoStop}%
%%CITATION = APNYA,125,35;%%
\bibitem [{\citenamefont {Choudhury}\ \emph {et~al.}(2004)\citenamefont
  {Choudhury}, \citenamefont {Joshi}, \citenamefont {Mahajan},\ and\
  \citenamefont {McKellar}}]{Choudhury:2002pu}%
  \BibitemOpen
  \bibfield  {author} {\bibinfo {author} {\bibnamefont {Choudhury},
  \bibfnamefont {S~R}}, \bibinfo {author} {\bibfnamefont {Girish~C.}\
  \bibnamefont {Joshi}}, \bibinfo {author} {\bibfnamefont {S.}~\bibnamefont
  {Mahajan}}, \ and\ \bibinfo {author} {\bibfnamefont {Bruce H.~J.}\
  \bibnamefont {McKellar}}} (\bibinfo {year} {2004}),\ \bibfield  {title}
  {\enquote {\bibinfo {title} {{Probing large distance higher dimensional
  gravity from lensing data}},}\ }\href {\doibase
  10.1016/j.astropartphys.2004.04.001} {\bibfield  {journal} {\bibinfo
  {journal} {Astropart. Phys.}\ }\textbf {\bibinfo {volume} {21}},\ \bibinfo
  {pages} {559--563}},\ \Eprint {http://arxiv.org/abs/hep-ph/0204161}
  {arXiv:hep-ph/0204161 [hep-ph]} \BibitemShut {NoStop}%
%%CITATION = HEP-PH/0204161;%%
\bibitem [{\citenamefont {Comelli}\ \emph {et~al.}(2015)\citenamefont
  {Comelli}, \citenamefont {Crisostomi}, \citenamefont {Koyama}, \citenamefont
  {Pilo},\ and\ \citenamefont {Tasinato}}]{Comelli:2015ksa}%
  \BibitemOpen
  \bibfield  {author} {\bibinfo {author} {\bibnamefont {Comelli}, \bibfnamefont
  {Denis}}, \bibinfo {author} {\bibfnamefont {Marco}\ \bibnamefont
  {Crisostomi}}, \bibinfo {author} {\bibfnamefont {Kazuya}\ \bibnamefont
  {Koyama}}, \bibinfo {author} {\bibfnamefont {Luigi}\ \bibnamefont {Pilo}}, \
  and\ \bibinfo {author} {\bibfnamefont {Gianmassimo}\ \bibnamefont
  {Tasinato}}} (\bibinfo {year} {2015}),\ \bibfield  {title} {\enquote
  {\bibinfo {title} {{New Branches of Massive Gravity}},}\ }\href {\doibase
  10.1103/PhysRevD.91.121502} {\bibfield  {journal} {\bibinfo  {journal} {Phys.
  Rev.}\ }\textbf {\bibinfo {volume} {D91}}~(\bibinfo {number} {12}),\ \bibinfo
  {pages} {121502}},\ \Eprint {http://arxiv.org/abs/1505.00632}
  {arXiv:1505.00632 [hep-th]} \BibitemShut {NoStop}%
%%CITATION = ARXIV:1505.00632;%%
\bibitem [{\citenamefont {Comelli}\ \emph {et~al.}(2011)\citenamefont
  {Comelli}, \citenamefont {Nesti},\ and\ \citenamefont
  {Pilo}}]{Comelli:2010bj}%
  \BibitemOpen
  \bibfield  {author} {\bibinfo {author} {\bibnamefont {Comelli}, \bibfnamefont
  {Denis}}, \bibinfo {author} {\bibfnamefont {Fabrizio}\ \bibnamefont {Nesti}},
  \ and\ \bibinfo {author} {\bibfnamefont {Luigi}\ \bibnamefont {Pilo}}}
  (\bibinfo {year} {2011}),\ \bibfield  {title} {\enquote {\bibinfo {title}
  {{Stars and (Furry) Black Holes in Lorentz Breaking Massive Gravity}},}\
  }\href {\doibase 10.1103/PhysRevD.83.084042} {\bibfield  {journal} {\bibinfo
  {journal} {Phys. Rev.}\ }\textbf {\bibinfo {volume} {D83}},\ \bibinfo {pages}
  {084042}},\ \Eprint {http://arxiv.org/abs/1010.4773} {arXiv:1010.4773
  [hep-th]} \BibitemShut {NoStop}%
%%CITATION = ARXIV:1010.4773;%%
\bibitem [{\citenamefont {Comelli}\ \emph {et~al.}(2014)\citenamefont
  {Comelli}, \citenamefont {Nesti},\ and\ \citenamefont
  {Pilo}}]{Comelli:2014xga}%
  \BibitemOpen
  \bibfield  {author} {\bibinfo {author} {\bibnamefont {Comelli}, \bibfnamefont
  {Denis}}, \bibinfo {author} {\bibfnamefont {Fabrizio}\ \bibnamefont {Nesti}},
  \ and\ \bibinfo {author} {\bibfnamefont {Luigi}\ \bibnamefont {Pilo}}}
  (\bibinfo {year} {2014}),\ \bibfield  {title} {\enquote {\bibinfo {title}
  {{Nonderivative Modified Gravity: a Classification}},}\ }\href {\doibase
  10.1088/1475-7516/2014/11/018} {\bibfield  {journal} {\bibinfo  {journal}
  {JCAP}\ }\textbf {\bibinfo {volume} {1411}}~(\bibinfo {number} {11}),\
  \bibinfo {pages} {018}},\ \Eprint {http://arxiv.org/abs/1407.4991}
  {arXiv:1407.4991 [hep-th]} \BibitemShut {NoStop}%
%%CITATION = ARXIV:1407.4991;%%
\bibitem [{\citenamefont {Cooray}\ and\ \citenamefont
  {Seto}(2004)}]{Cooray:2003cv}%
  \BibitemOpen
  \bibfield  {author} {\bibinfo {author} {\bibnamefont {Cooray}, \bibfnamefont
  {Asantha}}, \ and\ \bibinfo {author} {\bibfnamefont {Naoki}\ \bibnamefont
  {Seto}}} (\bibinfo {year} {2004}),\ \bibfield  {title} {\enquote {\bibinfo
  {title} {{Graviton mass from close white dwarf binaries detectable with
  LISA}},}\ }\href {\doibase 10.1103/PhysRevD.69.103502} {\bibfield  {journal}
  {\bibinfo  {journal} {Phys. Rev.}\ }\textbf {\bibinfo {volume} {D69}},\
  \bibinfo {pages} {103502}},\ \Eprint {http://arxiv.org/abs/astro-ph/0311054}
  {arXiv:astro-ph/0311054 [astro-ph]} \BibitemShut {NoStop}%
%%CITATION = ASTRO-PH/0311054;%%
\bibitem [{\citenamefont {Crill}\ \emph {et~al.}(2008)\citenamefont {Crill}
  \emph {et~al.}}]{Crill:2008rd}%
  \BibitemOpen
  \bibfield  {author} {\bibinfo {author} {\bibnamefont {Crill}, \bibfnamefont
  {B~P}},  \emph {et~al.}} (\bibinfo {year} {2008}),\ \bibfield  {title}
  {\enquote {\bibinfo {title} {{SPIDER: A Balloon-borne Large-scale CMB
  Polarimeter}},}\ }\href {\doibase 10.1117/12.787446} {\bibfield  {journal}
  {\bibinfo  {journal} {Proc. SPIE Int. Soc. Opt. Eng.}\ }\textbf {\bibinfo
  {volume} {7010}},\ \bibinfo {pages} {2P}},\ \Eprint
  {http://arxiv.org/abs/0807.1548} {arXiv:0807.1548 [astro-ph]} \BibitemShut
  {NoStop}%
%%CITATION = ARXIV:0807.1548;%%
\bibitem [{\citenamefont {Cusin}\ \emph {et~al.}(2016)\citenamefont {Cusin},
  \citenamefont {Foffa}, \citenamefont {Maggiore},\ and\ \citenamefont
  {Mancarella}}]{Cusin:2016nzi}%
  \BibitemOpen
  \bibfield  {author} {\bibinfo {author} {\bibnamefont {Cusin}, \bibfnamefont
  {Giulia}}, \bibinfo {author} {\bibfnamefont {Stefano}\ \bibnamefont {Foffa}},
  \bibinfo {author} {\bibfnamefont {Michele}\ \bibnamefont {Maggiore}}, \ and\
  \bibinfo {author} {\bibfnamefont {Michele}\ \bibnamefont {Mancarella}}}
  (\bibinfo {year} {2016}),\ \bibfield  {title} {\enquote {\bibinfo {title}
  {{Conformal symmetry and nonlinear extensions of nonlocal gravity}},}\ }\href
  {\doibase 10.1103/PhysRevD.93.083008} {\bibfield  {journal} {\bibinfo
  {journal} {Phys. Rev.}\ }\textbf {\bibinfo {volume} {D93}}~(\bibinfo {number}
  {8}),\ \bibinfo {pages} {083008}},\ \Eprint {http://arxiv.org/abs/1602.01078}
  {arXiv:1602.01078 [hep-th]} \BibitemShut {NoStop}%
%%CITATION = ARXIV:1602.01078;%%
\bibitem [{\citenamefont {Cutler}\ \emph {et~al.}(2003)\citenamefont {Cutler},
  \citenamefont {Hiscock},\ and\ \citenamefont {Larson}}]{Cutler:2002ef}%
  \BibitemOpen
  \bibfield  {author} {\bibinfo {author} {\bibnamefont {Cutler}, \bibfnamefont
  {Curt}}, \bibinfo {author} {\bibfnamefont {William~A.}\ \bibnamefont
  {Hiscock}}, \ and\ \bibinfo {author} {\bibfnamefont {Shane~L.}\ \bibnamefont
  {Larson}}} (\bibinfo {year} {2003}),\ \bibfield  {title} {\enquote {\bibinfo
  {title} {{LISA, binary stars, and the mass of the graviton}},}\ }\href
  {\doibase 10.1103/PhysRevD.67.024015} {\bibfield  {journal} {\bibinfo
  {journal} {Phys. Rev.}\ }\textbf {\bibinfo {volume} {D67}},\ \bibinfo {pages}
  {024015}},\ \Eprint {http://arxiv.org/abs/gr-qc/0209101} {arXiv:gr-qc/0209101
  [gr-qc]} \BibitemShut {NoStop}%
%%CITATION = GR-QC/0209101;%%
\bibitem [{\citenamefont {van Dam}\ and\ \citenamefont
  {Veltman}(1970)}]{vanDam:1970vg}%
  \BibitemOpen
  \bibfield  {author} {\bibinfo {author} {\bibnamefont {van Dam}, \bibfnamefont
  {H}}, \ and\ \bibinfo {author} {\bibfnamefont {M.~J.~G.}\ \bibnamefont
  {Veltman}}} (\bibinfo {year} {1970}),\ \bibfield  {title} {\enquote {\bibinfo
  {title} {{Massive and massless Yang-Mills and gravitational fields}},}\
  }\href {\doibase 10.1016/0550-3213(70)90416-5} {\bibfield  {journal}
  {\bibinfo  {journal} {Nucl. Phys.}\ }\textbf {\bibinfo {volume} {B22}},\
  \bibinfo {pages} {397--411}}\BibitemShut {NoStop}%
%%CITATION = NUPHA,B22,397;%%
\bibitem [{\citenamefont {D'Amico}\ \emph {et~al.}(2011)\citenamefont
  {D'Amico}, \citenamefont {de~Rham}, \citenamefont {Dubovsky}, \citenamefont
  {Gabadadze}, \citenamefont {Pirtskhalava},\ and\ \citenamefont
  {Tolley}}]{D'Amico:2011jj}%
  \BibitemOpen
  \bibfield  {author} {\bibinfo {author} {\bibnamefont {D'Amico}, \bibfnamefont
  {G}}, \bibinfo {author} {\bibfnamefont {C.}~\bibnamefont {de~Rham}}, \bibinfo
  {author} {\bibfnamefont {S.}~\bibnamefont {Dubovsky}}, \bibinfo {author}
  {\bibfnamefont {G.}~\bibnamefont {Gabadadze}}, \bibinfo {author}
  {\bibfnamefont {D.}~\bibnamefont {Pirtskhalava}}, \ and\ \bibinfo {author}
  {\bibfnamefont {A.~J.}\ \bibnamefont {Tolley}}} (\bibinfo {year} {2011}),\
  \bibfield  {title} {\enquote {\bibinfo {title} {{Massive Cosmologies}},}\
  }\href {\doibase 10.1103/PhysRevD.84.124046} {\bibfield  {journal} {\bibinfo
  {journal} {Phys. Rev.}\ }\textbf {\bibinfo {volume} {D84}},\ \bibinfo {pages}
  {124046}},\ \Eprint {http://arxiv.org/abs/1108.5231} {arXiv:1108.5231
  [hep-th]} \BibitemShut {NoStop}%
%%CITATION = ARXIV:1108.5231;%%
\bibitem [{\citenamefont {D'Amico}\ \emph {et~al.}(2013)\citenamefont
  {D'Amico}, \citenamefont {Gabadadze}, \citenamefont {Hui},\ and\
  \citenamefont {Pirtskhalava}}]{D'Amico:2012zv}%
  \BibitemOpen
  \bibfield  {author} {\bibinfo {author} {\bibnamefont {D'Amico}, \bibfnamefont
  {Guido}}, \bibinfo {author} {\bibfnamefont {Gregory}\ \bibnamefont
  {Gabadadze}}, \bibinfo {author} {\bibfnamefont {Lam}\ \bibnamefont {Hui}}, \
  and\ \bibinfo {author} {\bibfnamefont {David}\ \bibnamefont {Pirtskhalava}}}
  (\bibinfo {year} {2013}),\ \bibfield  {title} {\enquote {\bibinfo {title}
  {{Quasidilaton: Theory and cosmology}},}\ }\href {\doibase
  10.1103/PhysRevD.87.064037} {\bibfield  {journal} {\bibinfo  {journal} {Phys.
  Rev.}\ }\textbf {\bibinfo {volume} {D87}},\ \bibinfo {pages} {064037}},\
  \Eprint {http://arxiv.org/abs/1206.4253} {arXiv:1206.4253 [hep-th]}
  \BibitemShut {NoStop}%
%%CITATION = ARXIV:1206.4253;%%
\bibitem [{\citenamefont {Dar}\ \emph {et~al.}(2016)\citenamefont {Dar},
  \citenamefont {de~Rham}, \citenamefont {Deskins}, \citenamefont {Giblin},\
  and\ \citenamefont {Tolley}}]{NumericalPulsar}%
  \BibitemOpen
  \bibfield  {author} {\bibinfo {author} {\bibnamefont {Dar}, \bibfnamefont
  {Furqan}}, \bibinfo {author} {\bibfnamefont {Claudia}\ \bibnamefont
  {de~Rham}}, \bibinfo {author} {\bibfnamefont {J.~Tate}\ \bibnamefont
  {Deskins}}, \bibinfo {author} {\bibfnamefont {John~T.}\ \bibnamefont
  {Giblin}}, \ and\ \bibinfo {author} {\bibfnamefont {Andrew~J.}\ \bibnamefont
  {Tolley}}} (\bibinfo {year} {2016}),\ \bibfield  {title} {\enquote {\bibinfo
  {title} {{Numerical Study of Galileon Radiation from Binary Pulsars}},}\
  }\href {\doibase To appear} {\ To appear}\BibitemShut {NoStop}%
\bibitem [{\citenamefont {De~Felice}\ \emph {et~al.}(2013)\citenamefont
  {De~Felice}, \citenamefont {Emir~Gumrukcuolu},\ and\ \citenamefont
  {Mukohyama}}]{DeFelice:2013dua}%
  \BibitemOpen
  \bibfield  {author} {\bibinfo {author} {\bibnamefont {De~Felice},
  \bibfnamefont {Antonio}}, \bibinfo {author} {\bibfnamefont {A.}~\bibnamefont
  {Emir~Gumrukcuolu}}, \ and\ \bibinfo {author} {\bibfnamefont {Shinji}\
  \bibnamefont {Mukohyama}}} (\bibinfo {year} {2013}),\ \bibfield  {title}
  {\enquote {\bibinfo {title} {{Generalized quasidilaton theory}},}\ }\href
  {\doibase 10.1103/PhysRevD.88.124006} {\bibfield  {journal} {\bibinfo
  {journal} {Phys. Rev.}\ }\textbf {\bibinfo {volume} {D88}}~(\bibinfo {number}
  {12}),\ \bibinfo {pages} {124006}},\ \Eprint {http://arxiv.org/abs/1309.3162}
  {arXiv:1309.3162 [hep-th]} \BibitemShut {NoStop}%
%%CITATION = ARXIV:1309.3162;%%
\bibitem [{\citenamefont {De~Felice}\ and\ \citenamefont
  {Mukohyama}(2016)}]{DeFelice:2015hla}%
  \BibitemOpen
  \bibfield  {author} {\bibinfo {author} {\bibnamefont {De~Felice},
  \bibfnamefont {Antonio}}, \ and\ \bibinfo {author} {\bibfnamefont {Shinji}\
  \bibnamefont {Mukohyama}}} (\bibinfo {year} {2016}),\ \bibfield  {title}
  {\enquote {\bibinfo {title} {{Minimal theory of massive gravity}},}\ }\href
  {\doibase 10.1016/j.physletb.2015.11.050} {\bibfield  {journal} {\bibinfo
  {journal} {Phys. Lett.}\ }\textbf {\bibinfo {volume} {B752}},\ \bibinfo
  {pages} {302--305}},\ \Eprint {http://arxiv.org/abs/1506.01594}
  {arXiv:1506.01594 [hep-th]} \BibitemShut {NoStop}%
%%CITATION = ARXIV:1506.01594;%%
\bibitem [{\citenamefont {Deffayet}\ \emph
  {et~al.}(2002{\natexlab{a}})\citenamefont {Deffayet}, \citenamefont {Dvali},\
  and\ \citenamefont {Gabadadze}}]{Deffayet:2001pu}%
  \BibitemOpen
  \bibfield  {author} {\bibinfo {author} {\bibnamefont {Deffayet},
  \bibfnamefont {Cedric}}, \bibinfo {author} {\bibfnamefont {G.~R.}\
  \bibnamefont {Dvali}}, \ and\ \bibinfo {author} {\bibfnamefont {Gregory}\
  \bibnamefont {Gabadadze}}} (\bibinfo {year} {2002}{\natexlab{a}}),\ \bibfield
   {title} {\enquote {\bibinfo {title} {{Accelerated universe from gravity
  leaking to extra dimensions}},}\ }\href {\doibase 10.1103/PhysRevD.65.044023}
  {\bibfield  {journal} {\bibinfo  {journal} {Phys. Rev.}\ }\textbf {\bibinfo
  {volume} {D65}},\ \bibinfo {pages} {044023}},\ \Eprint
  {http://arxiv.org/abs/astro-ph/0105068} {arXiv:astro-ph/0105068 [astro-ph]}
  \BibitemShut {NoStop}%
%%CITATION = ASTRO-PH/0105068;%%
\bibitem [{\citenamefont {Deffayet}\ \emph
  {et~al.}(2002{\natexlab{b}})\citenamefont {Deffayet}, \citenamefont {Dvali},
  \citenamefont {Gabadadze},\ and\ \citenamefont
  {Vainshtein}}]{Deffayet:2001uk}%
  \BibitemOpen
  \bibfield  {author} {\bibinfo {author} {\bibnamefont {Deffayet},
  \bibfnamefont {Cedric}}, \bibinfo {author} {\bibfnamefont {G.~R.}\
  \bibnamefont {Dvali}}, \bibinfo {author} {\bibfnamefont {Gregory}\
  \bibnamefont {Gabadadze}}, \ and\ \bibinfo {author} {\bibfnamefont
  {Arkady~I.}\ \bibnamefont {Vainshtein}}} (\bibinfo {year}
  {2002}{\natexlab{b}}),\ \bibfield  {title} {\enquote {\bibinfo {title}
  {{Nonperturbative continuity in graviton mass versus perturbative
  discontinuity}},}\ }\href {\doibase 10.1103/PhysRevD.65.044026} {\bibfield
  {journal} {\bibinfo  {journal} {Phys. Rev.}\ }\textbf {\bibinfo {volume}
  {D65}},\ \bibinfo {pages} {044026}},\ \Eprint
  {http://arxiv.org/abs/hep-th/0106001} {arXiv:hep-th/0106001 [hep-th]}
  \BibitemShut {NoStop}%
%%CITATION = HEP-TH/0106001;%%
\bibitem [{\citenamefont {Del~Pozzo}\ \emph {et~al.}(2011)\citenamefont
  {Del~Pozzo}, \citenamefont {Veitch},\ and\ \citenamefont
  {Vecchio}}]{DelPozzo:2011pg}%
  \BibitemOpen
  \bibfield  {author} {\bibinfo {author} {\bibnamefont {Del~Pozzo},
  \bibfnamefont {Walter}}, \bibinfo {author} {\bibfnamefont {John}\
  \bibnamefont {Veitch}}, \ and\ \bibinfo {author} {\bibfnamefont {Alberto}\
  \bibnamefont {Vecchio}}} (\bibinfo {year} {2011}),\ \bibfield  {title}
  {\enquote {\bibinfo {title} {{Testing General Relativity using Bayesian model
  selection: Applications to observations of gravitational waves from compact
  binary systems}},}\ }\href {\doibase 10.1103/PhysRevD.83.082002} {\bibfield
  {journal} {\bibinfo  {journal} {Phys. Rev.}\ }\textbf {\bibinfo {volume}
  {D83}},\ \bibinfo {pages} {082002}},\ \Eprint
  {http://arxiv.org/abs/1101.1391} {arXiv:1101.1391 [gr-qc]} \BibitemShut
  {NoStop}%
%%CITATION = ARXIV:1101.1391;%%
\bibitem [{\citenamefont {Dirian}\ \emph {et~al.}(2014)\citenamefont {Dirian},
  \citenamefont {Foffa}, \citenamefont {Khosravi}, \citenamefont {Kunz},\ and\
  \citenamefont {Maggiore}}]{Dirian:2014ara}%
  \BibitemOpen
  \bibfield  {author} {\bibinfo {author} {\bibnamefont {Dirian}, \bibfnamefont
  {Yves}}, \bibinfo {author} {\bibfnamefont {Stefano}\ \bibnamefont {Foffa}},
  \bibinfo {author} {\bibfnamefont {Nima}\ \bibnamefont {Khosravi}}, \bibinfo
  {author} {\bibfnamefont {Martin}\ \bibnamefont {Kunz}}, \ and\ \bibinfo
  {author} {\bibfnamefont {Michele}\ \bibnamefont {Maggiore}}} (\bibinfo {year}
  {2014}),\ \bibfield  {title} {\enquote {\bibinfo {title} {{Cosmological
  perturbations and structure formation in nonlocal infrared modifications of
  general relativity}},}\ }\href {\doibase 10.1088/1475-7516/2014/06/033}
  {\bibfield  {journal} {\bibinfo  {journal} {JCAP}\ }\textbf {\bibinfo
  {volume} {1406}},\ \bibinfo {pages} {033}},\ \Eprint
  {http://arxiv.org/abs/1403.6068} {arXiv:1403.6068 [astro-ph.CO]} \BibitemShut
  {NoStop}%
%%CITATION = ARXIV:1403.6068;%%
\bibitem [{\citenamefont {Dubovsky}\ \emph {et~al.}(2005)\citenamefont
  {Dubovsky}, \citenamefont {Tinyakov},\ and\ \citenamefont
  {Tkachev}}]{Dubovsky:2004ud}%
  \BibitemOpen
  \bibfield  {author} {\bibinfo {author} {\bibnamefont {Dubovsky},
  \bibfnamefont {S~L}}, \bibinfo {author} {\bibfnamefont {P.~G.}\ \bibnamefont
  {Tinyakov}}, \ and\ \bibinfo {author} {\bibfnamefont {I.~I.}\ \bibnamefont
  {Tkachev}}} (\bibinfo {year} {2005}),\ \bibfield  {title} {\enquote {\bibinfo
  {title} {{Massive graviton as a testable cold dark matter candidate}},}\
  }\href {\doibase 10.1103/PhysRevLett.94.181102} {\bibfield  {journal}
  {\bibinfo  {journal} {Phys. Rev. Lett.}\ }\textbf {\bibinfo {volume} {94}},\
  \bibinfo {pages} {181102}},\ \Eprint {http://arxiv.org/abs/hep-th/0411158}
  {arXiv:hep-th/0411158 [hep-th]} \BibitemShut {NoStop}%
%%CITATION = HEP-TH/0411158;%%
\bibitem [{\citenamefont {Dubovsky}\ \emph {et~al.}(2010)\citenamefont
  {Dubovsky}, \citenamefont {Flauger}, \citenamefont {Starobinsky},\ and\
  \citenamefont {Tkachev}}]{Dubovsky:2009xk}%
  \BibitemOpen
  \bibfield  {author} {\bibinfo {author} {\bibnamefont {Dubovsky},
  \bibfnamefont {Sergei}}, \bibinfo {author} {\bibfnamefont {Raphael}\
  \bibnamefont {Flauger}}, \bibinfo {author} {\bibfnamefont {Alexei}\
  \bibnamefont {Starobinsky}}, \ and\ \bibinfo {author} {\bibfnamefont {Igor}\
  \bibnamefont {Tkachev}}} (\bibinfo {year} {2010}),\ \bibfield  {title}
  {\enquote {\bibinfo {title} {{Signatures of a Graviton Mass in the Cosmic
  Microwave Background}},}\ }\href {\doibase 10.1103/PhysRevD.81.023523}
  {\bibfield  {journal} {\bibinfo  {journal} {Phys. Rev.}\ }\textbf {\bibinfo
  {volume} {D81}},\ \bibinfo {pages} {023523}},\ \Eprint
  {http://arxiv.org/abs/0907.1658} {arXiv:0907.1658 [astro-ph.CO]} \BibitemShut
  {NoStop}%
%%CITATION = ARXIV:0907.1658;%%
\bibitem [{\citenamefont {Dubovsky}(2004)}]{Dubovsky:2004sg}%
  \BibitemOpen
  \bibfield  {author} {\bibinfo {author} {\bibnamefont {Dubovsky},
  \bibfnamefont {SL}}} (\bibinfo {year} {2004}),\ \bibfield  {title} {\enquote
  {\bibinfo {title} {{Phases of massive gravity}},}\ }\href {\doibase
  10.1088/1126-6708/2004/10/076} {\bibfield  {journal} {\bibinfo  {journal}
  {JHEP}\ }\textbf {\bibinfo {volume} {0410}},\ \bibinfo {pages} {076}},\
  \Eprint {http://arxiv.org/abs/hep-th/0409124} {arXiv:hep-th/0409124 [hep-th]}
  \BibitemShut {NoStop}%
%%CITATION = HEP-TH/0409124;%%
\bibitem [{\citenamefont {Dvali}\ \emph {et~al.}(2003)\citenamefont {Dvali},
  \citenamefont {Gruzinov},\ and\ \citenamefont {Zaldarriaga}}]{Dvali:2002vf}%
  \BibitemOpen
  \bibfield  {author} {\bibinfo {author} {\bibnamefont {Dvali}, \bibfnamefont
  {Gia}}, \bibinfo {author} {\bibfnamefont {Andrei}\ \bibnamefont {Gruzinov}},
  \ and\ \bibinfo {author} {\bibfnamefont {Matias}\ \bibnamefont
  {Zaldarriaga}}} (\bibinfo {year} {2003}),\ \bibfield  {title} {\enquote
  {\bibinfo {title} {{The Accelerated universe and the moon}},}\ }\href
  {\doibase 10.1103/PhysRevD.68.024012} {\bibfield  {journal} {\bibinfo
  {journal} {Phys. Rev.}\ }\textbf {\bibinfo {volume} {D68}},\ \bibinfo {pages}
  {024012}},\ \Eprint {http://arxiv.org/abs/hep-ph/0212069}
  {arXiv:hep-ph/0212069 [hep-ph]} \BibitemShut {NoStop}%
%%CITATION = HEP-PH/0212069;%%
\bibitem [{\citenamefont {Dvali}\ \emph
  {et~al.}(2000{\natexlab{a}})\citenamefont {Dvali}, \citenamefont
  {Gabadadze},\ and\ \citenamefont {Porrati}}]{Dvali:2000rv}%
  \BibitemOpen
  \bibfield  {author} {\bibinfo {author} {\bibnamefont {Dvali}, \bibfnamefont
  {GR}}, \bibinfo {author} {\bibfnamefont {G.}~\bibnamefont {Gabadadze}}, \
  and\ \bibinfo {author} {\bibfnamefont {M.}~\bibnamefont {Porrati}}} (\bibinfo
  {year} {2000}{\natexlab{a}}),\ \bibfield  {title} {\enquote {\bibinfo {title}
  {{Metastable gravitons and infinite volume extra dimensions}},}\ }\href
  {\doibase 10.1016/S0370-2693(00)00631-6} {\bibfield  {journal} {\bibinfo
  {journal} {Phys.Lett.}\ }\textbf {\bibinfo {volume} {B484}},\ \bibinfo
  {pages} {112--118}},\ \Eprint {http://arxiv.org/abs/hep-th/0002190}
  {arXiv:hep-th/0002190 [hep-th]} \BibitemShut {NoStop}%
%%CITATION = HEP-TH/0002190;%%
\bibitem [{\citenamefont {Dvali}\ \emph
  {et~al.}(2000{\natexlab{b}})\citenamefont {Dvali}, \citenamefont
  {Gabadadze},\ and\ \citenamefont {Porrati}}]{Dvali:2000hr}%
  \BibitemOpen
  \bibfield  {author} {\bibinfo {author} {\bibnamefont {Dvali}, \bibfnamefont
  {GR}}, \bibinfo {author} {\bibfnamefont {Gregory}\ \bibnamefont {Gabadadze}},
  \ and\ \bibinfo {author} {\bibfnamefont {Massimo}\ \bibnamefont {Porrati}}}
  (\bibinfo {year} {2000}{\natexlab{b}}),\ \bibfield  {title} {\enquote
  {\bibinfo {title} {{4-D gravity on a brane in 5-D Minkowski space}},}\ }\href
  {\doibase 10.1016/S0370-2693(00)00669-9} {\bibfield  {journal} {\bibinfo
  {journal} {Phys.Lett.}\ }\textbf {\bibinfo {volume} {B485}},\ \bibinfo
  {pages} {208--214}},\ \Eprint {http://arxiv.org/abs/hep-th/0005016}
  {arXiv:hep-th/0005016 [hep-th]} \BibitemShut {NoStop}%
%%CITATION = HEP-TH/0005016;%%
\bibitem [{\citenamefont {Eglseer}\ \emph {et~al.}(2015)\citenamefont
  {Eglseer}, \citenamefont {Niedermann},\ and\ \citenamefont
  {Schneider}}]{Eglseer:2015xla}%
  \BibitemOpen
  \bibfield  {author} {\bibinfo {author} {\bibnamefont {Eglseer}, \bibfnamefont
  {Ludwig}}, \bibinfo {author} {\bibfnamefont {Florian}\ \bibnamefont
  {Niedermann}}, \ and\ \bibinfo {author} {\bibfnamefont {Robert}\ \bibnamefont
  {Schneider}}} (\bibinfo {year} {2015}),\ \bibfield  {title} {\enquote
  {\bibinfo {title} {{Brane induced gravity: Ghosts and naturalness}},}\ }\href
  {\doibase 10.1103/PhysRevD.92.084029} {\bibfield  {journal} {\bibinfo
  {journal} {Phys. Rev.}\ }\textbf {\bibinfo {volume} {D92}}~(\bibinfo {number}
  {8}),\ \bibinfo {pages} {084029}},\ \Eprint {http://arxiv.org/abs/1506.02666}
  {arXiv:1506.02666 [gr-qc]} \BibitemShut {NoStop}%
%%CITATION = ARXIV:1506.02666;%%
\bibitem [{\citenamefont {Fasiello}\ and\ \citenamefont
  {Tolley}(2012)}]{Fasiello:2012rw}%
  \BibitemOpen
  \bibfield  {author} {\bibinfo {author} {\bibnamefont {Fasiello},
  \bibfnamefont {Matteo}}, \ and\ \bibinfo {author} {\bibfnamefont {Andrew~J.}\
  \bibnamefont {Tolley}}} (\bibinfo {year} {2012}),\ \bibfield  {title}
  {\enquote {\bibinfo {title} {{Cosmological perturbations in Massive Gravity
  and the Higuchi bound}},}\ }\href {\doibase 10.1088/1475-7516/2012/11/035}
  {\bibfield  {journal} {\bibinfo  {journal} {JCAP}\ }\textbf {\bibinfo
  {volume} {1211}},\ \bibinfo {pages} {035}},\ \Eprint
  {http://arxiv.org/abs/1206.3852} {arXiv:1206.3852 [hep-th]} \BibitemShut
  {NoStop}%
%%CITATION = ARXIV:1206.3852;%%
\bibitem [{\citenamefont {Fierz}\ and\ \citenamefont
  {Pauli}(1939)}]{Fierz:1939ix}%
  \BibitemOpen
  \bibfield  {author} {\bibinfo {author} {\bibnamefont {Fierz}, \bibfnamefont
  {M}}, \ and\ \bibinfo {author} {\bibfnamefont {W.}~\bibnamefont {Pauli}}}
  (\bibinfo {year} {1939}),\ \bibfield  {title} {\enquote {\bibinfo {title}
  {{On relativistic wave equations for particles of arbitrary spin in an
  electromagnetic field}},}\ }\href {\doibase 10.1098/rspa.1939.0140}
  {\bibfield  {journal} {\bibinfo  {journal} {Proc.Roy.Soc.Lond.}\ }\textbf
  {\bibinfo {volume} {A173}},\ \bibinfo {pages} {211--232}}\BibitemShut
  {NoStop}%
%%CITATION = PRSLA,A173,211;%%
\bibitem [{\citenamefont {Finn}\ and\ \citenamefont
  {Sutton}(2002)}]{Finn:2001qi}%
  \BibitemOpen
  \bibfield  {author} {\bibinfo {author} {\bibnamefont {Finn}, \bibfnamefont
  {Lee~Samuel}}, \ and\ \bibinfo {author} {\bibfnamefont {Patrick~J.}\
  \bibnamefont {Sutton}}} (\bibinfo {year} {2002}),\ \bibfield  {title}
  {\enquote {\bibinfo {title} {{Bounding the mass of the graviton using binary
  pulsar observations}},}\ }\href {\doibase 10.1103/PhysRevD.65.044022}
  {\bibfield  {journal} {\bibinfo  {journal} {Phys. Rev.}\ }\textbf {\bibinfo
  {volume} {D65}},\ \bibinfo {pages} {044022}},\ \Eprint
  {http://arxiv.org/abs/gr-qc/0109049} {arXiv:gr-qc/0109049 [gr-qc]}
  \BibitemShut {NoStop}%
%%CITATION = GR-QC/0109049;%%
\bibitem [{\citenamefont {Foffa}\ \emph {et~al.}(2014)\citenamefont {Foffa},
  \citenamefont {Maggiore},\ and\ \citenamefont {Mitsou}}]{Foffa:2013vma}%
  \BibitemOpen
  \bibfield  {author} {\bibinfo {author} {\bibnamefont {Foffa}, \bibfnamefont
  {Stefano}}, \bibinfo {author} {\bibfnamefont {Michele}\ \bibnamefont
  {Maggiore}}, \ and\ \bibinfo {author} {\bibfnamefont {Ermis}\ \bibnamefont
  {Mitsou}}} (\bibinfo {year} {2014}),\ \bibfield  {title} {\enquote {\bibinfo
  {title} {{Cosmological dynamics and dark energy from nonlocal infrared
  modifications of gravity}},}\ }\href {\doibase 10.1142/S0217751X14501164}
  {\bibfield  {journal} {\bibinfo  {journal} {Int. J. Mod. Phys.}\ }\textbf
  {\bibinfo {volume} {A29}},\ \bibinfo {pages} {1450116}},\ \Eprint
  {http://arxiv.org/abs/1311.3435} {arXiv:1311.3435 [hep-th]} \BibitemShut
  {NoStop}%
%%CITATION = ARXIV:1311.3435;%%
\bibitem [{\citenamefont {Gabadadze}\ and\ \citenamefont
  {Shifman}(2004)}]{Gabadadze:2003ck}%
  \BibitemOpen
  \bibfield  {author} {\bibinfo {author} {\bibnamefont {Gabadadze},
  \bibfnamefont {G}}, \ and\ \bibinfo {author} {\bibfnamefont {M.}~\bibnamefont
  {Shifman}}} (\bibinfo {year} {2004}),\ \bibfield  {title} {\enquote {\bibinfo
  {title} {{Softly massive gravity}},}\ }\href {\doibase
  10.1103/PhysRevD.69.124032} {\bibfield  {journal} {\bibinfo  {journal} {Phys.
  Rev.}\ }\textbf {\bibinfo {volume} {D69}},\ \bibinfo {pages} {124032}},\
  \Eprint {http://arxiv.org/abs/hep-th/0312289} {arXiv:hep-th/0312289 [hep-th]}
  \BibitemShut {NoStop}%
%%CITATION = HEP-TH/0312289;%%
\bibitem [{\citenamefont {Gabadadze}\ \emph {et~al.}(2013)\citenamefont
  {Gabadadze}, \citenamefont {Hinterbichler}, \citenamefont {Pirtskhalava},\
  and\ \citenamefont {Shang}}]{Gabadadze:2013ria}%
  \BibitemOpen
  \bibfield  {author} {\bibinfo {author} {\bibnamefont {Gabadadze},
  \bibfnamefont {Gregory}}, \bibinfo {author} {\bibfnamefont {Kurt}\
  \bibnamefont {Hinterbichler}}, \bibinfo {author} {\bibfnamefont {David}\
  \bibnamefont {Pirtskhalava}}, \ and\ \bibinfo {author} {\bibfnamefont
  {Yanwen}\ \bibnamefont {Shang}}} (\bibinfo {year} {2013}),\ \bibfield
  {title} {\enquote {\bibinfo {title} {{Potential for general relativity and
  its geometry}},}\ }\href {\doibase 10.1103/PhysRevD.88.084003} {\bibfield
  {journal} {\bibinfo  {journal} {Phys. Rev.}\ }\textbf {\bibinfo {volume}
  {D88}}~(\bibinfo {number} {8}),\ \bibinfo {pages} {084003}},\ \Eprint
  {http://arxiv.org/abs/1307.2245} {arXiv:1307.2245} \BibitemShut {NoStop}%
%%CITATION = ARXIV:1307.2245;%%
\bibitem [{\citenamefont {Goldhaber}\ and\ \citenamefont
  {Nieto}(1974)}]{Goldhaber:1974wg}%
  \BibitemOpen
  \bibfield  {author} {\bibinfo {author} {\bibnamefont {Goldhaber},
  \bibfnamefont {A~S}}, \ and\ \bibinfo {author} {\bibfnamefont {M.~M.}\
  \bibnamefont {Nieto}}} (\bibinfo {year} {1974}),\ \bibfield  {title}
  {\enquote {\bibinfo {title} {{Mass of the graviton}},}\ }\href {\doibase
  10.1103/PhysRevD.9.1119} {\bibfield  {journal} {\bibinfo  {journal} {Phys.
  Rev.}\ }\textbf {\bibinfo {volume} {D9}},\ \bibinfo {pages}
  {1119--1121}}\BibitemShut {NoStop}%
%%CITATION = PHRVA,D9,1119;%%
\bibitem [{\citenamefont {Goldhaber}\ and\ \citenamefont
  {Nieto}(2010)}]{Goldhaber:2008xy}%
  \BibitemOpen
  \bibfield  {author} {\bibinfo {author} {\bibnamefont {Goldhaber},
  \bibfnamefont {Alfred~Scharff}}, \ and\ \bibinfo {author} {\bibfnamefont
  {Michael~Martin}\ \bibnamefont {Nieto}}} (\bibinfo {year} {2010}),\ \bibfield
   {title} {\enquote {\bibinfo {title} {{Photon and Graviton Mass Limits}},}\
  }\href {\doibase 10.1103/RevModPhys.82.939} {\bibfield  {journal} {\bibinfo
  {journal} {Rev. Mod. Phys.}\ }\textbf {\bibinfo {volume} {82}},\ \bibinfo
  {pages} {939--979}},\ \Eprint {http://arxiv.org/abs/0809.1003}
  {arXiv:0809.1003 [hep-ph]} \BibitemShut {NoStop}%
%%CITATION = ARXIV:0809.1003;%%
\bibitem [{\citenamefont {Gregory}\ \emph {et~al.}(2000)\citenamefont
  {Gregory}, \citenamefont {Rubakov},\ and\ \citenamefont
  {Sibiryakov}}]{Gregory:2000jc}%
  \BibitemOpen
  \bibfield  {author} {\bibinfo {author} {\bibnamefont {Gregory}, \bibfnamefont
  {Ruth}}, \bibinfo {author} {\bibfnamefont {V.~A.}\ \bibnamefont {Rubakov}}, \
  and\ \bibinfo {author} {\bibfnamefont {Sergei~M.}\ \bibnamefont
  {Sibiryakov}}} (\bibinfo {year} {2000}),\ \bibfield  {title} {\enquote
  {\bibinfo {title} {{Opening up extra dimensions at ultra large scales}},}\
  }\href {\doibase 10.1103/PhysRevLett.84.5928} {\bibfield  {journal} {\bibinfo
   {journal} {Phys. Rev. Lett.}\ }\textbf {\bibinfo {volume} {84}},\ \bibinfo
  {pages} {5928--5931}},\ \Eprint {http://arxiv.org/abs/hep-th/0002072}
  {arXiv:hep-th/0002072 [hep-th]} \BibitemShut {NoStop}%
%%CITATION = HEP-TH/0002072;%%
\bibitem [{\citenamefont {Gruzinov}(2005)}]{Gruzinov:2001hp}%
  \BibitemOpen
  \bibfield  {author} {\bibinfo {author} {\bibnamefont {Gruzinov},
  \bibfnamefont {Andrei}}} (\bibinfo {year} {2005}),\ \bibfield  {title}
  {\enquote {\bibinfo {title} {{On the graviton mass}},}\ }\href {\doibase
  10.1016/j.newast.2004.12.001} {\bibfield  {journal} {\bibinfo  {journal} {New
  Astron.}\ }\textbf {\bibinfo {volume} {10}},\ \bibinfo {pages} {311--314}},\
  \Eprint {http://arxiv.org/abs/astro-ph/0112246} {arXiv:astro-ph/0112246
  [astro-ph]} \BibitemShut {NoStop}%
%%CITATION = ASTRO-PH/0112246;%%
\bibitem [{\citenamefont {Gruzinov}\ and\ \citenamefont
  {Mirbabayi}(2011)}]{Gruzinov:2011mm}%
  \BibitemOpen
  \bibfield  {author} {\bibinfo {author} {\bibnamefont {Gruzinov},
  \bibfnamefont {Andrei}}, \ and\ \bibinfo {author} {\bibfnamefont {Mehrdad}\
  \bibnamefont {Mirbabayi}}} (\bibinfo {year} {2011}),\ \bibfield  {title}
  {\enquote {\bibinfo {title} {{Stars and Black Holes in Massive Gravity}},}\
  }\href {\doibase 10.1103/PhysRevD.84.124019} {\bibfield  {journal} {\bibinfo
  {journal} {Phys. Rev.}\ }\textbf {\bibinfo {volume} {D84}},\ \bibinfo {pages}
  {124019}},\ \Eprint {http://arxiv.org/abs/1106.2551} {arXiv:1106.2551
  [hep-th]} \BibitemShut {NoStop}%
%%CITATION = ARXIV:1106.2551;%%
\bibitem [{\citenamefont {Gumrukcuoglu}\ \emph {et~al.}(2012)\citenamefont
  {Gumrukcuoglu}, \citenamefont {Kuroyanagi}, \citenamefont {Lin},
  \citenamefont {Mukohyama},\ and\ \citenamefont
  {Tanahashi}}]{Gumrukcuoglu:2012wt}%
  \BibitemOpen
  \bibfield  {author} {\bibinfo {author} {\bibnamefont {Gumrukcuoglu},
  \bibfnamefont {A~Emir}}, \bibinfo {author} {\bibfnamefont {Sachiko}\
  \bibnamefont {Kuroyanagi}}, \bibinfo {author} {\bibfnamefont {Chunshan}\
  \bibnamefont {Lin}}, \bibinfo {author} {\bibfnamefont {Shinji}\ \bibnamefont
  {Mukohyama}}, \ and\ \bibinfo {author} {\bibfnamefont {Norihiro}\
  \bibnamefont {Tanahashi}}} (\bibinfo {year} {2012}),\ \bibfield  {title}
  {\enquote {\bibinfo {title} {{Gravitational wave signal from massive
  gravity}},}\ }\href {\doibase 10.1088/0264-9381/29/23/235026} {\bibfield
  {journal} {\bibinfo  {journal} {Class. Quant. Grav.}\ }\textbf {\bibinfo
  {volume} {29}},\ \bibinfo {pages} {235026}},\ \Eprint
  {http://arxiv.org/abs/1208.5975} {arXiv:1208.5975 [hep-th]} \BibitemShut
  {NoStop}%
%%CITATION = ARXIV:1208.5975;%%
\bibitem [{\citenamefont {Hare}(1973)}]{Hare:1973px}%
  \BibitemOpen
  \bibfield  {author} {\bibinfo {author} {\bibnamefont {Hare}, \bibfnamefont
  {M~G}}} (\bibinfo {year} {1973}),\ \bibfield  {title} {\enquote {\bibinfo
  {title} {{Mass of the graviton}},}\ }\href {\doibase 10.1139/p73-056}
  {\bibfield  {journal} {\bibinfo  {journal} {Can. J. Phys.}\ }\textbf
  {\bibinfo {volume} {51}},\ \bibinfo {pages} {431--433}}\BibitemShut {NoStop}%
%%CITATION = CJPHA,51,431;%%
\bibitem [{\citenamefont {Harry}(2010)}]{Harry:2010zz}%
  \BibitemOpen
  \bibfield  {author} {\bibinfo {author} {\bibnamefont {Harry}, \bibfnamefont
  {Gregory~M}} (\bibinfo {collaboration} {LIGO Scientific})} (\bibinfo {year}
  {2010}),\ \bibfield  {title} {\enquote {\bibinfo {title} {{Advanced LIGO: The
  next generation of gravitational wave detectors}},}\ }\bibfield  {booktitle}
  {\emph {\bibinfo {booktitle} {{Gravitational waves. Proceedings, 8th Edoardo
  Amaldi Conference, Amaldi 8, New York, USA, June 22-26, 2009}}},\ }\href
  {\doibase 10.1088/0264-9381/27/8/084006} {\bibfield  {journal} {\bibinfo
  {journal} {Class. Quant. Grav.}\ }\textbf {\bibinfo {volume} {27}},\ \bibinfo
  {pages} {084006}}\BibitemShut {NoStop}%
%%CITATION = CQGRD,27,084006;%%
\bibitem [{\citenamefont {Hassan}\ and\ \citenamefont
  {Rosen}(2012{\natexlab{a}})}]{Hassan:2011zd}%
  \BibitemOpen
  \bibfield  {author} {\bibinfo {author} {\bibnamefont {Hassan}, \bibfnamefont
  {S~F}}, \ and\ \bibinfo {author} {\bibfnamefont {Rachel~A.}\ \bibnamefont
  {Rosen}}} (\bibinfo {year} {2012}{\natexlab{a}}),\ \bibfield  {title}
  {\enquote {\bibinfo {title} {{Bimetric Gravity from Ghost-free Massive
  Gravity}},}\ }\href {\doibase 10.1007/JHEP02(2012)126} {\bibfield  {journal}
  {\bibinfo  {journal} {JHEP}\ }\textbf {\bibinfo {volume} {02}},\ \bibinfo
  {pages} {126}},\ \Eprint {http://arxiv.org/abs/1109.3515} {arXiv:1109.3515
  [hep-th]} \BibitemShut {NoStop}%
%%CITATION = ARXIV:1109.3515;%%
\bibitem [{\citenamefont {Hassan}\ and\ \citenamefont
  {Rosen}(2012{\natexlab{b}})}]{Hassan:2011ea}%
  \BibitemOpen
  \bibfield  {author} {\bibinfo {author} {\bibnamefont {Hassan}, \bibfnamefont
  {S~F}}, \ and\ \bibinfo {author} {\bibfnamefont {Rachel~A.}\ \bibnamefont
  {Rosen}}} (\bibinfo {year} {2012}{\natexlab{b}}),\ \bibfield  {title}
  {\enquote {\bibinfo {title} {{Confirmation of the Secondary Constraint and
  Absence of Ghost in Massive Gravity and Bimetric Gravity}},}\ }\href
  {\doibase 10.1007/JHEP04(2012)123} {\bibfield  {journal} {\bibinfo  {journal}
  {JHEP}\ }\textbf {\bibinfo {volume} {04}},\ \bibinfo {pages} {123}},\ \Eprint
  {http://arxiv.org/abs/1111.2070} {arXiv:1111.2070 [hep-th]} \BibitemShut
  {NoStop}%
%%CITATION = ARXIV:1111.2070;%%
\bibitem [{\citenamefont {Hassan}\ and\ \citenamefont
  {Rosen}(2012{\natexlab{c}})}]{Hassan:2011hr}%
  \BibitemOpen
  \bibfield  {author} {\bibinfo {author} {\bibnamefont {Hassan}, \bibfnamefont
  {S~F}}, \ and\ \bibinfo {author} {\bibfnamefont {Rachel~A.}\ \bibnamefont
  {Rosen}}} (\bibinfo {year} {2012}{\natexlab{c}}),\ \bibfield  {title}
  {\enquote {\bibinfo {title} {{Resolving the Ghost Problem in non-Linear
  Massive Gravity}},}\ }\href {\doibase 10.1103/PhysRevLett.108.041101}
  {\bibfield  {journal} {\bibinfo  {journal} {Phys. Rev. Lett.}\ }\textbf
  {\bibinfo {volume} {108}},\ \bibinfo {pages} {041101}},\ \Eprint
  {http://arxiv.org/abs/1106.3344} {arXiv:1106.3344 [hep-th]} \BibitemShut
  {NoStop}%
%%CITATION = ARXIV:1106.3344;%%
\bibitem [{\citenamefont {Hinterbichler}(2012)}]{Hinterbichler:2011tt}%
  \BibitemOpen
  \bibfield  {author} {\bibinfo {author} {\bibnamefont {Hinterbichler},
  \bibfnamefont {Kurt}}} (\bibinfo {year} {2012}),\ \bibfield  {title}
  {\enquote {\bibinfo {title} {{Theoretical Aspects of Massive Gravity}},}\
  }\href {\doibase 10.1103/RevModPhys.84.671} {\bibfield  {journal} {\bibinfo
  {journal} {Rev. Mod. Phys.}\ }\textbf {\bibinfo {volume} {84}},\ \bibinfo
  {pages} {671--710}},\ \Eprint {http://arxiv.org/abs/1105.3735}
  {arXiv:1105.3735 [hep-th]} \BibitemShut {NoStop}%
%%CITATION = ARXIV:1105.3735;%%
\bibitem [{\citenamefont {Hinterbichler}\ and\ \citenamefont
  {Rosen}(2012)}]{Hinterbichler:2012cn}%
  \BibitemOpen
  \bibfield  {author} {\bibinfo {author} {\bibnamefont {Hinterbichler},
  \bibfnamefont {Kurt}}, \ and\ \bibinfo {author} {\bibfnamefont {Rachel~A.}\
  \bibnamefont {Rosen}}} (\bibinfo {year} {2012}),\ \bibfield  {title}
  {\enquote {\bibinfo {title} {{Interacting Spin-2 Fields}},}\ }\href {\doibase
  10.1007/JHEP07(2012)047} {\bibfield  {journal} {\bibinfo  {journal} {JHEP}\
  }\textbf {\bibinfo {volume} {07}},\ \bibinfo {pages} {047}},\ \Eprint
  {http://arxiv.org/abs/1203.5783} {arXiv:1203.5783 [hep-th]} \BibitemShut
  {NoStop}%
%%CITATION = ARXIV:1203.5783;%%
\bibitem [{\citenamefont {Hobbs}\ \emph {et~al.}(2010)\citenamefont {Hobbs}
  \emph {et~al.}}]{Hobbs:2009yy}%
  \BibitemOpen
  \bibfield  {author} {\bibinfo {author} {\bibnamefont {Hobbs}, \bibfnamefont
  {G}},  \emph {et~al.}} (\bibinfo {year} {2010}),\ \bibfield  {title}
  {\enquote {\bibinfo {title} {{The international pulsar timing array project:
  using pulsars as a gravitational wave detector}},}\ }\bibfield  {booktitle}
  {\emph {\bibinfo {booktitle} {{Gravitational waves. Proceedings, 8th Edoardo
  Amaldi Conference, Amaldi 8, New York, USA, June 22-26, 2009}}},\ }\href
  {\doibase 10.1088/0264-9381/27/8/084013} {\bibfield  {journal} {\bibinfo
  {journal} {Class. Quant. Grav.}\ }\textbf {\bibinfo {volume} {27}},\ \bibinfo
  {pages} {084013}},\ \Eprint {http://arxiv.org/abs/0911.5206} {arXiv:0911.5206
  [astro-ph.SR]} \BibitemShut {NoStop}%
%%CITATION = ARXIV:0911.5206;%%
\bibitem [{\citenamefont {Huang}\ \emph {et~al.}(2012)\citenamefont {Huang},
  \citenamefont {Piao},\ and\ \citenamefont {Zhou}}]{Huang:2012pe}%
  \BibitemOpen
  \bibfield  {author} {\bibinfo {author} {\bibnamefont {Huang}, \bibfnamefont
  {Qing-Guo}}, \bibinfo {author} {\bibfnamefont {Yun-Song}\ \bibnamefont
  {Piao}}, \ and\ \bibinfo {author} {\bibfnamefont {Shuang-Yong}\ \bibnamefont
  {Zhou}}} (\bibinfo {year} {2012}),\ \bibfield  {title} {\enquote {\bibinfo
  {title} {{Mass-Varying Massive Gravity}},}\ }\href {\doibase
  10.1103/PhysRevD.86.124014} {\bibfield  {journal} {\bibinfo  {journal} {Phys.
  Rev.}\ }\textbf {\bibinfo {volume} {D86}},\ \bibinfo {pages} {124014}},\
  \Eprint {http://arxiv.org/abs/1206.5678} {arXiv:1206.5678 [hep-th]}
  \BibitemShut {NoStop}%
%%CITATION = ARXIV:1206.5678;%%
\bibitem [{\citenamefont {Huang}\ \emph {et~al.}(2015)\citenamefont {Huang},
  \citenamefont {Ribeiro}, \citenamefont {Xing}, \citenamefont {Zhang},\ and\
  \citenamefont {Zhou}}]{Huang:2015yga}%
  \BibitemOpen
  \bibfield  {author} {\bibinfo {author} {\bibnamefont {Huang}, \bibfnamefont
  {Qing-Guo}}, \bibinfo {author} {\bibfnamefont {Raquel~H.}\ \bibnamefont
  {Ribeiro}}, \bibinfo {author} {\bibfnamefont {Yu-Hang}\ \bibnamefont {Xing}},
  \bibinfo {author} {\bibfnamefont {Ke-Chao}\ \bibnamefont {Zhang}}, \ and\
  \bibinfo {author} {\bibfnamefont {Shuang-Yong}\ \bibnamefont {Zhou}}}
  (\bibinfo {year} {2015}),\ \bibfield  {title} {\enquote {\bibinfo {title}
  {{On the uniqueness of the non-minimal matter coupling in massive gravity and
  bigravity}},}\ }\href {\doibase 10.1016/j.physletb.2015.07.003} {\bibfield
  {journal} {\bibinfo  {journal} {Phys. Lett.}\ }\textbf {\bibinfo {volume}
  {B748}},\ \bibinfo {pages} {356--360}},\ \bibinfo {note} {[Phys.
  Lett.B748,356(2015)]},\ \Eprint {http://arxiv.org/abs/1505.02616}
  {arXiv:1505.02616 [hep-th]} \BibitemShut {NoStop}%
%%CITATION = ARXIV:1505.02616;%%
\bibitem [{\citenamefont {Huang}\ \emph {et~al.}(2013)\citenamefont {Huang},
  \citenamefont {Zhang},\ and\ \citenamefont {Zhou}}]{Huang:2013mha}%
  \BibitemOpen
  \bibfield  {author} {\bibinfo {author} {\bibnamefont {Huang}, \bibfnamefont
  {Qing-Guo}}, \bibinfo {author} {\bibfnamefont {Ke-Chao}\ \bibnamefont
  {Zhang}}, \ and\ \bibinfo {author} {\bibfnamefont {Shuang-Yong}\ \bibnamefont
  {Zhou}}} (\bibinfo {year} {2013}),\ \bibfield  {title} {\enquote {\bibinfo
  {title} {{Generalized massive gravity in arbitrary dimensions and its
  Hamiltonian formulation}},}\ }\href {\doibase 10.1088/1475-7516/2013/08/050}
  {\bibfield  {journal} {\bibinfo  {journal} {JCAP}\ }\textbf {\bibinfo
  {volume} {1308}},\ \bibinfo {pages} {050}},\ \Eprint
  {http://arxiv.org/abs/1306.4740} {arXiv:1306.4740 [hep-th]} \BibitemShut
  {NoStop}%
%%CITATION = ARXIV:1306.4740;%%
\bibitem [{\citenamefont {Hulse}\ and\ \citenamefont
  {Taylor}(1975)}]{Hulse:1974eb}%
  \BibitemOpen
  \bibfield  {author} {\bibinfo {author} {\bibnamefont {Hulse}, \bibfnamefont
  {R~A}}, \ and\ \bibinfo {author} {\bibfnamefont {J.~H.}\ \bibnamefont
  {Taylor}}} (\bibinfo {year} {1975}),\ \bibfield  {title} {\enquote {\bibinfo
  {title} {{Discovery of a pulsar in a binary system}},}\ }\href {\doibase
  10.1086/181708} {\bibfield  {journal} {\bibinfo  {journal} {Astrophys. J.}\
  }\textbf {\bibinfo {volume} {195}},\ \bibinfo {pages} {L51--L53}}\BibitemShut
  {NoStop}%
%%CITATION = ASJOA,195,L51;%%
\bibitem [{\citenamefont {Huwyler}\ \emph {et~al.}(2012)\citenamefont
  {Huwyler}, \citenamefont {Klein},\ and\ \citenamefont
  {Jetzer}}]{Huwyler:2011iq}%
  \BibitemOpen
  \bibfield  {author} {\bibinfo {author} {\bibnamefont {Huwyler}, \bibfnamefont
  {Cedric}}, \bibinfo {author} {\bibfnamefont {Antoine}\ \bibnamefont {Klein}},
  \ and\ \bibinfo {author} {\bibfnamefont {Philippe}\ \bibnamefont {Jetzer}}}
  (\bibinfo {year} {2012}),\ \bibfield  {title} {\enquote {\bibinfo {title}
  {{Testing General Relativity with LISA including Spin Precession and Higher
  Harmonics in the Waveform}},}\ }\href {\doibase 10.1103/PhysRevD.86.084028}
  {\bibfield  {journal} {\bibinfo  {journal} {Phys. Rev.}\ }\textbf {\bibinfo
  {volume} {D86}},\ \bibinfo {pages} {084028}},\ \Eprint
  {http://arxiv.org/abs/1108.1826} {arXiv:1108.1826 [gr-qc]} \BibitemShut
  {NoStop}%
%%CITATION = ARXIV:1108.1826;%%
\bibitem [{\citenamefont {Iwasaki}(1970)}]{Iwasaki:1971uz}%
  \BibitemOpen
  \bibfield  {author} {\bibinfo {author} {\bibnamefont {Iwasaki}, \bibfnamefont
  {Y}}} (\bibinfo {year} {1970}),\ \bibfield  {title} {\enquote {\bibinfo
  {title} {{Consistency condition for propagators}},}\ }\href {\doibase
  10.1103/PhysRevD.2.2255} {\bibfield  {journal} {\bibinfo  {journal} {Phys.
  Rev.}\ }\textbf {\bibinfo {volume} {D2}},\ \bibinfo {pages}
  {2255--2256}}\BibitemShut {NoStop}%
%%CITATION = PHRVA,D2,2255;%%
\bibitem [{\citenamefont {Jaccard}\ \emph {et~al.}(2013)\citenamefont
  {Jaccard}, \citenamefont {Maggiore},\ and\ \citenamefont
  {Mitsou}}]{Jaccard:2013gla}%
  \BibitemOpen
  \bibfield  {author} {\bibinfo {author} {\bibnamefont {Jaccard}, \bibfnamefont
  {Maud}}, \bibinfo {author} {\bibfnamefont {Michele}\ \bibnamefont
  {Maggiore}}, \ and\ \bibinfo {author} {\bibfnamefont {Ermis}\ \bibnamefont
  {Mitsou}}} (\bibinfo {year} {2013}),\ \bibfield  {title} {\enquote {\bibinfo
  {title} {{Nonlocal theory of massive gravity}},}\ }\href {\doibase
  10.1103/PhysRevD.88.044033} {\bibfield  {journal} {\bibinfo  {journal} {Phys.
  Rev.}\ }\textbf {\bibinfo {volume} {D88}}~(\bibinfo {number} {4}),\ \bibinfo
  {pages} {044033}},\ \Eprint {http://arxiv.org/abs/1305.3034} {arXiv:1305.3034
  [hep-th]} \BibitemShut {NoStop}%
%%CITATION = ARXIV:1305.3034;%%
\bibitem [{\citenamefont {Jones}(2004)}]{Jones:2004uy}%
  \BibitemOpen
  \bibfield  {author} {\bibinfo {author} {\bibnamefont {Jones}, \bibfnamefont
  {D~I}}} (\bibinfo {year} {2004}),\ \bibfield  {title} {\enquote {\bibinfo
  {title} {{Bounding the mass of the graviton using eccentric binaries}},}\
  }\href {\doibase 10.1086/427773} {\bibfield  {journal} {\bibinfo  {journal}
  {Astrophys. J.}\ }\textbf {\bibinfo {volume} {618}},\ \bibinfo {pages}
  {L115--L118}},\ \Eprint {http://arxiv.org/abs/gr-qc/0411123}
  {arXiv:gr-qc/0411123 [gr-qc]} \BibitemShut {NoStop}%
%%CITATION = GR-QC/0411123;%%
\bibitem [{\citenamefont {Kahniashvili}\ \emph {et~al.}(2015)\citenamefont
  {Kahniashvili}, \citenamefont {Kar}, \citenamefont {Lavrelashvili},
  \citenamefont {Agarwal}, \citenamefont {Heisenberg},\ and\ \citenamefont
  {Kosowsky}}]{Kahniashvili:2014wua}%
  \BibitemOpen
  \bibfield  {author} {\bibinfo {author} {\bibnamefont {Kahniashvili},
  \bibfnamefont {Tina}}, \bibinfo {author} {\bibfnamefont {Arjun}\ \bibnamefont
  {Kar}}, \bibinfo {author} {\bibfnamefont {George}\ \bibnamefont
  {Lavrelashvili}}, \bibinfo {author} {\bibfnamefont {Nishant}\ \bibnamefont
  {Agarwal}}, \bibinfo {author} {\bibfnamefont {Lavinia}\ \bibnamefont
  {Heisenberg}}, \ and\ \bibinfo {author} {\bibfnamefont {Arthur}\ \bibnamefont
  {Kosowsky}}} (\bibinfo {year} {2015}),\ \bibfield  {title} {\enquote
  {\bibinfo {title} {{Cosmic Expansion in Extended Quasidilaton Massive
  Gravity}},}\ }\href {\doibase 10.1103/PhysRevD.91.041301} {\bibfield
  {journal} {\bibinfo  {journal} {Phys. Rev.}\ }\textbf {\bibinfo {volume}
  {D91}}~(\bibinfo {number} {4}),\ \bibinfo {pages} {041301}},\ \Eprint
  {http://arxiv.org/abs/1412.4300} {arXiv:1412.4300 [astro-ph.CO]} \BibitemShut
  {NoStop}%
%%CITATION = ARXIV:1412.4300;%%
\bibitem [{\citenamefont {Kehagias}\ and\ \citenamefont
  {Maggiore}(2014)}]{Kehagias:2014sda}%
  \BibitemOpen
  \bibfield  {author} {\bibinfo {author} {\bibnamefont {Kehagias},
  \bibfnamefont {Alex}}, \ and\ \bibinfo {author} {\bibfnamefont {Michele}\
  \bibnamefont {Maggiore}}} (\bibinfo {year} {2014}),\ \bibfield  {title}
  {\enquote {\bibinfo {title} {{Spherically symmetric static solutions in a
  nonlocal infrared modification of General Relativity}},}\ }\href {\doibase
  10.1007/JHEP08(2014)029} {\bibfield  {journal} {\bibinfo  {journal} {JHEP}\
  }\textbf {\bibinfo {volume} {08}},\ \bibinfo {pages} {029}},\ \Eprint
  {http://arxiv.org/abs/1401.8289} {arXiv:1401.8289 [hep-th]} \BibitemShut
  {NoStop}%
%%CITATION = ARXIV:1401.8289;%%
\bibitem [{\citenamefont {Keppel}\ and\ \citenamefont
  {Ajith}(2010)}]{Keppel:2010qu}%
  \BibitemOpen
  \bibfield  {author} {\bibinfo {author} {\bibnamefont {Keppel}, \bibfnamefont
  {Drew}}, \ and\ \bibinfo {author} {\bibfnamefont {P.}~\bibnamefont {Ajith}}}
  (\bibinfo {year} {2010}),\ \bibfield  {title} {\enquote {\bibinfo {title}
  {{Constraining the mass of the graviton using coalescing black-hole
  binaries}},}\ }\href {\doibase 10.1103/PhysRevD.82.122001} {\bibfield
  {journal} {\bibinfo  {journal} {Phys. Rev.}\ }\textbf {\bibinfo {volume}
  {D82}},\ \bibinfo {pages} {122001}},\ \Eprint
  {http://arxiv.org/abs/1004.0284} {arXiv:1004.0284 [gr-qc]} \BibitemShut
  {NoStop}%
%%CITATION = ARXIV:1004.0284;%%
\bibitem [{\citenamefont {Khoury}\ and\ \citenamefont
  {Wyman}(2009)}]{Khoury:2009tk}%
  \BibitemOpen
  \bibfield  {author} {\bibinfo {author} {\bibnamefont {Khoury}, \bibfnamefont
  {Justin}}, \ and\ \bibinfo {author} {\bibfnamefont {Mark}\ \bibnamefont
  {Wyman}}} (\bibinfo {year} {2009}),\ \bibfield  {title} {\enquote {\bibinfo
  {title} {{N-Body Simulations of DGP and Degravitation Theories}},}\ }\href
  {\doibase 10.1103/PhysRevD.80.064023} {\bibfield  {journal} {\bibinfo
  {journal} {Phys. Rev.}\ }\textbf {\bibinfo {volume} {D80}},\ \bibinfo {pages}
  {064023}},\ \Eprint {http://arxiv.org/abs/0903.1292} {arXiv:0903.1292
  [astro-ph.CO]} \BibitemShut {NoStop}%
%%CITATION = ARXIV:0903.1292;%%
\bibitem [{\citenamefont {Kiyota}\ and\ \citenamefont
  {Yamamoto}(2015)}]{Kiyota:2015dla}%
  \BibitemOpen
  \bibfield  {author} {\bibinfo {author} {\bibnamefont {Kiyota}, \bibfnamefont
  {Satoshi}}, \ and\ \bibinfo {author} {\bibfnamefont {Kazuhiro}\ \bibnamefont
  {Yamamoto}}} (\bibinfo {year} {2015}),\ \bibfield  {title} {\enquote
  {\bibinfo {title} {{Constraint on modified dispersion relations for
  gravitational waves from gravitational Cherenkov radiation}},}\ }\href
  {\doibase 10.1103/PhysRevD.92.104036} {\bibfield  {journal} {\bibinfo
  {journal} {Phys. Rev.}\ }\textbf {\bibinfo {volume} {D92}}~(\bibinfo {number}
  {10}),\ \bibinfo {pages} {104036}},\ \Eprint
  {http://arxiv.org/abs/1509.00610} {arXiv:1509.00610 [gr-qc]} \BibitemShut
  {NoStop}%
%%CITATION = ARXIV:1509.00610;%%
\bibitem [{\citenamefont {Kocsis}\ \emph {et~al.}(2008)\citenamefont {Kocsis},
  \citenamefont {Haiman},\ and\ \citenamefont {Menou}}]{Kocsis:2007yu}%
  \BibitemOpen
  \bibfield  {author} {\bibinfo {author} {\bibnamefont {Kocsis}, \bibfnamefont
  {Bence}}, \bibinfo {author} {\bibfnamefont {Zoltan}\ \bibnamefont {Haiman}},
  \ and\ \bibinfo {author} {\bibfnamefont {Kristen}\ \bibnamefont {Menou}}}
  (\bibinfo {year} {2008}),\ \bibfield  {title} {\enquote {\bibinfo {title}
  {{Pre-Merger Localization of Gravitational-Wave Standard Sirens With LISA:
  Triggered Search for an Electromagnetic Counterpart}},}\ }\href {\doibase
  10.1086/590230} {\bibfield  {journal} {\bibinfo  {journal} {Astrophys. J.}\
  }\textbf {\bibinfo {volume} {684}},\ \bibinfo {pages} {870--888}},\ \Eprint
  {http://arxiv.org/abs/0712.1144} {arXiv:0712.1144 [astro-ph]} \BibitemShut
  {NoStop}%
%%CITATION = ARXIV:0712.1144;%%
\bibitem [{\citenamefont {Kostelecky}\ and\ \citenamefont
  {Russell}(2011)}]{Kostelecky:2008ts}%
  \BibitemOpen
  \bibfield  {author} {\bibinfo {author} {\bibnamefont {Kostelecky},
  \bibfnamefont {V~Alan}}, \ and\ \bibinfo {author} {\bibfnamefont {Neil}\
  \bibnamefont {Russell}}} (\bibinfo {year} {2011}),\ \bibfield  {title}
  {\enquote {\bibinfo {title} {{Data Tables for Lorentz and CPT Violation}},}\
  }\href {\doibase 10.1103/RevModPhys.83.11} {\bibfield  {journal} {\bibinfo
  {journal} {Rev. Mod. Phys.}\ }\textbf {\bibinfo {volume} {83}},\ \bibinfo
  {pages} {11--31}},\ \Eprint {http://arxiv.org/abs/0801.0287} {arXiv:0801.0287
  [hep-ph]} \BibitemShut {NoStop}%
%%CITATION = ARXIV:0801.0287;%%
\bibitem [{\citenamefont {Kostelecký}\ and\ \citenamefont
  {Mewes}(2016)}]{Kostelecky:2016kfm}%
  \BibitemOpen
  \bibfield  {author} {\bibinfo {author} {\bibnamefont {Kostelecký},
  \bibfnamefont {V~Alan}}, \ and\ \bibinfo {author} {\bibfnamefont {Matthew}\
  \bibnamefont {Mewes}}} (\bibinfo {year} {2016}),\ \bibfield  {title}
  {\enquote {\bibinfo {title} {{Testing local Lorentz invariance with
  gravitational waves}},}\ }\href {\doibase 10.1016/j.physletb.2016.04.040}
  {\bibfield  {journal} {\bibinfo  {journal} {Phys. Lett.}\ }\textbf {\bibinfo
  {volume} {B757}},\ \bibinfo {pages} {510--514}},\ \Eprint
  {http://arxiv.org/abs/1602.04782} {arXiv:1602.04782 [gr-qc]} \BibitemShut
  {NoStop}%
%%CITATION = ARXIV:1602.04782;%%
\bibitem [{\citenamefont {Koyama}(2005)}]{Koyama:2005tx}%
  \BibitemOpen
  \bibfield  {author} {\bibinfo {author} {\bibnamefont {Koyama}, \bibfnamefont
  {Kazuya}}} (\bibinfo {year} {2005}),\ \bibfield  {title} {\enquote {\bibinfo
  {title} {{Are there ghosts in the self-accelerating brane universe?}}}\
  }\href {\doibase 10.1103/PhysRevD.72.123511} {\bibfield  {journal} {\bibinfo
  {journal} {Phys. Rev.}\ }\textbf {\bibinfo {volume} {D72}},\ \bibinfo {pages}
  {123511}},\ \Eprint {http://arxiv.org/abs/hep-th/0503191}
  {arXiv:hep-th/0503191 [hep-th]} \BibitemShut {NoStop}%
%%CITATION = HEP-TH/0503191;%%
\bibitem [{\citenamefont {Könnig}\ \emph {et~al.}(2016)\citenamefont {Könnig},
  \citenamefont {Nersisyan}, \citenamefont {Akrami}, \citenamefont {Amendola},\
  and\ \citenamefont {Zumalacárregui}}]{Konnig:2016idp}%
  \BibitemOpen
  \bibfield  {author} {\bibinfo {author} {\bibnamefont {Könnig}, \bibfnamefont
  {Frank}}, \bibinfo {author} {\bibfnamefont {Henrik}\ \bibnamefont
  {Nersisyan}}, \bibinfo {author} {\bibfnamefont {Yashar}\ \bibnamefont
  {Akrami}}, \bibinfo {author} {\bibfnamefont {Luca}\ \bibnamefont {Amendola}},
  \ and\ \bibinfo {author} {\bibfnamefont {Miguel}\ \bibnamefont
  {Zumalacárregui}}} (\bibinfo {year} {2016}),\ \bibfield  {title} {\enquote
  {\bibinfo {title} {{A spectre is haunting the cosmos: Quantum stability of
  massive gravity with ghosts}},}\ }\href {\doibase 10.1007/JHEP11(2016)118}
  {\bibfield  {journal} {\bibinfo  {journal} {JHEP}\ }\textbf {\bibinfo
  {volume} {11}},\ \bibinfo {pages} {118}},\ \Eprint
  {http://arxiv.org/abs/1605.08757} {arXiv:1605.08757 [gr-qc]} \BibitemShut
  {NoStop}%
%%CITATION = ARXIV:1605.08757;%%
\bibitem [{\citenamefont {Larson}\ and\ \citenamefont
  {Hiscock}(2000)}]{Larson:1999kg}%
  \BibitemOpen
  \bibfield  {author} {\bibinfo {author} {\bibnamefont {Larson}, \bibfnamefont
  {Shane~L}}, \ and\ \bibinfo {author} {\bibfnamefont {William~A.}\
  \bibnamefont {Hiscock}}} (\bibinfo {year} {2000}),\ \bibfield  {title}
  {\enquote {\bibinfo {title} {{Using binary stars to bound the mass of the
  graviton}},}\ }\href {\doibase 10.1103/PhysRevD.61.104008} {\bibfield
  {journal} {\bibinfo  {journal} {Phys. Rev.}\ }\textbf {\bibinfo {volume}
  {D61}},\ \bibinfo {pages} {104008}},\ \Eprint
  {http://arxiv.org/abs/gr-qc/9912102} {arXiv:gr-qc/9912102 [gr-qc]}
  \BibitemShut {NoStop}%
%%CITATION = GR-QC/9912102;%%
\bibitem [{\citenamefont {Lee}(2013)}]{Lee:2014awa}%
  \BibitemOpen
  \bibfield  {author} {\bibinfo {author} {\bibnamefont {Lee}, \bibfnamefont
  {K~J}}} (\bibinfo {year} {2013}),\ \bibfield  {title} {\enquote {\bibinfo
  {title} {{Pulsar Timing Arrays and Gravity Tests in the Radiative Regime}},}\
  }\href {\doibase 10.1088/0264-9381/30/22/224016} {\bibfield  {journal}
  {\bibinfo  {journal} {Class. Quant. Grav.}\ }\textbf {\bibinfo {volume}
  {30}},\ \bibinfo {pages} {224016}},\ \Eprint {http://arxiv.org/abs/1404.2090}
  {arXiv:1404.2090 [astro-ph.CO]} \BibitemShut {NoStop}%
%%CITATION = ARXIV:1404.2090;%%
\bibitem [{\citenamefont {Lee}\ \emph {et~al.}(2010)\citenamefont {Lee},
  \citenamefont {Jenet}, \citenamefont {Price}, \citenamefont {Wex},\ and\
  \citenamefont {Kramer}}]{Lee:2010cg}%
  \BibitemOpen
  \bibfield  {author} {\bibinfo {author} {\bibnamefont {Lee}, \bibfnamefont
  {Kejia}}, \bibinfo {author} {\bibfnamefont {Fredrick~A.}\ \bibnamefont
  {Jenet}}, \bibinfo {author} {\bibfnamefont {Richard~H.}\ \bibnamefont
  {Price}}, \bibinfo {author} {\bibfnamefont {Norbert}\ \bibnamefont {Wex}}, \
  and\ \bibinfo {author} {\bibfnamefont {Michael}\ \bibnamefont {Kramer}}}
  (\bibinfo {year} {2010}),\ \bibfield  {title} {\enquote {\bibinfo {title}
  {{Detecting massive gravitons using pulsar timing arrays}},}\ }\href
  {\doibase 10.1088/0004-637X/722/2/1589} {\bibfield  {journal} {\bibinfo
  {journal} {Astrophys. J.}\ }\textbf {\bibinfo {volume} {722}},\ \bibinfo
  {pages} {1589--1597}},\ \Eprint {http://arxiv.org/abs/1008.2561}
  {arXiv:1008.2561 [astro-ph.HE]} \BibitemShut {NoStop}%
%%CITATION = ARXIV:1008.2561;%%
\bibitem [{\citenamefont {Lin}(2013)}]{Lin:2013aha}%
  \BibitemOpen
  \bibfield  {author} {\bibinfo {author} {\bibnamefont {Lin}, \bibfnamefont
  {Chunshan}}} (\bibinfo {year} {2013}),\ \bibfield  {title} {\enquote
  {\bibinfo {title} {{SO(3) massive gravity}},}\ }\href {\doibase
  10.1016/j.physletb.2013.10.031} {\bibfield  {journal} {\bibinfo  {journal}
  {Phys. Lett.}\ }\textbf {\bibinfo {volume} {B727}},\ \bibinfo {pages}
  {31--36}},\ \Eprint {http://arxiv.org/abs/1305.2069} {arXiv:1305.2069
  [hep-th]} \BibitemShut {NoStop}%
%%CITATION = ARXIV:1305.2069;%%
\bibitem [{\citenamefont {Lin}\ and\ \citenamefont
  {Ishak}(2016)}]{Lin:2016gve}%
  \BibitemOpen
  \bibfield  {author} {\bibinfo {author} {\bibnamefont {Lin}, \bibfnamefont
  {Weikang}}, \ and\ \bibinfo {author} {\bibfnamefont {Mustapha}\ \bibnamefont
  {Ishak}}} (\bibinfo {year} {2016}),\ \bibfield  {title} {\enquote {\bibinfo
  {title} {{Testing gravity theories using tensor perturbations}},}\ }\href
  {\doibase 10.1103/PhysRevD.94.123011} {\bibfield  {journal} {\bibinfo
  {journal} {Phys. Rev.}\ }\textbf {\bibinfo {volume} {D94}}~(\bibinfo {number}
  {12}),\ \bibinfo {pages} {123011}},\ \Eprint
  {http://arxiv.org/abs/1605.03504} {arXiv:1605.03504 [astro-ph.CO]}
  \BibitemShut {NoStop}%
%%CITATION = ARXIV:1605.03504;%%
\bibitem [{\citenamefont {Lue}\ and\ \citenamefont
  {Starkman}(2003)}]{Lue:2002sw}%
  \BibitemOpen
  \bibfield  {author} {\bibinfo {author} {\bibnamefont {Lue}, \bibfnamefont
  {Arthur}}, \ and\ \bibinfo {author} {\bibfnamefont {Glenn}\ \bibnamefont
  {Starkman}}} (\bibinfo {year} {2003}),\ \bibfield  {title} {\enquote
  {\bibinfo {title} {{Gravitational leakage into extra dimensions: Probing dark
  energy using local gravity}},}\ }\href {\doibase 10.1103/PhysRevD.67.064002}
  {\bibfield  {journal} {\bibinfo  {journal} {Phys. Rev.}\ }\textbf {\bibinfo
  {volume} {D67}},\ \bibinfo {pages} {064002}},\ \Eprint
  {http://arxiv.org/abs/astro-ph/0212083} {arXiv:astro-ph/0212083 [astro-ph]}
  \BibitemShut {NoStop}%
%%CITATION = ASTRO-PH/0212083;%%
\bibitem [{\citenamefont {Luty}\ \emph {et~al.}(2003)\citenamefont {Luty},
  \citenamefont {Porrati},\ and\ \citenamefont {Rattazzi}}]{Luty:2003vm}%
  \BibitemOpen
  \bibfield  {author} {\bibinfo {author} {\bibnamefont {Luty}, \bibfnamefont
  {Markus~A}}, \bibinfo {author} {\bibfnamefont {Massimo}\ \bibnamefont
  {Porrati}}, \ and\ \bibinfo {author} {\bibfnamefont {Riccardo}\ \bibnamefont
  {Rattazzi}}} (\bibinfo {year} {2003}),\ \bibfield  {title} {\enquote
  {\bibinfo {title} {{Strong interactions and stability in the DGP model}},}\
  }\href {\doibase 10.1088/1126-6708/2003/09/029} {\bibfield  {journal}
  {\bibinfo  {journal} {JHEP}\ }\textbf {\bibinfo {volume} {09}},\ \bibinfo
  {pages} {029}},\ \Eprint {http://arxiv.org/abs/hep-th/0303116}
  {arXiv:hep-th/0303116 [hep-th]} \BibitemShut {NoStop}%
%%CITATION = HEP-TH/0303116;%%
\bibitem [{\citenamefont {Maggiore}(2014)}]{Maggiore:2013mea}%
  \BibitemOpen
  \bibfield  {author} {\bibinfo {author} {\bibnamefont {Maggiore},
  \bibfnamefont {Michele}}} (\bibinfo {year} {2014}),\ \bibfield  {title}
  {\enquote {\bibinfo {title} {{Phantom dark energy from nonlocal infrared
  modifications of general relativity}},}\ }\href {\doibase
  10.1103/PhysRevD.89.043008} {\bibfield  {journal} {\bibinfo  {journal} {Phys.
  Rev.}\ }\textbf {\bibinfo {volume} {D89}}~(\bibinfo {number} {4}),\ \bibinfo
  {pages} {043008}},\ \Eprint {http://arxiv.org/abs/1307.3898} {arXiv:1307.3898
  [hep-th]} \BibitemShut {NoStop}%
%%CITATION = ARXIV:1307.3898;%%
\bibitem [{\citenamefont {Malsawmtluangi}\ and\ \citenamefont
  {Suresh}(2016)}]{Malsawmtluangi:2016agy}%
  \BibitemOpen
  \bibfield  {author} {\bibinfo {author} {\bibnamefont {Malsawmtluangi},
  \bibfnamefont {N}}, \ and\ \bibinfo {author} {\bibfnamefont {P.~K.}\
  \bibnamefont {Suresh}}} (\bibinfo {year} {2016}),\ \bibfield  {title}
  {\enquote {\bibinfo {title} {{BB mode angular power spectrum of CMB from
  massive gravity}},}\ }\href@noop {} {\ }\Eprint
  {http://arxiv.org/abs/1603.05836} {arXiv:1603.05836 [gr-qc]} \BibitemShut
  {NoStop}%
%%CITATION = ARXIV:1603.05836;%%
\bibitem [{\citenamefont {Matas}(2016)}]{Matas:2015qxa}%
  \BibitemOpen
  \bibfield  {author} {\bibinfo {author} {\bibnamefont {Matas}, \bibfnamefont
  {Andrew}}} (\bibinfo {year} {2016}),\ \bibfield  {title} {\enquote {\bibinfo
  {title} {{Cutoff for Extensions of Massive Gravity and Bi-Gravity}},}\ }\href
  {\doibase 10.1088/0264-9381/33/7/075004} {\bibfield  {journal} {\bibinfo
  {journal} {Class. Quant. Grav.}\ }\textbf {\bibinfo {volume} {33}}~(\bibinfo
  {number} {7}),\ \bibinfo {pages} {075004}},\ \Eprint
  {http://arxiv.org/abs/1506.00666} {arXiv:1506.00666 [hep-th]} \BibitemShut
  {NoStop}%
%%CITATION = ARXIV:1506.00666;%%
\bibitem [{\citenamefont {Merkowitz}(2010)}]{Merkowitz:2010kka}%
  \BibitemOpen
  \bibfield  {author} {\bibinfo {author} {\bibnamefont {Merkowitz},
  \bibfnamefont {Stephen~M}}} (\bibinfo {year} {2010}),\ \bibfield  {title}
  {\enquote {\bibinfo {title} {{Tests of Gravity Using Lunar Laser Ranging}},}\
  }\href {\doibase 10.12942/lrr-2010-7} {\bibfield  {journal} {\bibinfo
  {journal} {Living Rev. Rel.}\ }\textbf {\bibinfo {volume} {13}},\ \bibinfo
  {pages} {7}}\BibitemShut {NoStop}%
%%CITATION = 00222,13,7;%%
\bibitem [{\citenamefont {Modesto}\ and\ \citenamefont
  {Tsujikawa}(2013)}]{Modesto:2013jea}%
  \BibitemOpen
  \bibfield  {author} {\bibinfo {author} {\bibnamefont {Modesto}, \bibfnamefont
  {Leonardo}}, \ and\ \bibinfo {author} {\bibfnamefont {Shinji}\ \bibnamefont
  {Tsujikawa}}} (\bibinfo {year} {2013}),\ \bibfield  {title} {\enquote
  {\bibinfo {title} {{Non-local massive gravity}},}\ }\href {\doibase
  10.1016/j.physletb.2013.10.037} {\bibfield  {journal} {\bibinfo  {journal}
  {Phys. Lett.}\ }\textbf {\bibinfo {volume} {B727}},\ \bibinfo {pages}
  {48--56}},\ \Eprint {http://arxiv.org/abs/1307.6968} {arXiv:1307.6968
  [hep-th]} \BibitemShut {NoStop}%
%%CITATION = ARXIV:1307.6968;%%
\bibitem [{\citenamefont {Moore}\ and\ \citenamefont
  {Nelson}(2001)}]{Moore:2001bv}%
  \BibitemOpen
  \bibfield  {author} {\bibinfo {author} {\bibnamefont {Moore}, \bibfnamefont
  {Guy~D}}, \ and\ \bibinfo {author} {\bibfnamefont {Ann~E.}\ \bibnamefont
  {Nelson}}} (\bibinfo {year} {2001}),\ \bibfield  {title} {\enquote {\bibinfo
  {title} {{Lower bound on the propagation speed of gravity from gravitational
  Cherenkov radiation}},}\ }\href {\doibase 10.1088/1126-6708/2001/09/023}
  {\bibfield  {journal} {\bibinfo  {journal} {JHEP}\ }\textbf {\bibinfo
  {volume} {09}},\ \bibinfo {pages} {023}},\ \Eprint
  {http://arxiv.org/abs/hep-ph/0106220} {arXiv:hep-ph/0106220 [hep-ph]}
  \BibitemShut {NoStop}%
%%CITATION = HEP-PH/0106220;%%
\bibitem [{\citenamefont {Mukohyama}(2014)}]{Mukohyama:2014rca}%
  \BibitemOpen
  \bibfield  {author} {\bibinfo {author} {\bibnamefont {Mukohyama},
  \bibfnamefont {Shinji}}} (\bibinfo {year} {2014}),\ \bibfield  {title}
  {\enquote {\bibinfo {title} {{A new quasidilaton theory of massive
  gravity}},}\ }\href {\doibase 10.1088/1475-7516/2014/12/011} {\bibfield
  {journal} {\bibinfo  {journal} {JCAP}\ }\textbf {\bibinfo {volume}
  {1412}}~(\bibinfo {number} {12}),\ \bibinfo {pages} {011}},\ \Eprint
  {http://arxiv.org/abs/1410.1996} {arXiv:1410.1996 [hep-th]} \BibitemShut
  {NoStop}%
%%CITATION = ARXIV:1410.1996;%%
\bibitem [{\citenamefont {Nicolis}\ and\ \citenamefont
  {Rattazzi}(2004)}]{Nicolis:2004qq}%
  \BibitemOpen
  \bibfield  {author} {\bibinfo {author} {\bibnamefont {Nicolis}, \bibfnamefont
  {Alberto}}, \ and\ \bibinfo {author} {\bibfnamefont {Riccardo}\ \bibnamefont
  {Rattazzi}}} (\bibinfo {year} {2004}),\ \bibfield  {title} {\enquote
  {\bibinfo {title} {{Classical and quantum consistency of the DGP model}},}\
  }\href {\doibase 10.1088/1126-6708/2004/06/059} {\bibfield  {journal}
  {\bibinfo  {journal} {JHEP}\ }\textbf {\bibinfo {volume} {06}},\ \bibinfo
  {pages} {059}},\ \Eprint {http://arxiv.org/abs/hep-th/0404159}
  {arXiv:hep-th/0404159 [hep-th]} \BibitemShut {NoStop}%
%%CITATION = HEP-TH/0404159;%%
\bibitem [{\citenamefont {Nicolis}\ \emph {et~al.}(2009)\citenamefont
  {Nicolis}, \citenamefont {Rattazzi},\ and\ \citenamefont
  {Trincherini}}]{Nicolis:2008in}%
  \BibitemOpen
  \bibfield  {author} {\bibinfo {author} {\bibnamefont {Nicolis}, \bibfnamefont
  {Alberto}}, \bibinfo {author} {\bibfnamefont {Riccardo}\ \bibnamefont
  {Rattazzi}}, \ and\ \bibinfo {author} {\bibfnamefont {Enrico}\ \bibnamefont
  {Trincherini}}} (\bibinfo {year} {2009}),\ \bibfield  {title} {\enquote
  {\bibinfo {title} {{The Galileon as a local modification of gravity}},}\
  }\href {\doibase 10.1103/PhysRevD.79.064036} {\bibfield  {journal} {\bibinfo
  {journal} {Phys. Rev.}\ }\textbf {\bibinfo {volume} {D79}},\ \bibinfo {pages}
  {064036}},\ \Eprint {http://arxiv.org/abs/0811.2197} {arXiv:0811.2197
  [hep-th]} \BibitemShut {NoStop}%
%%CITATION = ARXIV:0811.2197;%%
\bibitem [{\citenamefont {Niedermann}\ \emph {et~al.}(2015)\citenamefont
  {Niedermann}, \citenamefont {Schneider}, \citenamefont {Hofmann},\ and\
  \citenamefont {Khoury}}]{Niedermann:2014bqa}%
  \BibitemOpen
  \bibfield  {author} {\bibinfo {author} {\bibnamefont {Niedermann},
  \bibfnamefont {Florian}}, \bibinfo {author} {\bibfnamefont {Robert}\
  \bibnamefont {Schneider}}, \bibinfo {author} {\bibfnamefont {Stefan}\
  \bibnamefont {Hofmann}}, \ and\ \bibinfo {author} {\bibfnamefont {Justin}\
  \bibnamefont {Khoury}}} (\bibinfo {year} {2015}),\ \bibfield  {title}
  {\enquote {\bibinfo {title} {{Universe as a cosmic string}},}\ }\href
  {\doibase 10.1103/PhysRevD.91.024002} {\bibfield  {journal} {\bibinfo
  {journal} {Phys. Rev.}\ }\textbf {\bibinfo {volume} {D91}}~(\bibinfo {number}
  {2}),\ \bibinfo {pages} {024002}},\ \Eprint {http://arxiv.org/abs/1410.0700}
  {arXiv:1410.0700 [gr-qc]} \BibitemShut {NoStop}%
%%CITATION = ARXIV:1410.0700;%%
\bibitem [{\citenamefont {Nieuwenhuizen}(2011)}]{Nieuwenhuizen:2011sq}%
  \BibitemOpen
  \bibfield  {author} {\bibinfo {author} {\bibnamefont {Nieuwenhuizen},
  \bibfnamefont {Th~M}}} (\bibinfo {year} {2011}),\ \bibfield  {title}
  {\enquote {\bibinfo {title} {{Exact Schwarzschild-de Sitter black holes in a
  family of massive gravity models}},}\ }\href {\doibase
  10.1103/PhysRevD.84.024038} {\bibfield  {journal} {\bibinfo  {journal} {Phys.
  Rev.}\ }\textbf {\bibinfo {volume} {D84}},\ \bibinfo {pages} {024038}},\
  \Eprint {http://arxiv.org/abs/1103.5912} {arXiv:1103.5912 [gr-qc]}
  \BibitemShut {NoStop}%
%%CITATION = ARXIV:1103.5912;%%
\bibitem [{\citenamefont {Nishizawa}\ and\ \citenamefont
  {Nakamura}(2014)}]{Nishizawa:2014zna}%
  \BibitemOpen
  \bibfield  {author} {\bibinfo {author} {\bibnamefont {Nishizawa},
  \bibfnamefont {Atsushi}}, \ and\ \bibinfo {author} {\bibfnamefont {Takashi}\
  \bibnamefont {Nakamura}}} (\bibinfo {year} {2014}),\ \bibfield  {title}
  {\enquote {\bibinfo {title} {{Measuring Speed of Gravitational Waves by
  Observations of Photons and Neutrinos from Compact Binary Mergers and
  Supernovae}},}\ }\href {\doibase 10.1103/PhysRevD.90.044048} {\bibfield
  {journal} {\bibinfo  {journal} {Phys. Rev.}\ }\textbf {\bibinfo {volume}
  {D90}}~(\bibinfo {number} {4}),\ \bibinfo {pages} {044048}},\ \Eprint
  {http://arxiv.org/abs/1406.5544} {arXiv:1406.5544 [gr-qc]} \BibitemShut
  {NoStop}%
%%CITATION = ARXIV:1406.5544;%%
\bibitem [{\citenamefont {Olive}\ \emph {et~al.}(2014)\citenamefont {Olive}
  \emph {et~al.}}]{Agashe:2014kda}%
  \BibitemOpen
  \bibfield  {author} {\bibinfo {author} {\bibnamefont {Olive}, \bibfnamefont
  {K~A}},  \emph {et~al.} (\bibinfo {collaboration} {Particle Data Group})}
  (\bibinfo {year} {2014}),\ \bibfield  {title} {\enquote {\bibinfo {title}
  {{Review of Particle Physics}},}\ }\href {\doibase
  10.1088/1674-1137/38/9/090001} {\bibfield  {journal} {\bibinfo  {journal}
  {Chin. Phys.}\ }\textbf {\bibinfo {volume} {C38}},\ \bibinfo {pages}
  {090001}}\BibitemShut {NoStop}%
%%CITATION = CHPHD,C38,090001;%%
\bibitem [{\citenamefont {Ondo}\ and\ \citenamefont
  {Tolley}(2013)}]{Ondo:2013wka}%
  \BibitemOpen
  \bibfield  {author} {\bibinfo {author} {\bibnamefont {Ondo}, \bibfnamefont
  {Nicholas~A}}, \ and\ \bibinfo {author} {\bibfnamefont {Andrew~J.}\
  \bibnamefont {Tolley}}} (\bibinfo {year} {2013}),\ \bibfield  {title}
  {\enquote {\bibinfo {title} {{Complete Decoupling Limit of Ghost-free Massive
  Gravity}},}\ }\href {\doibase 10.1007/JHEP11(2013)059} {\bibfield  {journal}
  {\bibinfo  {journal} {JHEP}\ }\textbf {\bibinfo {volume} {11}},\ \bibinfo
  {pages} {059}},\ \Eprint {http://arxiv.org/abs/1307.4769} {arXiv:1307.4769
  [hep-th]} \BibitemShut {NoStop}%
%%CITATION = ARXIV:1307.4769;%%
\bibitem [{\citenamefont {Park}\ and\ \citenamefont
  {Wyman}(2015)}]{Park:2014aga}%
  \BibitemOpen
  \bibfield  {author} {\bibinfo {author} {\bibnamefont {Park}, \bibfnamefont
  {Youngsoo}}, \ and\ \bibinfo {author} {\bibfnamefont {Mark}\ \bibnamefont
  {Wyman}}} (\bibinfo {year} {2015}),\ \bibfield  {title} {\enquote {\bibinfo
  {title} {{Detectability of Weak Lensing Modifications under Galileon
  Theories}},}\ }\href {\doibase 10.1103/PhysRevD.91.064012} {\bibfield
  {journal} {\bibinfo  {journal} {Phys. Rev.}\ }\textbf {\bibinfo {volume}
  {D91}}~(\bibinfo {number} {6}),\ \bibinfo {pages} {064012}},\ \Eprint
  {http://arxiv.org/abs/1408.4773} {arXiv:1408.4773 [astro-ph.CO]} \BibitemShut
  {NoStop}%
%%CITATION = ARXIV:1408.4773;%%
\bibitem [{\citenamefont {Porrati}(2002)}]{Porrati:2001gx}%
  \BibitemOpen
  \bibfield  {author} {\bibinfo {author} {\bibnamefont {Porrati}, \bibfnamefont
  {Massimo}}} (\bibinfo {year} {2002}),\ \bibfield  {title} {\enquote {\bibinfo
  {title} {{Mass and gauge invariance 4. Holography for the Karch-Randall
  model}},}\ }\href {\doibase 10.1103/PhysRevD.65.044015} {\bibfield  {journal}
  {\bibinfo  {journal} {Phys. Rev.}\ }\textbf {\bibinfo {volume} {D65}},\
  \bibinfo {pages} {044015}},\ \Eprint {http://arxiv.org/abs/hep-th/0109017}
  {arXiv:hep-th/0109017 [hep-th]} \BibitemShut {NoStop}%
%%CITATION = HEP-TH/0109017;%%
\bibitem [{\citenamefont {Raveri}\ \emph {et~al.}(2015)\citenamefont {Raveri},
  \citenamefont {Baccigalupi}, \citenamefont {Silvestri},\ and\ \citenamefont
  {Zhou}}]{Raveri:2014eea}%
  \BibitemOpen
  \bibfield  {author} {\bibinfo {author} {\bibnamefont {Raveri}, \bibfnamefont
  {Marco}}, \bibinfo {author} {\bibfnamefont {Carlo}\ \bibnamefont
  {Baccigalupi}}, \bibinfo {author} {\bibfnamefont {Alessandra}\ \bibnamefont
  {Silvestri}}, \ and\ \bibinfo {author} {\bibfnamefont {Shuang-Yong}\
  \bibnamefont {Zhou}}} (\bibinfo {year} {2015}),\ \bibfield  {title} {\enquote
  {\bibinfo {title} {{Measuring the speed of cosmological gravitational
  waves}},}\ }\href {\doibase 10.1103/PhysRevD.91.061501} {\bibfield  {journal}
  {\bibinfo  {journal} {Phys. Rev.}\ }\textbf {\bibinfo {volume}
  {D91}}~(\bibinfo {number} {6}),\ \bibinfo {pages} {061501}},\ \Eprint
  {http://arxiv.org/abs/1405.7974} {arXiv:1405.7974 [astro-ph.CO]} \BibitemShut
  {NoStop}%
%%CITATION = ARXIV:1405.7974;%%
\bibitem [{\citenamefont {de~Rham}(2014)}]{deRham:2014zqa}%
  \BibitemOpen
  \bibfield  {author} {\bibinfo {author} {\bibnamefont {de~Rham}, \bibfnamefont
  {Claudia}}} (\bibinfo {year} {2014}),\ \bibfield  {title} {\enquote {\bibinfo
  {title} {{Massive Gravity}},}\ }\href {\doibase 10.12942/lrr-2014-7}
  {\bibfield  {journal} {\bibinfo  {journal} {Living Rev. Rel.}\ }\textbf
  {\bibinfo {volume} {17}},\ \bibinfo {pages} {7}},\ \Eprint
  {http://arxiv.org/abs/1401.4173} {arXiv:1401.4173 [hep-th]} \BibitemShut
  {NoStop}%
%%CITATION = ARXIV:1401.4173;%%
\bibitem [{\citenamefont {de~Rham}\ \emph
  {et~al.}(2008{\natexlab{a}})\citenamefont {de~Rham}, \citenamefont {Dvali},
  \citenamefont {Hofmann}, \citenamefont {Khoury}, \citenamefont {Pujolas},
  \citenamefont {Redi},\ and\ \citenamefont {Tolley}}]{deRham:2008zz}%
  \BibitemOpen
  \bibfield  {author} {\bibinfo {author} {\bibnamefont {de~Rham}, \bibfnamefont
  {Claudia}}, \bibinfo {author} {\bibfnamefont {Gia}\ \bibnamefont {Dvali}},
  \bibinfo {author} {\bibfnamefont {Stefan}\ \bibnamefont {Hofmann}}, \bibinfo
  {author} {\bibfnamefont {Justin}\ \bibnamefont {Khoury}}, \bibinfo {author}
  {\bibfnamefont {Oriol}\ \bibnamefont {Pujolas}}, \bibinfo {author}
  {\bibfnamefont {Michele}\ \bibnamefont {Redi}}, \ and\ \bibinfo {author}
  {\bibfnamefont {Andrew~J.}\ \bibnamefont {Tolley}}} (\bibinfo {year}
  {2008}{\natexlab{a}}),\ \bibfield  {title} {\enquote {\bibinfo {title}
  {{Cascading gravity: Extending the Dvali-Gabadadze-Porrati model to higher
  dimension}},}\ }\href {\doibase 10.1103/PhysRevLett.100.251603} {\bibfield
  {journal} {\bibinfo  {journal} {Phys. Rev. Lett.}\ }\textbf {\bibinfo
  {volume} {100}},\ \bibinfo {pages} {251603}},\ \Eprint
  {http://arxiv.org/abs/0711.2072} {arXiv:0711.2072 [hep-th]} \BibitemShut
  {NoStop}%
%%CITATION = ARXIV:0711.2072;%%
\bibitem [{\citenamefont {de~Rham}\ and\ \citenamefont
  {Gabadadze}(2010)}]{deRham:2010ik}%
  \BibitemOpen
  \bibfield  {author} {\bibinfo {author} {\bibnamefont {de~Rham}, \bibfnamefont
  {Claudia}}, \ and\ \bibinfo {author} {\bibfnamefont {Gregory}\ \bibnamefont
  {Gabadadze}}} (\bibinfo {year} {2010}),\ \bibfield  {title} {\enquote
  {\bibinfo {title} {{Generalization of the Fierz-Pauli Action}},}\ }\href
  {\doibase 10.1103/PhysRevD.82.044020} {\bibfield  {journal} {\bibinfo
  {journal} {Phys.Rev.}\ }\textbf {\bibinfo {volume} {D82}},\ \bibinfo {pages}
  {044020}},\ \Eprint {http://arxiv.org/abs/1007.0443} {arXiv:1007.0443
  [hep-th]} \BibitemShut {NoStop}%
%%CITATION = ARXIV:1007.0443;%%
\bibitem [{\citenamefont {de~Rham}\ \emph
  {et~al.}(2013{\natexlab{a}})\citenamefont {de~Rham}, \citenamefont
  {Gabadadze}, \citenamefont {Heisenberg},\ and\ \citenamefont
  {Pirtskhalava}}]{deRham:2012ew}%
  \BibitemOpen
  \bibfield  {author} {\bibinfo {author} {\bibnamefont {de~Rham}, \bibfnamefont
  {Claudia}}, \bibinfo {author} {\bibfnamefont {Gregory}\ \bibnamefont
  {Gabadadze}}, \bibinfo {author} {\bibfnamefont {Lavinia}\ \bibnamefont
  {Heisenberg}}, \ and\ \bibinfo {author} {\bibfnamefont {David}\ \bibnamefont
  {Pirtskhalava}}} (\bibinfo {year} {2013}{\natexlab{a}}),\ \bibfield  {title}
  {\enquote {\bibinfo {title} {{Nonrenormalization and naturalness in a class
  of scalar-tensor theories}},}\ }\href {\doibase 10.1103/PhysRevD.87.085017}
  {\bibfield  {journal} {\bibinfo  {journal} {Phys. Rev.}\ }\textbf {\bibinfo
  {volume} {D87}}~(\bibinfo {number} {8}),\ \bibinfo {pages} {085017}},\
  \Eprint {http://arxiv.org/abs/1212.4128} {arXiv:1212.4128} \BibitemShut
  {NoStop}%
%%CITATION = ARXIV:1212.4128;%%
\bibitem [{\citenamefont {de~Rham}\ \emph
  {et~al.}(2011{\natexlab{a}})\citenamefont {de~Rham}, \citenamefont
  {Gabadadze}, \citenamefont {Pirtskhalava}, \citenamefont {Tolley},\ and\
  \citenamefont {Yavin}}]{deRham:2011ca}%
  \BibitemOpen
  \bibfield  {author} {\bibinfo {author} {\bibnamefont {de~Rham}, \bibfnamefont
  {Claudia}}, \bibinfo {author} {\bibfnamefont {Gregory}\ \bibnamefont
  {Gabadadze}}, \bibinfo {author} {\bibfnamefont {David}\ \bibnamefont
  {Pirtskhalava}}, \bibinfo {author} {\bibfnamefont {Andrew~J.}\ \bibnamefont
  {Tolley}}, \ and\ \bibinfo {author} {\bibfnamefont {Itay}\ \bibnamefont
  {Yavin}}} (\bibinfo {year} {2011}{\natexlab{a}}),\ \bibfield  {title}
  {\enquote {\bibinfo {title} {{Nonlinear Dynamics of 3D Massive Gravity}},}\
  }\href {\doibase 10.1007/JHEP06(2011)028} {\bibfield  {journal} {\bibinfo
  {journal} {JHEP}\ }\textbf {\bibinfo {volume} {06}},\ \bibinfo {pages}
  {028}},\ \Eprint {http://arxiv.org/abs/1103.1351} {arXiv:1103.1351 [hep-th]}
  \BibitemShut {NoStop}%
%%CITATION = ARXIV:1103.1351;%%
\bibitem [{\citenamefont {de~Rham}\ \emph
  {et~al.}(2011{\natexlab{b}})\citenamefont {de~Rham}, \citenamefont
  {Gabadadze},\ and\ \citenamefont {Tolley}}]{deRham:2010kj}%
  \BibitemOpen
  \bibfield  {author} {\bibinfo {author} {\bibnamefont {de~Rham}, \bibfnamefont
  {Claudia}}, \bibinfo {author} {\bibfnamefont {Gregory}\ \bibnamefont
  {Gabadadze}}, \ and\ \bibinfo {author} {\bibfnamefont {Andrew~J.}\
  \bibnamefont {Tolley}}} (\bibinfo {year} {2011}{\natexlab{b}}),\ \bibfield
  {title} {\enquote {\bibinfo {title} {{Resummation of Massive Gravity}},}\
  }\href {\doibase 10.1103/PhysRevLett.106.231101} {\bibfield  {journal}
  {\bibinfo  {journal} {Phys.Rev.Lett.}\ }\textbf {\bibinfo {volume} {106}},\
  \bibinfo {pages} {231101}},\ \Eprint {http://arxiv.org/abs/1011.1232}
  {arXiv:1011.1232 [hep-th]} \BibitemShut {NoStop}%
%%CITATION = ARXIV:1011.1232;%%
\bibitem [{\citenamefont {de~Rham}\ \emph
  {et~al.}(2013{\natexlab{b}})\citenamefont {de~Rham}, \citenamefont
  {Heisenberg},\ and\ \citenamefont {Ribeiro}}]{deRham:2013qqa}%
  \BibitemOpen
  \bibfield  {author} {\bibinfo {author} {\bibnamefont {de~Rham}, \bibfnamefont
  {Claudia}}, \bibinfo {author} {\bibfnamefont {Lavinia}\ \bibnamefont
  {Heisenberg}}, \ and\ \bibinfo {author} {\bibfnamefont {Raquel~H.}\
  \bibnamefont {Ribeiro}}} (\bibinfo {year} {2013}{\natexlab{b}}),\ \bibfield
  {title} {\enquote {\bibinfo {title} {{Quantum Corrections in Massive
  Gravity}},}\ }\href {\doibase 10.1103/PhysRevD.88.084058} {\bibfield
  {journal} {\bibinfo  {journal} {Phys. Rev.}\ }\textbf {\bibinfo {volume}
  {D88}},\ \bibinfo {pages} {084058}},\ \Eprint
  {http://arxiv.org/abs/1307.7169} {arXiv:1307.7169 [hep-th]} \BibitemShut
  {NoStop}%
%%CITATION = ARXIV:1307.7169;%%
\bibitem [{\citenamefont {de~Rham}\ \emph
  {et~al.}(2014{\natexlab{a}})\citenamefont {de~Rham}, \citenamefont
  {Heisenberg},\ and\ \citenamefont {Ribeiro}}]{deRham:2014fha}%
  \BibitemOpen
  \bibfield  {author} {\bibinfo {author} {\bibnamefont {de~Rham}, \bibfnamefont
  {Claudia}}, \bibinfo {author} {\bibfnamefont {Lavinia}\ \bibnamefont
  {Heisenberg}}, \ and\ \bibinfo {author} {\bibfnamefont {Raquel~H.}\
  \bibnamefont {Ribeiro}}} (\bibinfo {year} {2014}{\natexlab{a}}),\ \bibfield
  {title} {\enquote {\bibinfo {title} {{Ghosts and matter couplings in massive
  gravity, bigravity and multigravity}},}\ }\href {\doibase
  10.1103/PhysRevD.90.124042} {\bibfield  {journal} {\bibinfo  {journal} {Phys.
  Rev.}\ }\textbf {\bibinfo {volume} {D90}},\ \bibinfo {pages} {124042}},\
  \Eprint {http://arxiv.org/abs/1409.3834} {arXiv:1409.3834 [hep-th]}
  \BibitemShut {NoStop}%
%%CITATION = ARXIV:1409.3834;%%
\bibitem [{\citenamefont {de~Rham}\ \emph
  {et~al.}(2015{\natexlab{a}})\citenamefont {de~Rham}, \citenamefont
  {Heisenberg},\ and\ \citenamefont {Ribeiro}}]{deRham:2014naa}%
  \BibitemOpen
  \bibfield  {author} {\bibinfo {author} {\bibnamefont {de~Rham}, \bibfnamefont
  {Claudia}}, \bibinfo {author} {\bibfnamefont {Lavinia}\ \bibnamefont
  {Heisenberg}}, \ and\ \bibinfo {author} {\bibfnamefont {Raquel~H.}\
  \bibnamefont {Ribeiro}}} (\bibinfo {year} {2015}{\natexlab{a}}),\ \bibfield
  {title} {\enquote {\bibinfo {title} {{On couplings to matter in massive
  (bi-)gravity}},}\ }\href {\doibase 10.1088/0264-9381/32/3/035022} {\bibfield
  {journal} {\bibinfo  {journal} {Class. Quant. Grav.}\ }\textbf {\bibinfo
  {volume} {32}},\ \bibinfo {pages} {035022}},\ \Eprint
  {http://arxiv.org/abs/1408.1678} {arXiv:1408.1678 [hep-th]} \BibitemShut
  {NoStop}%
%%CITATION = ARXIV:1408.1678;%%
\bibitem [{\citenamefont {de~Rham}\ \emph
  {et~al.}(2008{\natexlab{b}})\citenamefont {de~Rham}, \citenamefont {Hofmann},
  \citenamefont {Khoury},\ and\ \citenamefont {Tolley}}]{deRham:2007rw}%
  \BibitemOpen
  \bibfield  {author} {\bibinfo {author} {\bibnamefont {de~Rham}, \bibfnamefont
  {Claudia}}, \bibinfo {author} {\bibfnamefont {Stefan}\ \bibnamefont
  {Hofmann}}, \bibinfo {author} {\bibfnamefont {Justin}\ \bibnamefont
  {Khoury}}, \ and\ \bibinfo {author} {\bibfnamefont {Andrew~J.}\ \bibnamefont
  {Tolley}}} (\bibinfo {year} {2008}{\natexlab{b}}),\ \bibfield  {title}
  {\enquote {\bibinfo {title} {{Cascading Gravity and Degravitation}},}\ }\href
  {\doibase 10.1088/1475-7516/2008/02/011} {\bibfield  {journal} {\bibinfo
  {journal} {JCAP}\ }\textbf {\bibinfo {volume} {0802}},\ \bibinfo {pages}
  {011}},\ \Eprint {http://arxiv.org/abs/0712.2821} {arXiv:0712.2821 [hep-th]}
  \BibitemShut {NoStop}%
%%CITATION = ARXIV:0712.2821;%%
\bibitem [{\citenamefont {de~Rham}\ \emph {et~al.}(2009)\citenamefont
  {de~Rham}, \citenamefont {Khoury},\ and\ \citenamefont
  {Tolley}}]{deRham:2009wb}%
  \BibitemOpen
  \bibfield  {author} {\bibinfo {author} {\bibnamefont {de~Rham}, \bibfnamefont
  {Claudia}}, \bibinfo {author} {\bibfnamefont {Justin}\ \bibnamefont
  {Khoury}}, \ and\ \bibinfo {author} {\bibfnamefont {Andrew~J.}\ \bibnamefont
  {Tolley}}} (\bibinfo {year} {2009}),\ \bibfield  {title} {\enquote {\bibinfo
  {title} {{Flat 3-Brane with Tension in Cascading Gravity}},}\ }\href
  {\doibase 10.1103/PhysRevLett.103.161601} {\bibfield  {journal} {\bibinfo
  {journal} {Phys. Rev. Lett.}\ }\textbf {\bibinfo {volume} {103}},\ \bibinfo
  {pages} {161601}},\ \Eprint {http://arxiv.org/abs/0907.0473} {arXiv:0907.0473
  [hep-th]} \BibitemShut {NoStop}%
%%CITATION = ARXIV:0907.0473;%%
\bibitem [{\citenamefont {de~Rham}\ \emph
  {et~al.}(2013{\natexlab{c}})\citenamefont {de~Rham}, \citenamefont {Matas},\
  and\ \citenamefont {Tolley}}]{deRham:2012fg}%
  \BibitemOpen
  \bibfield  {author} {\bibinfo {author} {\bibnamefont {de~Rham}, \bibfnamefont
  {Claudia}}, \bibinfo {author} {\bibfnamefont {Andrew}\ \bibnamefont {Matas}},
  \ and\ \bibinfo {author} {\bibfnamefont {Andrew~J.}\ \bibnamefont {Tolley}}}
  (\bibinfo {year} {2013}{\natexlab{c}}),\ \bibfield  {title} {\enquote
  {\bibinfo {title} {{Galileon Radiation from Binary Systems}},}\ }\href
  {\doibase 10.1103/PhysRevD.87.064024} {\bibfield  {journal} {\bibinfo
  {journal} {Phys. Rev.}\ }\textbf {\bibinfo {volume} {D87}}~(\bibinfo {number}
  {6}),\ \bibinfo {pages} {064024}},\ \Eprint {http://arxiv.org/abs/1212.5212}
  {arXiv:1212.5212 [hep-th]} \BibitemShut {NoStop}%
%%CITATION = ARXIV:1212.5212;%%
\bibitem [{\citenamefont {de~Rham}\ \emph
  {et~al.}(2014{\natexlab{b}})\citenamefont {de~Rham}, \citenamefont {Matas},\
  and\ \citenamefont {Tolley}}]{deRham:2013tfa}%
  \BibitemOpen
  \bibfield  {author} {\bibinfo {author} {\bibnamefont {de~Rham}, \bibfnamefont
  {Claudia}}, \bibinfo {author} {\bibfnamefont {Andrew}\ \bibnamefont {Matas}},
  \ and\ \bibinfo {author} {\bibfnamefont {Andrew~J.}\ \bibnamefont {Tolley}}}
  (\bibinfo {year} {2014}{\natexlab{b}}),\ \bibfield  {title} {\enquote
  {\bibinfo {title} {{New Kinetic Interactions for Massive Gravity?}}}\ }\href
  {\doibase 10.1088/0264-9381/31/16/165004} {\bibfield  {journal} {\bibinfo
  {journal} {Class. Quant. Grav.}\ }\textbf {\bibinfo {volume} {31}},\ \bibinfo
  {pages} {165004}},\ \Eprint {http://arxiv.org/abs/1311.6485} {arXiv:1311.6485
  [hep-th]} \BibitemShut {NoStop}%
%%CITATION = ARXIV:1311.6485;%%
\bibitem [{\citenamefont {de~Rham}\ \emph
  {et~al.}(2015{\natexlab{b}})\citenamefont {de~Rham}, \citenamefont {Matas},\
  and\ \citenamefont {Tolley}}]{deRham:2015rxa}%
  \BibitemOpen
  \bibfield  {author} {\bibinfo {author} {\bibnamefont {de~Rham}, \bibfnamefont
  {Claudia}}, \bibinfo {author} {\bibfnamefont {Andrew}\ \bibnamefont {Matas}},
  \ and\ \bibinfo {author} {\bibfnamefont {Andrew~J.}\ \bibnamefont {Tolley}}}
  (\bibinfo {year} {2015}{\natexlab{b}}),\ \bibfield  {title} {\enquote
  {\bibinfo {title} {{New Kinetic Terms for Massive Gravity and Multi-gravity:
  A No-Go in Vielbein Form}},}\ }\href {\doibase
  10.1088/0264-9381/32/21/215027} {\bibfield  {journal} {\bibinfo  {journal}
  {Class. Quant. Grav.}\ }\textbf {\bibinfo {volume} {32}}~(\bibinfo {number}
  {21}),\ \bibinfo {pages} {215027}},\ \Eprint
  {http://arxiv.org/abs/1505.00831} {arXiv:1505.00831 [hep-th]} \BibitemShut
  {NoStop}%
%%CITATION = ARXIV:1505.00831;%%
\bibitem [{\citenamefont {de~Rham}\ and\ \citenamefont
  {Tolley}(2015)}]{deRham:2015cha}%
  \BibitemOpen
  \bibfield  {author} {\bibinfo {author} {\bibnamefont {de~Rham}, \bibfnamefont
  {Claudia}}, \ and\ \bibinfo {author} {\bibfnamefont {Andrew~J.}\ \bibnamefont
  {Tolley}}} (\bibinfo {year} {2015}),\ \bibfield  {title} {\enquote {\bibinfo
  {title} {{Vielbein to the rescue? Breaking the symmetric vielbein condition
  in massive gravity and multigravity}},}\ }\href {\doibase
  10.1103/PhysRevD.92.024024} {\bibfield  {journal} {\bibinfo  {journal} {Phys.
  Rev.}\ }\textbf {\bibinfo {volume} {D92}}~(\bibinfo {number} {2}),\ \bibinfo
  {pages} {024024}},\ \Eprint {http://arxiv.org/abs/1505.01450}
  {arXiv:1505.01450 [hep-th]} \BibitemShut {NoStop}%
%%CITATION = ARXIV:1505.01450;%%
\bibitem [{\citenamefont {de~Rham}\ \emph
  {et~al.}(2013{\natexlab{d}})\citenamefont {de~Rham}, \citenamefont {Tolley},\
  and\ \citenamefont {Wesley}}]{deRham:2012fw}%
  \BibitemOpen
  \bibfield  {author} {\bibinfo {author} {\bibnamefont {de~Rham}, \bibfnamefont
  {Claudia}}, \bibinfo {author} {\bibfnamefont {Andrew~J.}\ \bibnamefont
  {Tolley}}, \ and\ \bibinfo {author} {\bibfnamefont {Daniel~H.}\ \bibnamefont
  {Wesley}}} (\bibinfo {year} {2013}{\natexlab{d}}),\ \bibfield  {title}
  {\enquote {\bibinfo {title} {{Vainshtein Mechanism in Binary Pulsars}},}\
  }\href {\doibase 10.1103/PhysRevD.87.044025} {\bibfield  {journal} {\bibinfo
  {journal} {Phys. Rev.}\ }\textbf {\bibinfo {volume} {D87}}~(\bibinfo {number}
  {4}),\ \bibinfo {pages} {044025}},\ \Eprint {http://arxiv.org/abs/1208.0580}
  {arXiv:1208.0580 [gr-qc]} \BibitemShut {NoStop}%
%%CITATION = ARXIV:1208.0580;%%
\bibitem [{\citenamefont {de~Rham}\ \emph
  {et~al.}(2016{\natexlab{a}})\citenamefont {de~Rham}, \citenamefont {Tolley},\
  and\ \citenamefont {Zhou}}]{deRham:2015ijs}%
  \BibitemOpen
  \bibfield  {author} {\bibinfo {author} {\bibnamefont {de~Rham}, \bibfnamefont
  {Claudia}}, \bibinfo {author} {\bibfnamefont {Andrew~J.}\ \bibnamefont
  {Tolley}}, \ and\ \bibinfo {author} {\bibfnamefont {Shuang-Yong}\
  \bibnamefont {Zhou}}} (\bibinfo {year} {2016}{\natexlab{a}}),\ \bibfield
  {title} {\enquote {\bibinfo {title} {{Non-compact nonlinear sigma models}},}\
  }\href {\doibase 10.1016/j.physletb.2016.07.035} {\bibfield  {journal}
  {\bibinfo  {journal} {Phys. Lett.}\ }\textbf {\bibinfo {volume} {B760}},\
  \bibinfo {pages} {579--583}},\ \Eprint {http://arxiv.org/abs/1512.06838}
  {arXiv:1512.06838 [hep-th]} \BibitemShut {NoStop}%
%%CITATION = ARXIV:1512.06838;%%
\bibitem [{\citenamefont {de~Rham}\ \emph
  {et~al.}(2016{\natexlab{b}})\citenamefont {de~Rham}, \citenamefont {Tolley},\
  and\ \citenamefont {Zhou}}]{deRham:2016plk}%
  \BibitemOpen
  \bibfield  {author} {\bibinfo {author} {\bibnamefont {de~Rham}, \bibfnamefont
  {Claudia}}, \bibinfo {author} {\bibfnamefont {Andrew~J.}\ \bibnamefont
  {Tolley}}, \ and\ \bibinfo {author} {\bibfnamefont {Shuang-Yong}\
  \bibnamefont {Zhou}}} (\bibinfo {year} {2016}{\natexlab{b}}),\ \bibfield
  {title} {\enquote {\bibinfo {title} {{The $\Lambda_{2}$ limit of massive
  gravity}},}\ }\href {\doibase 10.1007/JHEP04(2016)188} {\bibfield  {journal}
  {\bibinfo  {journal} {JHEP}\ }\textbf {\bibinfo {volume} {04}},\ \bibinfo
  {pages} {188}},\ \Eprint {http://arxiv.org/abs/1602.03721} {arXiv:1602.03721
  [hep-th]} \BibitemShut {NoStop}%
%%CITATION = ARXIV:1602.03721;%%
\bibitem [{\citenamefont {Rubakov}\ and\ \citenamefont
  {Tinyakov}(2008)}]{Rubakov:2008nh}%
  \BibitemOpen
  \bibfield  {author} {\bibinfo {author} {\bibnamefont {Rubakov}, \bibfnamefont
  {V~A}}, \ and\ \bibinfo {author} {\bibfnamefont {P.~G.}\ \bibnamefont
  {Tinyakov}}} (\bibinfo {year} {2008}),\ \bibfield  {title} {\enquote
  {\bibinfo {title} {{Infrared-modified gravities and massive gravitons}},}\
  }\href {\doibase 10.1070/PU2008v051n08ABEH006600} {\bibfield  {journal}
  {\bibinfo  {journal} {Phys. Usp.}\ }\textbf {\bibinfo {volume} {51}},\
  \bibinfo {pages} {759--792}},\ \Eprint {http://arxiv.org/abs/0802.4379}
  {arXiv:0802.4379 [hep-th]} \BibitemShut {NoStop}%
%%CITATION = ARXIV:0802.4379;%%
\bibitem [{\citenamefont {Rubakov}(2004)}]{Rubakov:2004eb}%
  \BibitemOpen
  \bibfield  {author} {\bibinfo {author} {\bibnamefont {Rubakov}, \bibfnamefont
  {VA}}} (\bibinfo {year} {2004}),\ \bibfield  {title} {\enquote {\bibinfo
  {title} {{Lorentz-violating graviton masses: Getting around ghosts, low
  strong coupling scale and VDVZ discontinuity}},}\ }\href@noop {} {\ }\Eprint
  {http://arxiv.org/abs/hep-th/0407104} {arXiv:hep-th/0407104 [hep-th]}
  \BibitemShut {NoStop}%
%%CITATION = HEP-TH/0407104;%%
\bibitem [{\citenamefont {Scharre}\ and\ \citenamefont
  {Will}(2002)}]{Scharre:2001hn}%
  \BibitemOpen
  \bibfield  {author} {\bibinfo {author} {\bibnamefont {Scharre}, \bibfnamefont
  {Paul~D}}, \ and\ \bibinfo {author} {\bibfnamefont {Clifford~M.}\
  \bibnamefont {Will}}} (\bibinfo {year} {2002}),\ \bibfield  {title} {\enquote
  {\bibinfo {title} {{Testing scalar tensor gravity using space gravitational
  wave interferometers}},}\ }\href {\doibase 10.1103/PhysRevD.65.042002}
  {\bibfield  {journal} {\bibinfo  {journal} {Phys. Rev.}\ }\textbf {\bibinfo
  {volume} {D65}},\ \bibinfo {pages} {042002}},\ \Eprint
  {http://arxiv.org/abs/gr-qc/0109044} {arXiv:gr-qc/0109044 [gr-qc]}
  \BibitemShut {NoStop}%
%%CITATION = GR-QC/0109044;%%
\bibitem [{\citenamefont {Schwinger}(1962)}]{PhysRev.125.397}%
  \BibitemOpen
  \bibfield  {author} {\bibinfo {author} {\bibnamefont {Schwinger},
  \bibfnamefont {Julian}}} (\bibinfo {year} {1962}),\ \bibfield  {title}
  {\enquote {\bibinfo {title} {Gauge invariance and mass},}\ }\href {\doibase
  10.1103/PhysRev.125.397} {\bibfield  {journal} {\bibinfo  {journal} {Phys.
  Rev.}\ }\textbf {\bibinfo {volume} {125}},\ \bibinfo {pages}
  {397--398}}\BibitemShut {NoStop}%
\bibitem [{\citenamefont {Stavridis}\ and\ \citenamefont
  {Will}(2009)}]{Stavridis:2009mb}%
  \BibitemOpen
  \bibfield  {author} {\bibinfo {author} {\bibnamefont {Stavridis},
  \bibfnamefont {Adamantios}}, \ and\ \bibinfo {author} {\bibfnamefont
  {Clifford~M.}\ \bibnamefont {Will}}} (\bibinfo {year} {2009}),\ \bibfield
  {title} {\enquote {\bibinfo {title} {{Bounding the mass of the graviton with
  gravitational waves: Effect of spin precessions in massive black hole
  binaries}},}\ }\href {\doibase 10.1103/PhysRevD.80.044002} {\bibfield
  {journal} {\bibinfo  {journal} {Phys. Rev.}\ }\textbf {\bibinfo {volume}
  {D80}},\ \bibinfo {pages} {044002}},\ \Eprint
  {http://arxiv.org/abs/0906.3602} {arXiv:0906.3602 [gr-qc]} \BibitemShut
  {NoStop}%
%%CITATION = ARXIV:0906.3602;%%
\bibitem [{\citenamefont {Talmadge}\ \emph {et~al.}(1988)\citenamefont
  {Talmadge}, \citenamefont {Berthias}, \citenamefont {Hellings},\ and\
  \citenamefont {Standish}}]{Talmadge:1988qz}%
  \BibitemOpen
  \bibfield  {author} {\bibinfo {author} {\bibnamefont {Talmadge},
  \bibfnamefont {C}}, \bibinfo {author} {\bibfnamefont {J.~P.}\ \bibnamefont
  {Berthias}}, \bibinfo {author} {\bibfnamefont {R.~W.}\ \bibnamefont
  {Hellings}}, \ and\ \bibinfo {author} {\bibfnamefont {E.~M.}\ \bibnamefont
  {Standish}}} (\bibinfo {year} {1988}),\ \bibfield  {title} {\enquote
  {\bibinfo {title} {{Model Independent Constraints on Possible Modifications
  of Newtonian Gravity}},}\ }\href {\doibase 10.1103/PhysRevLett.61.1159}
  {\bibfield  {journal} {\bibinfo  {journal} {Phys. Rev. Lett.}\ }\textbf
  {\bibinfo {volume} {61}},\ \bibinfo {pages} {1159--1162}}\BibitemShut
  {NoStop}%
%%CITATION = PRLTA,61,1159;%%
\bibitem [{\citenamefont {Taylor}\ and\ \citenamefont
  {Weisberg}(1989)}]{Taylor:1989sw}%
  \BibitemOpen
  \bibfield  {author} {\bibinfo {author} {\bibnamefont {Taylor}, \bibfnamefont
  {Joseph~H}}, \ and\ \bibinfo {author} {\bibfnamefont {J.~M.}\ \bibnamefont
  {Weisberg}}} (\bibinfo {year} {1989}),\ \bibfield  {title} {\enquote
  {\bibinfo {title} {{Further experimental tests of relativistic gravity using
  the binary pulsar PSR 1913+16}},}\ }\href {\doibase 10.1086/167917}
  {\bibfield  {journal} {\bibinfo  {journal} {Astrophys. J.}\ }\textbf
  {\bibinfo {volume} {345}},\ \bibinfo {pages} {434--450}}\BibitemShut
  {NoStop}%
%%CITATION = ASJOA,345,434;%%
\bibitem [{\citenamefont {Tolley}\ \emph {et~al.}(2015)\citenamefont {Tolley},
  \citenamefont {Wu},\ and\ \citenamefont {Zhou}}]{Tolley:2015ywa}%
  \BibitemOpen
  \bibfield  {author} {\bibinfo {author} {\bibnamefont {Tolley}, \bibfnamefont
  {Andrew~J}}, \bibinfo {author} {\bibfnamefont {De-Jun}\ \bibnamefont {Wu}}, \
  and\ \bibinfo {author} {\bibfnamefont {Shuang-Yong}\ \bibnamefont {Zhou}}}
  (\bibinfo {year} {2015}),\ \bibfield  {title} {\enquote {\bibinfo {title}
  {{Hairy black holes in scalar extended massive gravity}},}\ }\href {\doibase
  10.1103/PhysRevD.92.124063} {\bibfield  {journal} {\bibinfo  {journal} {Phys.
  Rev.}\ }\textbf {\bibinfo {volume} {D92}}~(\bibinfo {number} {12}),\ \bibinfo
  {pages} {124063}},\ \Eprint {http://arxiv.org/abs/1510.05208}
  {arXiv:1510.05208 [hep-th]} \BibitemShut {NoStop}%
%%CITATION = ARXIV:1510.05208;%%
\bibitem [{\citenamefont {Trodden}\ and\ \citenamefont
  {Hinterbichler}(2011)}]{Trodden:2011xh}%
  \BibitemOpen
  \bibfield  {author} {\bibinfo {author} {\bibnamefont {Trodden}, \bibfnamefont
  {Mark}}, \ and\ \bibinfo {author} {\bibfnamefont {Kurt}\ \bibnamefont
  {Hinterbichler}}} (\bibinfo {year} {2011}),\ \bibfield  {title} {\enquote
  {\bibinfo {title} {{Generalizing Galileons}},}\ }\href {\doibase
  10.1088/0264-9381/28/20/204003} {\bibfield  {journal} {\bibinfo  {journal}
  {Class. Quant. Grav.}\ }\textbf {\bibinfo {volume} {28}},\ \bibinfo {pages}
  {204003}},\ \Eprint {http://arxiv.org/abs/1104.2088} {arXiv:1104.2088
  [hep-th]} \BibitemShut {NoStop}%
%%CITATION = ARXIV:1104.2088;%%
\bibitem [{\citenamefont {Vainshtein}(1972)}]{Vainshtein:1972sx}%
  \BibitemOpen
  \bibfield  {author} {\bibinfo {author} {\bibnamefont {Vainshtein},
  \bibfnamefont {A~I}}} (\bibinfo {year} {1972}),\ \bibfield  {title} {\enquote
  {\bibinfo {title} {{To the problem of nonvanishing gravitation mass}},}\
  }\href {\doibase 10.1016/0370-2693(72)90147-5} {\bibfield  {journal}
  {\bibinfo  {journal} {Phys. Lett.}\ }\textbf {\bibinfo {volume} {B39}},\
  \bibinfo {pages} {393--394}}\BibitemShut {NoStop}%
%%CITATION = PHLTA,B39,393;%%
\bibitem [{\citenamefont {Van~Waerbeke}\ \emph {et~al.}(2001)\citenamefont
  {Van~Waerbeke} \emph {et~al.}}]{VanWaerbeke:2001nh}%
  \BibitemOpen
  \bibfield  {author} {\bibinfo {author} {\bibnamefont {Van~Waerbeke},
  \bibfnamefont {Ludovic}},  \emph {et~al.}} (\bibinfo {year} {2001}),\
  \bibfield  {title} {\enquote {\bibinfo {title} {{Cosmic shear statistics and
  cosmology}},}\ }\href {\doibase 10.1051/0004-6361:20010766} {\bibfield
  {journal} {\bibinfo  {journal} {Astron. Astrophys.}\ }\textbf {\bibinfo
  {volume} {374}},\ \bibinfo {pages} {757--769}},\ \Eprint
  {http://arxiv.org/abs/astro-ph/0101511} {arXiv:astro-ph/0101511 [astro-ph]}
  \BibitemShut {NoStop}%
%%CITATION = ASTRO-PH/0101511;%%
\bibitem [{\citenamefont {Visser}(1998)}]{Visser:1997hd}%
  \BibitemOpen
  \bibfield  {author} {\bibinfo {author} {\bibnamefont {Visser}, \bibfnamefont
  {Matt}}} (\bibinfo {year} {1998}),\ \bibfield  {title} {\enquote {\bibinfo
  {title} {{Mass for the graviton}},}\ }\href {\doibase
  10.1023/A:1026611026766} {\bibfield  {journal} {\bibinfo  {journal} {Gen.
  Rel. Grav.}\ }\textbf {\bibinfo {volume} {30}},\ \bibinfo {pages}
  {1717--1728}},\ \Eprint {http://arxiv.org/abs/gr-qc/9705051}
  {arXiv:gr-qc/9705051 [gr-qc]} \BibitemShut {NoStop}%
%%CITATION = GR-QC/9705051;%%
\bibitem [{\citenamefont {Volkov}(2015)}]{Volkov:2014ooa}%
  \BibitemOpen
  \bibfield  {author} {\bibinfo {author} {\bibnamefont {Volkov}, \bibfnamefont
  {Mikhail~S}}} (\bibinfo {year} {2015}),\ \bibfield  {title} {\enquote
  {\bibinfo {title} {{Hairy black holes in theories with massive gravitons}},}\
  }\bibfield  {booktitle} {\emph {\bibinfo {booktitle} {{Proceedings of the 7th
  Aegean Summer School : Beyond Einstein's theory of gravity. Modifications of
  Einstein's Theory of Gravity at Large Distances.}}},\ }\href {\doibase
  10.1007/978-3-319-10070-8_6} {\bibfield  {journal} {\bibinfo  {journal}
  {Lect. Notes Phys.}\ }\textbf {\bibinfo {volume} {892}},\ \bibinfo {pages}
  {161--180}},\ \Eprint {http://arxiv.org/abs/1405.1742} {arXiv:1405.1742
  [hep-th]} \BibitemShut {NoStop}%
%%CITATION = ARXIV:1405.1742;%%
\bibitem [{\citenamefont {Weinberg}(2005)}]{Weinberg:1995mt}%
  \BibitemOpen
  \bibfield  {author} {\bibinfo {author} {\bibnamefont {Weinberg},
  \bibfnamefont {Steven}}} (\bibinfo {year} {2005}),\ \href@noop {} {\emph
  {\bibinfo {title} {{The Quantum theory of fields. Vol. 1: Foundations}}}}\
  (\bibinfo  {publisher} {Cambridge University Press})\BibitemShut {NoStop}%
%%CITATION = INSPIRE-406190;%%
\bibitem [{\citenamefont {Weisberg}\ and\ \citenamefont
  {Taylor}(2005)}]{Weisberg:2004hi}%
  \BibitemOpen
  \bibfield  {author} {\bibinfo {author} {\bibnamefont {Weisberg},
  \bibfnamefont {Joel~M}}, \ and\ \bibinfo {author} {\bibfnamefont {Joseph~H.}\
  \bibnamefont {Taylor}}} (\bibinfo {year} {2005}),\ \bibfield  {title}
  {\enquote {\bibinfo {title} {{Relativistic binary pulsar B1913+16: Thirty
  years of observations and analysis}},}\ }\bibfield  {booktitle} {\emph
  {\bibinfo {booktitle} {{Aspen Winter Conference on Astrophysics: Binary Radio
  Pulsars Aspen, Colorado, January 11-17, 2004}}},\ }\href@noop {} {\bibfield
  {journal} {\bibinfo  {journal} {ASP Conf. Ser.}\ }\textbf {\bibinfo {volume}
  {328}},\ \bibinfo {pages} {25}},\ \Eprint
  {http://arxiv.org/abs/astro-ph/0407149} {arXiv:astro-ph/0407149 [astro-ph]}
  \BibitemShut {NoStop}%
%%CITATION = ASTRO-PH/0407149;%%
\bibitem [{\citenamefont {Will}(1998)}]{Will:1997bb}%
  \BibitemOpen
  \bibfield  {author} {\bibinfo {author} {\bibnamefont {Will}, \bibfnamefont
  {Clifford~M}}} (\bibinfo {year} {1998}),\ \bibfield  {title} {\enquote
  {\bibinfo {title} {{Bounding the mass of the graviton using gravitational
  wave observations of inspiralling compact binaries}},}\ }\href {\doibase
  10.1103/PhysRevD.57.2061} {\bibfield  {journal} {\bibinfo  {journal} {Phys.
  Rev.}\ }\textbf {\bibinfo {volume} {D57}},\ \bibinfo {pages} {2061--2068}},\
  \Eprint {http://arxiv.org/abs/gr-qc/9709011} {arXiv:gr-qc/9709011 [gr-qc]}
  \BibitemShut {NoStop}%
%%CITATION = GR-QC/9709011;%%
\bibitem [{\citenamefont {Will}(2014)}]{Will:2014kxa}%
  \BibitemOpen
  \bibfield  {author} {\bibinfo {author} {\bibnamefont {Will}, \bibfnamefont
  {Clifford~M}}} (\bibinfo {year} {2014}),\ \bibfield  {title} {\enquote
  {\bibinfo {title} {{The Confrontation between General Relativity and
  Experiment}},}\ }\href {\doibase 10.12942/lrr-2014-4} {\bibfield  {journal}
  {\bibinfo  {journal} {Living Rev. Rel.}\ }\textbf {\bibinfo {volume} {17}},\
  \bibinfo {pages} {4}},\ \Eprint {http://arxiv.org/abs/1403.7377}
  {arXiv:1403.7377 [gr-qc]} \BibitemShut {NoStop}%
%%CITATION = ARXIV:1403.7377;%%
\bibitem [{\citenamefont {Will}\ and\ \citenamefont
  {Yunes}(2004)}]{Will:2004xi}%
  \BibitemOpen
  \bibfield  {author} {\bibinfo {author} {\bibnamefont {Will}, \bibfnamefont
  {Clifford~M}}, \ and\ \bibinfo {author} {\bibfnamefont {Nicolas}\
  \bibnamefont {Yunes}}} (\bibinfo {year} {2004}),\ \bibfield  {title}
  {\enquote {\bibinfo {title} {{Testing alternative theories of gravity using
  LISA}},}\ }\href {\doibase 10.1088/0264-9381/21/18/006} {\bibfield  {journal}
  {\bibinfo  {journal} {Class. Quant. Grav.}\ }\textbf {\bibinfo {volume}
  {21}},\ \bibinfo {pages} {4367}},\ \Eprint
  {http://arxiv.org/abs/gr-qc/0403100} {arXiv:gr-qc/0403100 [gr-qc]}
  \BibitemShut {NoStop}%
%%CITATION = GR-QC/0403100;%%
\bibitem [{\citenamefont {Williams}\ \emph {et~al.}(2004)\citenamefont
  {Williams}, \citenamefont {Turyshev},\ and\ \citenamefont
  {Boggs}}]{Williams:2004qba}%
  \BibitemOpen
  \bibfield  {author} {\bibinfo {author} {\bibnamefont {Williams},
  \bibfnamefont {James~G}}, \bibinfo {author} {\bibfnamefont {Slava~G.}\
  \bibnamefont {Turyshev}}, \ and\ \bibinfo {author} {\bibfnamefont {Dale~H.}\
  \bibnamefont {Boggs}}} (\bibinfo {year} {2004}),\ \bibfield  {title}
  {\enquote {\bibinfo {title} {{Progress in lunar laser ranging tests of
  relativistic gravity}},}\ }\href {\doibase 10.1103/PhysRevLett.93.261101}
  {\bibfield  {journal} {\bibinfo  {journal} {Phys.Rev.Lett.}\ }\textbf
  {\bibinfo {volume} {93}},\ \bibinfo {pages} {261101}},\ \Eprint
  {http://arxiv.org/abs/gr-qc/0411113} {arXiv:gr-qc/0411113 [gr-qc]}
  \BibitemShut {NoStop}%
%%CITATION = GR-QC/0411113;%%
\bibitem [{\citenamefont {Wu}\ and\ \citenamefont {Zhou}(2016)}]{Wu:2016jfw}%
  \BibitemOpen
  \bibfield  {author} {\bibinfo {author} {\bibnamefont {Wu}, \bibfnamefont
  {De-Jun}}, \ and\ \bibinfo {author} {\bibfnamefont {Shuang-Yong}\
  \bibnamefont {Zhou}}} (\bibinfo {year} {2016}),\ \bibfield  {title} {\enquote
  {\bibinfo {title} {{No hair theorem in quasi-dilaton massive gravity}},}\
  }\href {\doibase 10.1016/j.physletb.2016.04.016} {\bibfield  {journal}
  {\bibinfo  {journal} {Phys. Lett.}\ }\textbf {\bibinfo {volume} {B757}},\
  \bibinfo {pages} {324--329}},\ \Eprint {http://arxiv.org/abs/1601.04399}
  {arXiv:1601.04399 [hep-th]} \BibitemShut {NoStop}%
%%CITATION = ARXIV:1601.04399;%%
\bibitem [{\citenamefont {Wyman}(2011)}]{Wyman:2011mp}%
  \BibitemOpen
  \bibfield  {author} {\bibinfo {author} {\bibnamefont {Wyman}, \bibfnamefont
  {Mark}}} (\bibinfo {year} {2011}),\ \bibfield  {title} {\enquote {\bibinfo
  {title} {{Galilean-invariant scalar fields can strengthen gravitational
  lensing}},}\ }\href {\doibase 10.1103/PhysRevLett.106.201102} {\bibfield
  {journal} {\bibinfo  {journal} {Phys. Rev. Lett.}\ }\textbf {\bibinfo
  {volume} {106}},\ \bibinfo {pages} {201102}},\ \Eprint
  {http://arxiv.org/abs/1101.1295} {arXiv:1101.1295 [astro-ph.CO]} \BibitemShut
  {NoStop}%
%%CITATION = ARXIV:1101.1295;%%
\bibitem [{\citenamefont {Yagi}(2013)}]{Yagi:2013du}%
  \BibitemOpen
  \bibfield  {author} {\bibinfo {author} {\bibnamefont {Yagi}, \bibfnamefont
  {Kent}}} (\bibinfo {year} {2013}),\ \bibfield  {title} {\enquote {\bibinfo
  {title} {{Scientific Potential of DECIGO Pathfinder and Testing GR with
  Space-Borne Gravitational Wave Interferometers}},}\ }\href {\doibase
  10.1142/S0218271813410137} {\bibfield  {journal} {\bibinfo  {journal} {Int.
  J. Mod. Phys.}\ }\textbf {\bibinfo {volume} {D22}},\ \bibinfo {pages}
  {1341013}},\ \Eprint {http://arxiv.org/abs/1302.2388} {arXiv:1302.2388
  [gr-qc]} \BibitemShut {NoStop}%
%%CITATION = ARXIV:1302.2388;%%
\bibitem [{\citenamefont {Yagi}\ \emph
  {et~al.}(2014{\natexlab{a}})\citenamefont {Yagi}, \citenamefont {Blas},
  \citenamefont {Barausse},\ and\ \citenamefont {Yunes}}]{Yagi:2013ava}%
  \BibitemOpen
  \bibfield  {author} {\bibinfo {author} {\bibnamefont {Yagi}, \bibfnamefont
  {Kent}}, \bibinfo {author} {\bibfnamefont {Diego}\ \bibnamefont {Blas}},
  \bibinfo {author} {\bibfnamefont {Enrico}\ \bibnamefont {Barausse}}, \ and\
  \bibinfo {author} {\bibfnamefont {Nicolas}\ \bibnamefont {Yunes}}} (\bibinfo
  {year} {2014}{\natexlab{a}}),\ \bibfield  {title} {\enquote {\bibinfo {title}
  {{Constraints on Einstein-Aether theory and Horava gravity from binary pulsar
  observations}},}\ }\href {\doibase 10.1103/PhysRevD.90.069902,
  10.1103/PhysRevD.90.069901, 10.1103/PhysRevD.89.084067} {\bibfield  {journal}
  {\bibinfo  {journal} {Phys. Rev.}\ }\textbf {\bibinfo {volume}
  {D89}}~(\bibinfo {number} {8}),\ \bibinfo {pages} {084067}},\ \bibinfo {note}
  {[Erratum: Phys. Rev.D90,no.6,069901(2014)]},\ \Eprint
  {http://arxiv.org/abs/1311.7144} {arXiv:1311.7144 [gr-qc]} \BibitemShut
  {NoStop}%
%%CITATION = ARXIV:1311.7144;%%
\bibitem [{\citenamefont {Yagi}\ \emph
  {et~al.}(2014{\natexlab{b}})\citenamefont {Yagi}, \citenamefont {Blas},
  \citenamefont {Yunes},\ and\ \citenamefont {Barausse}}]{Yagi:2013qpa}%
  \BibitemOpen
  \bibfield  {author} {\bibinfo {author} {\bibnamefont {Yagi}, \bibfnamefont
  {Kent}}, \bibinfo {author} {\bibfnamefont {Diego}\ \bibnamefont {Blas}},
  \bibinfo {author} {\bibfnamefont {Nicolas}\ \bibnamefont {Yunes}}, \ and\
  \bibinfo {author} {\bibfnamefont {Enrico}\ \bibnamefont {Barausse}}}
  (\bibinfo {year} {2014}{\natexlab{b}}),\ \bibfield  {title} {\enquote
  {\bibinfo {title} {{Strong Binary Pulsar Constraints on Lorentz Violation in
  Gravity}},}\ }\href {\doibase 10.1103/PhysRevLett.112.161101} {\bibfield
  {journal} {\bibinfo  {journal} {Phys. Rev. Lett.}\ }\textbf {\bibinfo
  {volume} {112}}~(\bibinfo {number} {16}),\ \bibinfo {pages} {161101}},\
  \Eprint {http://arxiv.org/abs/1307.6219} {arXiv:1307.6219 [gr-qc]}
  \BibitemShut {NoStop}%
%%CITATION = ARXIV:1307.6219;%%
\bibitem [{\citenamefont {Yagi}\ and\ \citenamefont
  {Stein}(2016)}]{Yagi:2016jml}%
  \BibitemOpen
  \bibfield  {author} {\bibinfo {author} {\bibnamefont {Yagi}, \bibfnamefont
  {Kent}}, \ and\ \bibinfo {author} {\bibfnamefont {Leo~C.}\ \bibnamefont
  {Stein}}} (\bibinfo {year} {2016}),\ \bibfield  {title} {\enquote {\bibinfo
  {title} {{Black Hole Based Tests of General Relativity}},}\ }\href {\doibase
  10.1088/0264-9381/33/5/054001} {\bibfield  {journal} {\bibinfo  {journal}
  {Class. Quant. Grav.}\ }\textbf {\bibinfo {volume} {33}}~(\bibinfo {number}
  {5}),\ \bibinfo {pages} {054001}},\ \Eprint {http://arxiv.org/abs/1602.02413}
  {arXiv:1602.02413 [gr-qc]} \BibitemShut {NoStop}%
%%CITATION = ARXIV:1602.02413;%%
\bibitem [{\citenamefont {Yagi}\ and\ \citenamefont
  {Tanaka}(2010{\natexlab{a}})}]{Yagi:2009zm}%
  \BibitemOpen
  \bibfield  {author} {\bibinfo {author} {\bibnamefont {Yagi}, \bibfnamefont
  {Kent}}, \ and\ \bibinfo {author} {\bibfnamefont {Takahiro}\ \bibnamefont
  {Tanaka}}} (\bibinfo {year} {2010}{\natexlab{a}}),\ \bibfield  {title}
  {\enquote {\bibinfo {title} {{Constraining alternative theories of gravity by
  gravitational waves from precessing eccentric compact binaries with LISA}},}\
  }\href {\doibase 10.1103/PhysRevD.81.109902, 10.1103/PhysRevD.81.064008}
  {\bibfield  {journal} {\bibinfo  {journal} {Phys. Rev.}\ }\textbf {\bibinfo
  {volume} {D81}},\ \bibinfo {pages} {064008}},\ \bibinfo {note} {[Erratum:
  Phys. Rev.D81,109902(2010)]},\ \Eprint {http://arxiv.org/abs/0906.4269}
  {arXiv:0906.4269 [gr-qc]} \BibitemShut {NoStop}%
%%CITATION = ARXIV:0906.4269;%%
\bibitem [{\citenamefont {Yagi}\ and\ \citenamefont
  {Tanaka}(2010{\natexlab{b}})}]{Yagi:2009zz}%
  \BibitemOpen
  \bibfield  {author} {\bibinfo {author} {\bibnamefont {Yagi}, \bibfnamefont
  {Kent}}, \ and\ \bibinfo {author} {\bibfnamefont {Takahiro}\ \bibnamefont
  {Tanaka}}} (\bibinfo {year} {2010}{\natexlab{b}}),\ \bibfield  {title}
  {\enquote {\bibinfo {title} {{DECIGO/BBO as a probe to constrain alternative
  theories of gravity}},}\ }\href {\doibase 10.1143/PTP.123.1069} {\bibfield
  {journal} {\bibinfo  {journal} {Prog. Theor. Phys.}\ }\textbf {\bibinfo
  {volume} {123}},\ \bibinfo {pages} {1069--1078}},\ \Eprint
  {http://arxiv.org/abs/0908.3283} {arXiv:0908.3283 [gr-qc]} \BibitemShut
  {NoStop}%
%%CITATION = ARXIV:0908.3283;%%
\bibitem [{\citenamefont {Yunes}\ \emph {et~al.}(2016)\citenamefont {Yunes},
  \citenamefont {Yagi},\ and\ \citenamefont {Pretorius}}]{Yunes:2016jcc}%
  \BibitemOpen
  \bibfield  {author} {\bibinfo {author} {\bibnamefont {Yunes}, \bibfnamefont
  {Nicolas}}, \bibinfo {author} {\bibfnamefont {Kent}\ \bibnamefont {Yagi}}, \
  and\ \bibinfo {author} {\bibfnamefont {Frans}\ \bibnamefont {Pretorius}}}
  (\bibinfo {year} {2016}),\ \bibfield  {title} {\enquote {\bibinfo {title}
  {{Theoretical Physics Implications of the Binary Black-Hole Mergers GW150914
  and GW151226}},}\ }\href {\doibase 10.1103/PhysRevD.94.084002} {\bibfield
  {journal} {\bibinfo  {journal} {Phys. Rev.}\ }\textbf {\bibinfo {volume}
  {D94}}~(\bibinfo {number} {8}),\ \bibinfo {pages} {084002}},\ \Eprint
  {http://arxiv.org/abs/1603.08955} {arXiv:1603.08955 [gr-qc]} \BibitemShut
  {NoStop}%
%%CITATION = ARXIV:1603.08955;%%
\bibitem [{\citenamefont {Zakharov}\ \emph {et~al.}(2016)\citenamefont
  {Zakharov}, \citenamefont {Jovanovic}, \citenamefont {Borka},\ and\
  \citenamefont {Jovanovic}}]{Zakharov:2016lzv}%
  \BibitemOpen
  \bibfield  {author} {\bibinfo {author} {\bibnamefont {Zakharov},
  \bibfnamefont {Alexander~F}}, \bibinfo {author} {\bibfnamefont {Predrag}\
  \bibnamefont {Jovanovic}}, \bibinfo {author} {\bibfnamefont {Dusko}\
  \bibnamefont {Borka}}, \ and\ \bibinfo {author} {\bibfnamefont {Vesna~Borka}\
  \bibnamefont {Jovanovic}}} (\bibinfo {year} {2016}),\ \bibfield  {title}
  {\enquote {\bibinfo {title} {{Constraining the range of Yukawa gravity
  interaction from S2 star orbits II: Bounds on graviton mass}},}\ }\href
  {\doibase 10.1088/1475-7516/2016/05/045} {\bibfield  {journal} {\bibinfo
  {journal} {JCAP}\ }\textbf {\bibinfo {volume} {1605}}~(\bibinfo {number}
  {05}),\ \bibinfo {pages} {045}},\ \Eprint {http://arxiv.org/abs/1605.00913}
  {arXiv:1605.00913 [gr-qc]} \BibitemShut {NoStop}%
%%CITATION = ARXIV:1605.00913;%%
\bibitem [{\citenamefont {Zakharov}(1970)}]{Zakharov:1970cc}%
  \BibitemOpen
  \bibfield  {author} {\bibinfo {author} {\bibnamefont {Zakharov},
  \bibfnamefont {V~I}}} (\bibinfo {year} {1970}),\ \bibfield  {title} {\enquote
  {\bibinfo {title} {{Linearized gravitation theory and the graviton mass}},}\
  }\href@noop {} {\bibfield  {journal} {\bibinfo  {journal} {JETP Lett.}\
  }\textbf {\bibinfo {volume} {12}},\ \bibinfo {pages} {312}},\ \bibinfo {note}
  {[Pisma Zh. Eksp. Teor. Fiz.12,447(1970)]}\BibitemShut {NoStop}%
%%CITATION = JTPLA,12,312;%%
\bibitem [{\citenamefont {Zu}\ \emph {et~al.}(2014)\citenamefont {Zu},
  \citenamefont {Weinberg}, \citenamefont {Jennings}, \citenamefont {Li},\ and\
  \citenamefont {Wyman}}]{Zu:2013joa}%
  \BibitemOpen
  \bibfield  {author} {\bibinfo {author} {\bibnamefont {Zu}, \bibfnamefont
  {Ying}}, \bibinfo {author} {\bibfnamefont {D.~H.}\ \bibnamefont {Weinberg}},
  \bibinfo {author} {\bibfnamefont {Elise}\ \bibnamefont {Jennings}}, \bibinfo
  {author} {\bibfnamefont {Baojiu}\ \bibnamefont {Li}}, \ and\ \bibinfo
  {author} {\bibfnamefont {Mark}\ \bibnamefont {Wyman}}} (\bibinfo {year}
  {2014}),\ \bibfield  {title} {\enquote {\bibinfo {title} {{Galaxy Infall
  Kinematics as a Test of Modified Gravity}},}\ }\href {\doibase
  10.1093/mnras/stu1739} {\bibfield  {journal} {\bibinfo  {journal} {Mon. Not.
  Roy. Astron. Soc.}\ }\textbf {\bibinfo {volume} {445}}~(\bibinfo {number}
  {2}),\ \bibinfo {pages} {1885--1897}},\ \Eprint
  {http://arxiv.org/abs/1310.6768} {arXiv:1310.6768 [astro-ph.CO]} \BibitemShut
  {NoStop}%
%%CITATION = ARXIV:1310.6768;%%
\end{thebibliography}%

\end{document}